\UseRawInputEncoding
\documentclass[11pt]{article}

\usepackage{amssymb,amsmath,amsthm,amsfonts,dsfont}
\usepackage[numbers,comma,sort&compress]{natbib}
\usepackage[colorlinks,hypertexnames=false]{hyperref}
\usepackage[capitalise]{cleveref}
\usepackage[all]{hypcap}
\usepackage{url}
\usepackage{graphicx}
\usepackage[font=small]{caption}
\usepackage[subrefformat=parens,labelformat=parens]{subcaption}
\usepackage{float}
\usepackage{footnote}
\usepackage{enumitem}
\usepackage[margin=1in]{geometry}

\newcommand{\abs}[1]{\left\lvert#1\right\rvert}
\newcommand{\norm}[1]{\left\lVert#1\right\rVert}
\DeclareMathOperator{\poly}{poly}
\DeclareMathOperator{\polylog}{polylog}

\newtheorem{theorem}{Theorem}

\newtheorem{lemma}{Lemma}

\newtheorem{corollary}{Corollary}

\newcommand{\eq}[1]{\cref{eq:#1}}
\newcommand{\thm}[1]{\hyperref[thm:#1]{Theorem~\ref*{thm:#1}}}
\newcommand{\defn}[1]{\hyperref[defn:#1]{Definition~\ref*{defn:#1}}}
\newcommand{\lem}[1]{\hyperref[lem:#1]{Lemma~\ref*{lem:#1}}}
\newcommand{\prop}[1]{\hyperref[prop:#1]{Proposition~\ref*{prop:#1}}}
\newcommand{\fig}[1]{\hyperref[fig:#1]{Figure~\ref*{fig:#1}}}
\newcommand{\tab}[1]{\hyperref[tab:#1]{Table~\ref*{tab:#1}}}
\renewcommand{\sec}[1]{\hyperref[sec:#1]{Section~\ref*{sec:#1}}}
\newcommand{\append}[1]{\hyperref[append:#1]{Appendix~\ref*{append:#1}}}
\newcommand{\cor}[1]{\hyperref[cor:#1]{Corollary~\ref*{cor:#1}}}
\newcommand{\obs}[1]{\hyperref[obs:#1]{Observation~\ref*{obs:#1}}}

\newcommand{\ket}[1]{|#1\rangle}
\newcommand{\bra}[1]{\langle#1|}
\newcommand{\ketbra}[2]{\ket{#1}\!\bra{#2}}

\newcommand\blfootnote[1]{%
	\begingroup
	\renewcommand\thefootnote{}\footnote{#1}%
	\addtocounter{footnote}{-1}%
	\endgroup
}

\usepackage{booktabs}
\usepackage{diagbox}
\DeclareMathOperator{\vol}{vol}
\newcommand{\vertiii}[1]{{\left\vert\kern-0.25ex\left\vert\kern-0.25ex\left\vert #1
		\right\vert\kern-0.25ex\right\vert\kern-0.25ex\right\vert}}
\newcommand{\acommtilde}{\beta_{\mathrm{comm}}}
\newcommand{\floor}[1]{\lfloor{#1}\rfloor}
\newcommand{\ceil}[1]{\lceil{#1}\rceil}
\usepackage{tikz}
\DeclareMathOperator{\cost}{cost}
\DeclareMathOperator{\rank}{rank}
\DeclareMathOperator{\im}{im}
\usetikzlibrary{quantikz}

\newtheorem{innercustomthm}{Theorem}
\newenvironment{customthm}[1]
{\renewcommand\theinnercustomthm{#1}\innercustomthm}
{\endinnercustomthm}

\newtheorem{innercustomcor}{Corollary}
\newenvironment{customcor}[1]
{\renewcommand\theinnercustomcor{#1}\innercustomcor}
{\endinnercustomcor}

\definecolor{tumbleweed}{rgb}{0.87, 0.67, 0.53}

\DeclareFontFamily{U}{matha}{\hyphenchar\font45}
\DeclareFontShape{U}{matha}{m}{n}{
	<5> <6> <7> <8> <9> <10> gen * matha
	<10.95> matha10 <12> <14.4> <17.28> <20.74> <24.88> matha12
}{}
\DeclareSymbolFont{matha}{U}{matha}{m}{n}
\DeclareFontFamily{U}{mathx}{\hyphenchar\font45}
\DeclareFontShape{U}{mathx}{m}{n}{
	<5> <6> <7> <8> <9> <10>
	<10.95> <12> <14.4> <17.28> <20.74> <24.88>
	mathx10
}{}
\DeclareSymbolFont{mathx}{U}{mathx}{m}{n}
\DeclareMathSymbol{\obot}         {2}{matha}{"6B}
\DeclareMathSymbol{\bigobot}       {1}{mathx}{"CB}

\usepackage{multirow}

\begin{document}
\title{\huge On the complexity of implementing Trotter steps}
\author
{Guang Hao Low,$^{1}$ Yuan Su,$^{1}$ 
	Yu Tong,$^{2,3}$ and Minh C.\ Tran$^{4,5}$\\
}
\date{\vspace{-10mm}}
\maketitle

\blfootnote{This is a slightly enhanced version of the article entitled \emph{Complexity of Implementing Trotter Steps} published in PRX Quantum \textbf{4} (2023), 020323.}
\blfootnote{$^{1}$Microsoft Quantum, Redmond, WA 98052, USA}
\blfootnote{$^{2}$Department of Mathematics, University of California, Berkeley, CA 94720, USA}
\blfootnote{$^{3}$Institute for Quantum Information and Matter, California Institute of Technology, Pasadena, CA 91125, USA}
\blfootnote{$^{4}$Center for Theoretical Physics, Massachusetts Institute of Technology, Cambridge, MA 02139, USA}
\blfootnote{$^{5}$Department of Physics, Harvard University, Cambridge, MA 02138, USA}

\begin{abstract}
Quantum dynamics can be simulated on a quantum computer by exponentiating elementary terms from the Hamiltonian in a sequential manner. However, such an implementation of Trotter steps has gate complexity depending on the total Hamiltonian term number, comparing unfavorably to algorithms using more advanced techniques.

We develop methods to perform faster Trotter steps with gate complexity sublinear in the number of terms. 
We focus on a class of $2$-local Hamiltonians in one spatial dimension whose interaction strength decays with distance $x$ according to power law $1/x^\alpha$ ($\alpha>0$), although we discuss extensions to higher spatial dimensions as well. 
Naively, Trotter steps for $n$-qubit power-law interactions require $\Omega(n^2)$ gates to implement.

Our first method is based on block encodings of power-law systems with efficiently computable coefficients. While block encodings typically slow down quantum simulation by a factor proportional to the $1$-norm of Hamiltonian coefficients, we overcome this barrier using a recursive Trotter decomposition to effectively reduce the norm. We obtain further improvements by simulating commuting Hamiltonian terms with an average combination cost. The resulting complexity is almost linear in the spacetime volume for $\alpha\geq 2$ and improves the state of the art for $\alpha<2$. 

We also show that Trotter steps can be implemented more efficiently when certain blocks of Hamiltonian coefficients exhibit low-rank properties. In particular, 
using a recursive low-rank decomposition,
we show that power-law Hamiltonians can be simulated with gate complexity nearly linear in the spacetime volume for all $\alpha\geq1$.

We apply our methods to simulate electronic structure Hamiltonians in second quantization in real space. Combining with a tighter error analysis, we show that it suffices to use $\left(\eta^{1/3}n^{1/3}+\frac{n^{2/3}}{\eta^{2/3}}\right)n^{1+o(1)}$ gates and $\mathcal{O}\left(\log(n)\right)$ ancilla qubits to simulate uniform electron gas in real space with $n$ spin orbitals and $\eta$ electrons, asymptotically improving the best results from previous work. We obtain an analogous result when the external potential of nuclei is introduced under the Born-Oppenheimer approximation.

We prove a circuit lower bound when the Hamiltonian coefficients take a continuum range of values. 
Specifically, we construct a class of $2$-local Hamiltonians with commuting terms that requires at least $\Omega(n^2)$ gates to evolve with accuracy $\epsilon=\Omega(1/\poly(n))$ for time $t=\Omega(\epsilon)$.
Our proof is based on a gate-efficient reduction from the approximate synthesis of diagonal unitaries within the Hamming weight-$2$ subspace, which may be of independent interest.

Our result thus suggests the use of Hamiltonian structural properties as both necessary and sufficient to implement Trotter steps with lower gate complexity. 
\end{abstract}

\newpage
{
	\thispagestyle{empty}
	\clearpage\tableofcontents
	\thispagestyle{empty}
}
\newpage

\section{Introduction}
\label{sec:intro}
Many-body Hamiltonians can be efficiently simulated on digital quantum computers using either product formulas (such as the Lie-Trotter-Suzuki formulas) or more advanced simulation algorithms. While short steps of product formulas (known as \emph{Trotter steps}) typically require an implementation cost proportional to the number of Hamiltonian terms, a host of techniques have been developed in recent years to implement other quantum simulation algorithms with significantly reduced complexities.

The purpose of this work is to develop methods for performing Trotter steps that go beyond the sequential circuit implementation mentioned above, leading to faster quantum simulation algorithms.
We focus on $2$-local Hamiltonians with power-law decaying interactions---all-to-all interactions whose strengths decay with distance $x$ according to a power law $1/x^\alpha$ ($\alpha>0$)---for concreteness. 
Many systems of physical relevance can be modeled by power-law interactions, such as trapped ions, Rydberg atoms, ultracold atoms and molecules, nitrogen-vacancy centers, and superconducting systems.

We first develop a block-encoding-based method to simulate power-law Hamiltonians with efficiently computable coefficients. While a block encoding introduces a slow-down factor proportional to the $1$-norm of Hamiltonian coefficients which is large for power-law interactions, we overcome this barrier by recursively decomposing the system using product formulas to effectively reduce the norm. We obtain further improvements by simulating commuting Hamiltonian terms with an average combination cost. This gives simulations with complexities almost linear in the spacetime volume for $\alpha\geq 2$ and improved scalings for $\alpha<2$. We also implement faster Trotter steps when certain off-diagonal blocks of Hamiltonian coefficients exhibit low-rank properties. In particular, we achieve a nearly linear spacetime volume scaling for all $\alpha\geq 1$ using a recursive low-rank decomposition of the Hamiltonian.
We extend our recursion techniques to $\kappa$-local Hamiltonians and more general fermionic models. In these cases, the exact complexity is determined by the tensor structure of the Hamiltonian coefficients and depends on the problem at hand.

We apply our methods to simulate electronic structure Hamiltonians in second quantization in real space. Combining with a tight Trotter error analysis, we show that $\left(\eta^{1/3}n^{1/3}+\frac{n^{2/3}}{\eta^{2/3}}\right)n^{1+o(1)}$ gates suffice to simulate uniform electron gas with $n$ spin orbitals and $\eta$ electrons, improving the best results from previous work. An analogous result holds when the external potential of nuclei is included from the Born-Oppenheimer approximation.

Performing faster Trotter steps for Hamiltonians with arbitrary coefficients is a challenging task in general. To confirm this intuition, we prove a gate-complexity lower bound. Specifically, we construct a class of $n$-qubit $2$-local Hamiltonians with commuting Pauli-$Z$ terms whose coefficients take a continuum range of values. We show that these Hamiltonians require at least $\Omega(n^2)$ gates to simulate with accuracy $\epsilon=\Omega(1/\poly(n))$ for time $t=\Omega(\epsilon)$; thus the best method one can hope for is to sequentially exponentiate all the $\Theta(n^2)$ terms. Our proof depends on a gate-efficient reduction from the approximate synthesis of diagonal unitaries within the Hamming weight-$2$ subspace, which we then address by adapting a volume-comparison technique from previous work. Our result thus suggests the use of structural properties of the target Hamiltonian as both necessary and sufficient to achieve lower gate complexity for implementing Trotter steps.

\subsection{Quantum algorithms for quantum simulation}
\label{sec:intro_algs}
Simulating many-body physical systems is one of the most promising applications of digital quantum computers. Indeed, the idea of quantum computing as originally proposed by Feynman~\cite{Fey82}, Manin~\cite{Manin80} and others is strongly motivated by quantum simulation. Efficient quantum simulations can be used to extract statical and dynamical properties of physical systems, which has potential applications in various areas such as condensed matter physics~\cite{CMNRS18}, chemistry~\cite{mcardle2020quantum,cao2019quantum,MottaEmerging2022}, and high-energy physics~\cite{HighEnergy22}. Meanwhile, recent developments in quantum simulation algorithms have also provided technical tools that influenced the design of other quantum algorithms~\cite{HHL09,AL19,LT20,GSLW19,chakraborty19} and proofs of other results in areas beyond quantum computing~\cite{anshu2020communication,kuwahara2021improved,Tran18,AreaLaw22,HuangTong22}.

There are many quantum algorithms one can use to perform quantum simulation. At a high level, these algorithms can be categorized according to their default input models. Common Hamiltonian input models include: (i) Linear Combinations of Hermitians (LCH), where the target Hamiltonian takes the form
\begin{equation}
    H=\sum_{\gamma=1}^\Gamma H_\gamma,
\end{equation}
with $H_\gamma$ Hermitian and the exponentials $e^{-itH_\gamma}$ implementable on a quantum computer. Simulation algorithms that work in this input model include one based on product formulas~\cite{Lloyd96,BACS05} (such as the Lie-Trotter formula and its higher-order extensions), as well as a more recent algorithm based on random sampling~\cite{Campbell18}; and (ii) Linear Combinations of Unitaries (LCU), where the target Hamiltonian takes the form
\begin{equation}
    H=\sum_{\gamma=1}^\Gamma \beta_\gamma U_\gamma,
\end{equation}
with $U_\gamma$ unitary, $\beta_\gamma>0$ and the controlled operators $\ketbra{0}{0}\otimes I+\ketbra{1}{1}\otimes U_\gamma$ implementable on a quantum computer. There are also various algorithms working in the LCU model, including one based on implementing truncated Taylor series~\cite{BCCKS14} and one based on qubitization~\cite{LC16}. For our purpose, we will mainly focus on the algorithm based on product formulas as well as the qubitization algorithm, which we review in more detail in \sec{prelim_pf} and \sec{prelim_qubitization}.

Naturally, there is no silver bullet method that solves all simulation problems of interest with the optimal gate complexity. Choices of algorithms should thus be made on a case-by-case basis. There are in fact a few common desirable features shared among many algorithms mentioned above. For instance, it has been well known that many LCU approaches such as the Taylor-series algorithm and qubitization have complexities (nearly) linear in the simulation time and logarithmic in the inverse accuracy. But similar scalings can be achieved using product formulas as well by taking linear combinations of formulas with different step sizes and repetition numbers~\cite{LKW19}. LCU-type algorithms also have the appealing feature that they can be used to design other functions of Hamiltonians~\cite{LC17b}, which applies to problems beyond quantum simulation~\cite{GSLW19,Martyn21} such as solving linear systems of equations~\cite{HHL09,Low18}, preparing ground states~\cite{Ge19,LT20,PW09}, and performing phase estimation~\cite{Rall2021fastercoherent,Martyn21}. Again, these problems can be well solved using algorithms in the LCH model (and can sometimes be more resource friendly) as demonstrated in recent work such as~\cite{LinTong22,DongLinTong22}.

However, there is one question concerning the gate complexities of implementing these approaches which has not been satisfactorily answered so far. View a sufficiently long-time Hamiltonian evolution as a concatenation of short-time steps, and consider the complexity of simulating each short evolution. Then a host of techniques are recently developed to implement LCU approaches with cost depending on the target system size as opposed to the number of Hamiltonian terms~\cite{Lee21,vonBurg21}; a similar goal may be realized for the randomized method via an importance sampling~\cite{Campbell18,CHRT21,WBC22}. By contrast, how to provably achieve a similar complexity for product formulas was somewhat under investigated: we review related work on the implementation of Trotter steps in \sec{intro_previous}.

For an $n$-qubit system with $2$-local interactions, one has that the total number of terms scales like $\Theta(n^2)$ but the system size is only linear in $n$. This comparison becomes $\Theta(n^\kappa)$ versus $n$ for a $\kappa$-local Hamiltonian. In light of this gap, it is natural to ask when and how one can implement product formulas with cost scaling better than the sequential method. We address this question by presenting necessary and sufficient conditions under which faster Trotter steps are possible. As an application, we give a simulation algorithm for electronic structure Hamiltonians in second quantization in real space with complexity $\left(\eta^{1/3}n^{1/3}+\frac{n^{2/3}}{\eta^{2/3}}\right)n^{1+o(1)}$---asymptotically the fastest real-space simulation to date.

\subsection{Previous related work}
\label{sec:intro_previous}
We now discuss prior work on implementing Trotter steps that are relevant to our paper.

First, there are previous studies on the so-called fast-forwardability of Hamiltonian evolution~\cite{Atia2017,Gu2021fastforwarding,FixedDepth22,AlgebraicCompression22}. Some of those techniques, such as introducing efficiently computable phases and diagonalizing quadratic Hamiltonians, can also be used to perform Trotter steps. However, their fundamental goal is quite different from ours. In quantum fast-forwarding, one is asked to simulate the target Hamiltonian for a sufficiently long time, and the goal is to reduce the scaling of time in the gate complexity (potentially at the cost of increasing the system-size scaling). In contrast, each Trotter step only approximates the ideal evolution for a short time, and so the scaling with time is no longer a key contribution to the complexity. Instead, our main goal here is to reduce the dependence on the size of the simulated system.

For electronic structure models represented under arbitrary basis, the number of terms in the Hamiltonian typically scales like $\Theta(n^4)$ for $n$ spin orbitals, so a sequential implementation of Trotter steps would have cost scaling $\Theta(n^4)$. To address this, recent work developed quantum circuits based on low-rank factorizations of such systems~\cite{Pou15,MYMLMBC18} (see~\cite{RLB21,Anand22} for more recent developments of such a method). Specifically, they apply product formulas to decompose the Hamiltonian into multiple components, each of which has coefficients with certain low-rank properties and can be further implemented by diagonalization. The gate complexity of the resulting circuits would then depend on the value of rank as opposed to the number of Hamiltonian terms, which significantly reduces the cost per time step. However, those work did not rigorously analyze the total complexity of the proposed methods, and it is unclear how much overall advantage their approach can offer. In fact, their factorization does not seem to preserve the commutation relations between Hamiltonian terms and could potentially introduce a Trotter error larger than the sequential approach (so more Trotter steps would be required to reach the same simulation accuracy).

Another related approach to reducing the complexity of quantum simulation is to truncate Hamiltonian terms of small sizes. Such truncations are useful for not only performing Trotter steps~\cite{CSTWZ19,XSSS20,Clinton22}, but also implementing more advanced quantum simulation algorithms~\cite{Meister2022tailoringterm,Berry2019qubitizationof,Ivanov22}. Generally speaking, the error introduced in the truncation will grow linearly with time, so the simulation is accurate only when the evolution is sufficiently short. In particular, for rapidly decaying power-law interactions with exponent $\alpha>2$, a truncation is possible only when $t=\mathcal{O}\left(n^{(\alpha-2)/(\alpha-1)}\right)$~\cite{CSTWZ19}. For simulations of chemistry and material models, truncation thresholds can often be determined empirically under certain assumptions of the model Hamiltonians.

Here, our work considers simulating $2$-local Hamiltonians with interaction strength decaying according to power law, and we study the cost of implementing one short Trotter step as well as the entire long-time simulation, using the so-called block-encoding technique and recursive/hierarchical low-rank decompositions \cite{Hackbusch1999,Grasedyck2003}. Our motivation for using the low-rank decomposition partly overlaps with that of a recent work by Nguyen, Kiani, and Lloyd~\cite{NKL22}, but the main problems we study are different. Instead of Hamiltonian simulation, they studied the block encoding of kernel matrices of the form
\begin{equation}
    K=\sum_{j,k=1}^n\beta_{j,k}\ketbra{j}{k},
\end{equation}
where $\beta_{j,k}$ can be power-law functions such as $1/|j-k|^\alpha$. The matrix $K$ is an $n$-dimensional operator and has spectral norm $\norm{K}=\norm{\beta}=\Theta(1)$ for $\alpha>1$.\footnotemark\
Instead, our problem centers around the simulation of 
\begin{equation}
    H=\sum_{j,k=1}^n\beta_{j,k}X_jY_k,
\end{equation}
which is a $2^n$-dimensional operator and has spectral norm generally scaling with the \emph{vector $1$-norm} of coefficients: $\norm{H}=\Theta(\norm{\beta}_1)=\Theta(n)$ for $\alpha>1$.
Thus, a naive block encoding of our $H$ will have an intrinsically worse normalization factor than that of their $K$; see \sec{prelim_qubitization} for further explanations of how such normalization factors affect the complexity of quantum simulation. Nevertheless, we overcome this technical obstacle by recursively decomposing the Hamiltonian using product formulas, which significantly reduces the $1$-norm while maintaining the overall scaling of gate count.
\footnotetext{Roughly speaking, they are implementing operators in first quantization. One may attempt to apply a similar approach to the Coulomb interaction to improve the electronic structure simulation. We are not aware of a simple realization of this idea. Existing study of this problem uses a computation model stronger than the circuit model~\cite{CLLLZ22}.}

Finally, we note that there is a large body of previous work analyzing and optimizing the concrete resources for implementing Trotter steps, for both near-term and fault-tolerant quantum computers (see Refs.~\cite{CMNRS18,kivlichan2020improved,Shaw2020quantumalgorithms} as well as other work citing and cited by these papers). We have not attempted to optimize the constant factors of the complexity of our methods, but we consider such optimizations to be an interesting subject for future investigation.

\subsection{Faster Trotter steps by recursion}
\label{sec:intro_step}
Consider a Hamiltonian $H$ with $\Gamma$ terms $H=\sum_{\gamma=1}^\Gamma H_\gamma$. If each Hamiltonian term can be exponentiated on a quantum computer with cost $\mathcal{O}(1)$, then one can simulate the evolution of $H$ for a short time using product formulas, and the complexity would scale like $\mathcal{O}(\Gamma)$. We will identify scenarios in which improved implementations of Trotter steps are possible with gate complexities sublinear in $\Gamma$.

We focus on a class of $2$-local Hamiltonians in one spatial dimension with all-to-all interactions and magnitude of the coefficients decaying with distance $x$ according to power law $1/x^\alpha\ (\alpha>0)$. We describe how our results can be extended to higher spatial dimensions in \append{generalize}, and to more general local and fermionic models (though the amount of improvement largely depends on the tensor structure of the Hamiltonian coefficients which has not been fully understood). We restrict to power-law models because product formulas are known to provide the fastest method for simulating this class of Hamiltonians, so we can directly compare our result with the state of the art. 
Examples of power-law interactions include the Coulomb interaction between charged particles and the dipole-dipole interaction between molecules, both of which are ubiquitous in quantum chemistry---a primary target application of quantum computation.
In physics, impressive controls in recent experiments with trapped ions~\cite{Britton2012,Kim2011}, Rydberg atoms~\cite{Saffman10}, and ultracold atoms and polar molecules~\cite{Douglas2015,Yan2013} have enabled the possibility to study new phases of matter with power-law interactions~\cite{PhysRevLett.109.267203,PhysRevLett.98.060404,PhysRevLett.101.073201,PhysRevLett.105.140401,PhysRevA.76.043604,PhysRevLett.125.010401} and contributed to a growing interest in simulating such systems.
In fact, we will describe a direct application of our method in \sec{app} for faster simulations of electronic structure Hamiltonians in real space.

Assume that the coefficients of the target Hamiltonian are efficiently computable. As explained above, there have been a host of techniques developed recently based on the notion of block encoding, which enables simulation with complexities depending only on the system size. One may ask if these techniques also lead to faster Trotter steps with a similar cost scaling. Unfortunately, the answer is negative in general. This is because a block encoding typically introduces a normalization factor proportional to the $1$-norm of the Hamiltonian coefficients, and we thus need to repeat a corresponding number of times to perform Trotter steps. For instance, one can block encode power-law Hamiltonians with gate complexity $\Theta(n)$, but this introduces a normalization factor proportional to the $1$-norm which is generally $\Theta(n)$ for power-law interactions with $\alpha>1$. Meanwhile, a Trotter step for the power-law models has an almost constant evolution time. So one roughly needs a total number of
\begin{equation}
    \underbrace{n}_{\text{block encoding}}\cdot\underbrace{n}_{\text{effective time}}=\underbrace{n^2}_{\text{Trotter step complexity}}
\end{equation}
gates to implement a single Trotter step, which has no benefit over the sequential implementation. See \sec{block_norm} for a more detailed explanation of this issue.\footnote{The fractional-query algorithm also implements a compressed version of Trotter steps using more advanced simulation techniques~\cite{FractionalQuery14}. That algorithm is superseded by the block-encoding method~\cite{kothari2014efficient}, so it suffers from the same normalization-factor issue pointed out here.}

We develop a method based on block encoding that overcomes the above technical issue. The key observation is that product formulas can be used to reduce the $1$-norm of Hamiltonian coefficients ``almost for free'': we apply product formulas to recursively decompose the Hamiltonian into multiple groups, but such a coarse-grained decomposition introduces a Trotter error no larger than the sequential approach. We choose the decomposition to significantly reduce the $1$-norm of each group while maintaining the overall scaling of the gate complexity, giving an efficient block-encoding circuit. The resulting simulation has gate complexity $(nt)^{1+o(1)}$ when $\alpha\geq 2$, and $n^{3-\alpha+o(1)}t^{1+o(1)}$ when $\alpha<2$. We formally state this theorem as \thm{block} in \sec{block} and preview it below.

\begin{customthm}{}[Faster Trotter steps using block encoding]
Consider $2$-local Hamiltonians $H=\sum_{\sigma,\sigma'\in\{i,x,y,z\}}\sum_{1\leq j<k\leq n}\beta^{(\sigma,\sigma')}_{j,k}P_j^{(\sigma)}P_k^{(\sigma')}$, where $\abs{\beta^{(\sigma,\sigma')}_{j,k}}\leq 1/|j-k|^\alpha$ for some constant $\alpha>0$ and $P^{(\sigma)}$ ($\sigma=i,x,y,z$) are the identity and Pauli matrices. Let $t>0$ be the simulation time and $\epsilon>0$ be the target accuracy. Assume that the coefficient oracle
\begin{equation}
    O_{\beta,\sigma,\sigma'}\ket{j,k,0}=\ket{j,k,\beta_{j,k}^{(\sigma,\sigma')}}
\end{equation}
can be implemented with gate complexity $\mathcal{O}\left(\polylog(nt/\epsilon)\right)$.
Then $H$ can be simulated using the algorithm of \sec{block_summary} with $\mathcal{O}\left(\log(nt/\epsilon)\right)$ ancilla qubits and gate complexity
\begin{equation}
    \begin{cases}
        nt\left(\frac{nt}{\epsilon}\right)^{o(1)},\quad&\alpha\geq2,\\
        n^{3-\alpha}t\left(\frac{nt}{\epsilon}\right)^{o(1)},&0<\alpha<2.
    \end{cases}
\end{equation}
\end{customthm}

The Hamiltonian decomposition we study has an additional feature that all the terms within each group commute with each other. Such terms can be simulated from block encodings of their subcomponents with an average combination cost, similar to the interaction-picture simulation~\cite{LW18,Low18} but without an exponentially growing factor. We leverage this observation to further improve the complexity to $n^{2-\alpha/2+o(1)}t^{1+o(1)}$ for $1\leq\alpha<2$ and $n^{5/2-\alpha+o(1)}t^{1+o(1)}$ for $\alpha<1$. We summarize our result in \thm{avgcost} previewed in the following and describe the details in \sec{avgcost}.

\begin{customthm}{}[Faster Trotter steps using average-cost simulation]
Consider $2$-local Hamiltonians $H=\sum_{\sigma,\sigma'\in\{i,x,y,z\}}\sum_{1\leq j<k\leq n}\beta^{(\sigma,\sigma')}_{j,k}P_j^{(\sigma)}P_k^{(\sigma')}$, where $\abs{\beta^{(\sigma,\sigma')}_{j,k}}\leq 1/|j-k|^\alpha$ for some constant $\alpha>0$ and $P^{(\sigma)}$ ($\sigma=i,x,y,z$) are the identity and Pauli matrices. Let $t>0$ be the simulation time and $\epsilon>0$ be the target accuracy. Assume that the coefficient oracle
\begin{equation}
    O_{\beta,\sigma,\sigma'}\ket{j,k,0}=\ket{j,k,\beta_{j,k}^{(\sigma,\sigma')}}
\end{equation}
can be implemented with gate complexity $\mathcal{O}\left(\polylog(nt/\epsilon)\right)$. Then $H$ can be simulated using the algorithm of \sec{avgcost_summary} with $\mathcal{O}\left(\log(nt/\epsilon)\right)$ ancilla qubits and gate complexity
\begin{equation}
    \begin{cases}
        n^{2-\alpha/2}t\left(\frac{nt}{\epsilon}\right)^{o(1)},\quad&1\leq\alpha<2,\\
        n^{5/2-\alpha}t\left(\frac{nt}{\epsilon}\right)^{o(1)},&0<\alpha<1.
    \end{cases}
\end{equation}
\end{customthm}

We also realize faster Trotter steps when certain blocks of Hamiltonian coefficients exhibit low-rank properties. Such assumptions were studied in the context of block encoding kernel matrices~\cite{NKL22}. Under the same hierarchical low-rank assumptions \cite{Hackbusch1999,Grasedyck2003}, we directly implement the diagonalization procedure without using block encoding to achieve gate complexity $(nt)^{1+o(1)}$ nearly linear in the spacetime volume for all $\alpha\geq 1$. This is summarized in \thm{rank} and restated below. See \sec{rank} and \append{master} for details. We summarize our improvements (in a simplified form) in \tab{result_summary} for simulating power-law Hamiltonians in general $d$ spatial dimensions.

\begin{customthm}{}[Faster Trotter steps using low-rank decomposition]
Consider $2$-local Hamiltonians $H=\sum_{\sigma,\sigma'\in\{i,x,y,z\}}\sum_{1\leq j<k\leq n}\beta^{(\sigma,\sigma')}_{j,k}P_j^{(\sigma)}P_k^{(\sigma')}$, where $\abs{\beta^{(\sigma,\sigma')}_{j,k}}\leq 1/|j-k|^\alpha$ for some constant $\alpha>0$ and $P^{(\sigma)}$ ($\sigma=i,x,y,z$) are the identity and Pauli matrices. Let $t>0$ be the simulation time and $\epsilon>0$ be the target accuracy. Then $H$ can be simulated using the algorithm of \sec{rank_summary} with $\mathcal{O}\left(\log(nt/\epsilon)\right)$ ancilla qubits and gate complexity
\begin{equation}
    \begin{cases}
        \rho nt\left(\frac{nt}{\epsilon}\right)^{o(1)},\quad&\alpha\geq1,\\
        \rho n^{2-\alpha}t\left(\frac{nt}{\epsilon}\right)^{o(1)},&0<\alpha<1.
    \end{cases}
\end{equation}
Here, $1\leq\rho\leq n$ defined in \eq{rho_rank} is the maximum truncation rank of certain off-diagonal blocks of coefficient matrices ($\rho=\mathcal{O}\left(\log(nt/\epsilon)\right)$ if the coefficient distribution exactly matches a power law in one spatial dimension).
\end{customthm}

\begin{table}
	\begin{center}
		\footnotesize
        \begin{tabular}{c|ccc}
			\toprule
			\diagbox{Method}{Hamiltonian} & $\alpha\geq 2d$ & $d\leq\alpha<2d$ & $0<\alpha<d$\\
			\midrule
			Sequential~\cite{CSTWZ19} & $n^{2+o(1)}t^{1+o(1)}$ & $n^{2+o(1)}t^{1+o(1)}$ & $n^{3-\frac{\alpha}{d}+o(1)}t^{1+o(1)}$\\
			\midrule
			Block-encoding (\sec{block}) & $(nt)^{1+o(1)}$ & $n^{3-\frac{\alpha}{d}+o(1)}t^{1+o(1)}$ & $n^{3-\frac{\alpha}{d}+o(1)}t^{1+o(1)}$\\
			Average-cost (\sec{avgcost}) & --- & $n^{2-\frac{\alpha}{2d}+o(1)}t^{1+o(1)}$ & $n^{\frac{5}{2}-\frac{\alpha}{d}+o(1)}t^{1+o(1)}$\\
			Low-rank (\sec{rank}) & $\rho(nt)^{1+o(1)}$ & $\rho(nt)^{1+o(1)}$ & $\rho n^{2-\frac{\alpha}{d}+o(1)}t^{1+o(1)}$\\
			\bottomrule
		\end{tabular}
	\end{center}
	\caption{Comparison of our results and the best previous results for simulating power-law interactions $1/x^{\alpha}$ in $d$ spatial dimensions. Upper bounds are used for some gate complexity expressions for presentational purpose. The truncation result of~\cite{CSTWZ19} only holds for a sufficiently short time and is thus not compared here. Both the block-encoding and the average-cost simulation method assume that the coefficients of the Hamiltonian are efficiently computable. The low-rank method has a dependence on the maximum truncation rank $1\leq\rho\leq n$ which is polylogarithmic in the input parameters for all power-law models in 1D and many power-law models such as Coulomb interactions in higher spatial dimensions.
	}
	\label{tab:result_summary}
\end{table}

Although we achieve various speedups for simulating power-law Hamiltonians based on different techniques, we have used recursion in the development of all our methods, and the core idea behind our improvements can all be understood through the so-called ``master theorem''~\cite{cormen2022introduction,roughgarden2017algorithms,Neapolitan2015,kuszmaul2021floors}. Specifically, to solve a problem of size $n$ using recursion, we divide the problem into $m$ subproblems, each of which can be seen as an instance of the original problem of size $n/m$, so
\begin{equation}
    \cost_{\text{rec}}(n)
    =m\cost_{\text{rec}}\left(\frac{n}{m}\right)
    +\cost(n),
\end{equation}
where $\cost(n)$ quantifies the additional cost to combine solutions of the subproblems in the current layer of recursion and $\cost_{\text{rec}}(n)$ denotes the total cost of the recursion. Then the master theorem asserts that, under certain assumptions of the cost function, the scaling of $\cost_{\text{rec}}(n)$ is the same as that of $\cost(n)$ up to a logarithmic factor, i.e.,
\begin{equation*}
    \cost_{\text{rec}}(n)=\mathcal{O}\left(\cost(n)\log(n)+n\right).
\end{equation*}
See \lem{master} of \sec{prelim_notation} for a more formal description of this result. However, performing the combination step can often be much simpler than directly solving the full problem, and one then expects to get a better $\cost_{\text{rec}}(n)$ when $\cost(n)$ is improved. We show that improved recursions are indeed possible for power-law systems by reducing the normalization factor of the Hamiltonian and by exploiting low-rank properties of certain blocks of the Hamiltonian coefficients, leading to faster quantum simulation by recursion.

The electronic structure Hamiltonian is one of the most widely studied candidate models in quantum simulation~\cite{mcardle2020quantum,cao2019quantum,MottaEmerging2022}. An efficient simulation of such Hamiltonians could provide insights to various problems in chemistry and material science. Here, we focus on a simulation in real space, an idea investigated by Kassal et al.~\cite{KJLMA08} and subsequently pursued by later work such as~\cite{Jones_2012,Kivlichan_2017,BWMMNC18,SBWRB21,Grid22}. Although the full Hamiltonian does not satisfy power laws, the Coulomb potential part can be represented in second quantization with the magnitude of coefficients decaying as $1/x$, to which our method applies. In particular, we can choose $\rho=\mathcal{O}\left(\log(nt/\epsilon)\right)$ in the low-rank decomposition~\cite{Greengard1987} to efficiently implement Trotter steps. Combining with an improved Trotter error analysis, we show that $\left(\eta^{1/3}n^{1/3}+\frac{n^{2/3}}{\eta^{2/3}}\right)n^{1+o(1)}$ gates suffice to simulate the uniform electron gas with $n$ spin orbitals and $\eta$ electrons; we obtain an analogous result when the external potential of nuclei is introduced under the Born-Oppenheimer approximation. This improves the best previous results for simulating electronic structures. We describe this application in \sec{app}, with the improved Trotter error analysis detailed in \append{trotter}.

\subsection{Circuit lower bound}
\label{sec:intro_lowerbound}
It is worth noting that all our above methods hold under certain additional assumptions on the target Hamiltonian: we have assumed that either the Hamiltonian coefficients are efficiently computable or certain blocks of them have low rank. If no such structural properties are available, we are then faced with Hamiltonians with arbitrary coefficients, and intuitively there would be no implementation of Trotter steps better than the sequential method.

We prove a circuit lower bound to justify this intuition. Specifically, we consider a class of $2$-local Hamiltonians of the form
\begin{equation}
    H=\sum_{1\leq j<k\leq n}\beta_{j,k}Z_jZ_k,
\end{equation}
where $Z_j$ denotes the Pauli-$Z$ operator on the $j$th qubit and coefficients are arbitrarily chosen from a continuum range of values $|\beta_{j,k}|\leq t$. Subclasses of such Hamiltonians are of interest in areas beyond quantum simulation~\cite{Hadfield21}. Even for such commuting $H$, we show the gate-complexity lower bound $\Omega\left(n^2/\log(b|\mathcal{K}|)\right)$ to evolve with accuracy $\epsilon=\Omega(1/\poly(n))$ for $t=\Omega(\epsilon)$, using quantum circuits of $b\geq n$ qubits with a gate set $\mathcal{K}$ of finite size $|\mathcal{K}|$. For circuits with a continuous gate set, we can first compile them using a finite universal gate set (say applying the Clifford+T synthesis~\cite{BRS15} or the Solovay-Kitaev theorem~\cite{Kitaev97}) and invoke the above bound. We discuss the circuit lower bounds in detail in \sec{lowerbound} with the result summarized in \cor{lowerbound_ham} and previewed here.

\begin{customcor}{}[Simulating $2$-local commuting Hamiltonians]
Consider $2$-local Hamiltonians $H=\sum_{1\leq j<k\leq n}\beta_{j,k}Z_jZ_k$, where coefficients take values up to $|\beta_{j,k}|\leq t$. Given accuracy $0<\epsilon<1/3$, number of qubits $b\geq n$, and $2$-qubit gate set $\mathcal{K}$ of finite size $|\mathcal{K}|$, if $t=\Omega(\epsilon)$,
\begin{equation}
\begin{aligned}
    &\min\left\{g\ |\ \forall\text{ $2$-local Hamiltonian\ }H,\exists\text{\ circuit\ }V\text{\ on $b$ qubits with $g$ gates from $\mathcal{K}$},\norm{e^{-iH}-V}\leq\epsilon\right\}\\
    &=\Omega\left(\frac{n^2}{\log\big(b|\mathcal{K}|\big)}-n\polylog\left(\frac{n}{\epsilon}\right)\right).
\end{aligned}
\end{equation}
Under the same assumption but choosing $\mathcal{K}$ to be the set of arbitrary $2$-qubit gates,
\begin{equation}
\begin{aligned}
    &\min\left\{g\ |\ \forall\text{ $2$-local Hamiltonian\ }H,\exists\text{\ circuit\ }V\text{\ on $b$ qubits with $g$ gates from $\mathcal{K}$},\norm{e^{-iH}-V}\leq\epsilon\right\}\\
    &=\Omega\left(\frac{n^2}{\log b}-n\polylog\left(\frac{n}{\epsilon}\right)\right).
\end{aligned}
\end{equation}
\end{customcor}

Underpinning our circuit lower bound proof is an efficient reduction from the approximate synthesis of diagonal unitaries within the Hamming weight-$2$ subspace, up to a gate overhead of $\mathcal{O}\left(n\polylog(n/\epsilon)\right)$. We describe this reduction in \sec{lowerbound_hamming} (with the circuit illustrated in \fig{reduction}). The Hamming weight-$2$ subspace has dimensions $\binom{n}{2}=\Theta(n^2)$, so our problem is reduced to studying diagonal operators
\begin{equation}
    \mathcal{D}_{\theta_{\max}}=\left\{\sum_xe^{i\theta_x}\ketbra{x}{x},\ |\theta_{x}|\leq\theta_{\max}\right\}
\end{equation}
for $\Theta(n^2)$ values of $x$. We then show in \sec{lowerbound_diag} that such diagonal unitaries require roughly $n^2$ gates to approximately implement, generalizing a previous lower bound for exact synthesis due to Bullock and Markov~\cite{BM04,Welch_2014}. See also recent work~\cite{Jia22} for a related bound expressed in terms of measure. Our argument is based on an adaption of a technique of Knill~\cite{Knill95}. Knill proved asymptotic circuit lower bounds for synthesizing the full unitary group using a volume-comparison technique, but here we only consider diagonal unitaries whose volume can be easily evaluated in closed form (see \sec{prelim_volume}). The volume of diagonal unitaries predominantly depends on the dimensionality $\Theta(n^2)$, which leads to our desired lower bound scaling.

Additionally, we prove a second lower bound in \append{lowerbound2} showing that any approximate realization of the coefficient oracle $O_\beta$ requires $\Omega(n^2)$ gates in the worst case. Combining both our upper and lower bounds, we conclude that the use of structural properties of the Hamiltonian is both necessary and sufficient to achieve lower gate complexity for implementing Trotter steps.

We briefly summarize in \sec{prelim} the preliminaries of our paper and present in \sec{discussion} a collection of questions related to our result for future work.

\section{Preliminaries}
\label{sec:prelim}
\subsection{Notation and terminology}
\label{sec:prelim_notation}
We now introduce notation and terminology to be used in the remainder of our paper.

We use lowercase Latin letters as well as the Greek alphabet (in both upper and lower cases) to denote scalars, and we save uppercase Latin letters for matrices and operators. For instance, we will use $n$ for the system size, $t$ for the evolution time (assuming $t>0$ without loss of generality), $\epsilon$ for the simulation accuracy, and we write $H$ to denote the target Hamiltonian, $X$, $Y$, $Z$ to denote Pauli operators, and $I$ to denote the identity matrix. When discussing fermionic Hamiltonians, we use $A^\dagger$, $A$, and $N$ to represent fermionic creation, annihilation, and occupation-number operators respectively~\cite{SHC21}. We write the commutator of matrices $B$ and $C$ as $[B,C]:=BC-CB$ when the multiplications are well defined. When multiplying noncommutative quantities, we use abbreviations like $\prod_{j=1}^n$ to denote the ordering where the smallest index appears on the right, e.g., $\prod_{j=1}^nU_j=U_n\cdots U_1$. We let a summation be zero and a product be one if their lower limits exceed the upper limits. We interchangeably use the decimal representation $\ket{x}$ ($x=0,\ldots,2^n-1$) and the binary representation $\ket{x_{n-1},\ldots,x_0}$ ($x_j=0,1$) of computational basis states if no ambiguity arises. 

We also construct vectors of scalars and operators and use subscripts to index them, e.g. $\beta_{j,k}$ and $Z_j$. We then define the transpose operation $\beta^\top_{j,k}:=\beta_{k,j}$. Our focus will be on simulating $2$-local systems throughout the paper, 
so the Hamiltonian coefficients will typically have no more than two subscript indices.
We assume that coefficients in an $n$-qubit Hamiltonian can be represented in binary using $\mathcal{O}(\log(nt/\epsilon))$ bits, for otherwise we may truncate the binary representation and simulate the truncated Hamiltonian with error at most $\mathcal{O}(\epsilon)$.

Our analysis requires various norms defined for vectors and matrices. For $[\beta_{j,k}]$, 
we define the vector \emph{$1$-norm} $\norm{\beta}_1:=\sum_{j,k}|\beta_{j,k}|$, the \emph{max-norm} $\norm{\beta}_{\max}:=\max_{j,k}|\beta_{j,k}|$, the \emph{Euclidean norm} $\norm{\beta}:=\sqrt{\sum_{j,k}|\beta_{j,k}|^2}$ and the \emph{induced $1$-norm} $\vertiii{\beta}_1:=\max_j\sum_k|\beta_{j,k}|$. We also need a restricted version of the induced $1$-norm defined as
\begin{equation}
    \vertiii{\beta}_{1,[\eta]}:=\max_{j}\max_{k_1<\cdots<k_\eta}\left(\abs{\beta_{j,k_1}}+\cdots+\abs{\beta_{j,k_\eta}}\right).
\end{equation}
By definition, this norm $\vertiii{\beta}_{1,[\eta]}$ only sums the largest $\eta$ elements in a row (maximized over all rows), and is thus always upper bounded by the induced $1$-norm. We will see later in the quantum chemistry application that the gap between these two norms can be significant. We use $\norm{B}$ to denote the \emph{operator norm} of $B$; this is also known as the \emph{spectral norm} and its value is given by the largest singular value of $B$.

We will use calligraphic uppercase letters to denote (un)structured sets. For instance, we use $\mathcal{H}$ to represent an arbitrary finite-dimensional Hilbert space with (normalized) quantum states $\ket{\psi}\in\mathcal{H}$ (we will write $\mathbb{C}^m$ or $\mathbb{R}^m$ if the dimensionality $m$ is explicitly provided). Given an underlying $n$-qubit system, we denote $\mathcal{W}_\eta$ to be the subspace spanned by computational basis states with Hamming weight $\eta$ ($\dim(\mathcal{W}_\eta)=\binom{n}{\eta}$); $\mathcal{W}_\eta$ also denotes the subspace spanned by states with $\eta$ particles for second-quantized fermionic systems. We may then define the operator norm restricted to the Hamming weight-$\eta$ subspace
\begin{equation}
    \norm{B}_{\mathcal{W}_\eta}:=\max_{\ket{\phi_\eta},\ket{\psi_\eta}\in\mathcal{W}_\eta}\abs{\bra{\phi_\eta}B\ket{\psi_\eta}}
\end{equation}
for an arbitrary $n$-qubit operator $B$. Given $[\beta_{j,k}]$ and a collection $\mathcal{B}\subseteq\mathbb{Z}\times\mathbb{Z}$ of pairs of indices $(u,v)$, we define the \emph{restricted max-norm} and \emph{$1$-norm} as
\begin{equation}
    \norm{\beta}_{\max,\mathcal{B}}:=\max_{(u,v)\in\mathcal{B}}\abs{\beta_{u,v}},\qquad
    \norm{\beta}_{1,\mathcal{B}}:=\sum_{(u,v)\in\mathcal{B}}\abs{\beta_{u,v}}.
\end{equation}
And as introduced earlier, the set $\mathcal{D}_{\theta_{\max}}$ is the set of diagonal unitaries with phase angles between $-\theta_{\max}$ and $\theta_{\max}$.

We say an operator $G:\mathcal{G}\rightarrow\mathcal{H}$ is an \emph{isometry} if $G^\dagger G=I$. By definition, $G$ is necessarily injective and $G^\dagger$ is necessarily surjective, whereas $GG^\dagger$ is an orthogonal projection on $\mathcal{H}$ with image $\im(GG^\dagger)=\im(G)$ and kernel $\ker(GG^\dagger)=\ker(G^\dagger)$. We thus obtain the Hilbert space isomorphism $\mathcal{G}\cong\im(GG^\dagger)\subseteq\mathcal{H}$ specified by the operators $G$ and $G^\dagger$. Choosing any orthonormal basis that respects this isomorphism, we have the matrix representation
\begin{equation}
    G=\begin{bmatrix}
    I\\
    0
    \end{bmatrix},\qquad
    G^\dagger=\begin{bmatrix}
    I & 0
    \end{bmatrix}.
\end{equation}
Examples of isometries include: (i) unitary operators $U$; (ii) quantum states $\ket{\psi}$; (iii) tensor product $G_1\otimes G_2$ if $G_1$ and $G_2$ are isometries; and (iv) composition $G_2G_1$ if $G_1$ and $G_2$ are isometries and the composition is well defined. Isometries will be used later in \sec{prelim_qubitization} to describe block encodings and the qubitization algorithm.

Finally, we use $\mathcal{O}(\cdot)$ and $\Omega(\cdot)$ to mean asymptotically bounded above and below respectively, write $\Theta(\cdot)$ if both relations hold, and use the tilde symbol to suppress polylogarithmic factors. This is similar to the notation of a previous work on Trotter error analysis~\cite{CSTWZ19}, except we do not need their $O(\cdot)$ for order conditions. To analyze the scaling of functions satisfying recurrence relations, we use the following version of the master theorem adapted from~\cite{cormen2022introduction,roughgarden2017algorithms,Neapolitan2015,kuszmaul2021floors}.
\begin{lemma}[Master theorem]
\label{lem:master}
Let $\cost,\cost_{\text{rec}}:\mathbb{Z}_{\geq 1}\rightarrow\mathbb{R}_{\geq0}$ be nonnegative functions defined for positive integers, such that there exist $c_0\geq0$, $m_1,m_2\in\mathbb{Z}_{\geq 0}$ not all zero, $n_0,m\in\mathbb{Z}_{\geq 2}$ for which
\begin{equation}
\begin{cases}
    \cost_{\text{rec}}(n)\leq c_0,\qquad&1\leq n< n_0,\\
    \cost_{\text{rec}}(n)\leq m_1\cost_{\text{rec}}\left(\floor{\frac{n}{m}}\right)+m_2\cost_{\text{rec}}\left(\ceil{\frac{n}{m}}\right)+\cost(n),&n\geq n_0.
\end{cases}
\end{equation}
If $\cost(n)=\mathcal{O}\left(n^\alpha\log^{k}(n)\right)$ for some $\alpha\geq0$ and $k\in\mathbb{Z}_{\geq0}$, then
\begin{equation}
    \cost_{\text{rec}}(n)=
    \begin{cases}
        \mathcal{O}\left(n^{\log_{m}(m_1+m_2)}\right),\qquad&0\leq \alpha<\log_{m}(m_1+m_2),\\
        \mathcal{O}\left(n^{\alpha}\log^{k+1}(n)\right),\qquad&\alpha=\log_{m}(m_1+m_2),\\
        \mathcal{O}\left(n^\alpha\log^{k}(n)\right),\qquad&\alpha>\log_{m}(m_1+m_2).
    \end{cases}
\end{equation}
Thus, we have $\cost_{\text{rec}}(n)=\mathcal{O}\left(\cost(n)\log(n)+n^{\log_{m}(m_1+m_2)}\right)$ in all the cases.
\end{lemma}

\subsection{Product formulas}
\label{sec:prelim_pf}
Consider a Hamiltonian given in the LCH form $H=\sum_{\gamma=1}^\Gamma H_\gamma$. We can well approximate the evolution under $H$ for a short time using \emph{product formulas} with error high order in time. A longer evolution can then be simulated by repeating the short-time steps. Product formulas provide a simple yet surprisingly efficient approach to quantum simulation. Indeed, recent work has shown that product formulas: (i) can simulate geometrically local lattice systems~\cite{haah2018quantum} with nearly optimal gate complexity~\cite{CS19}; (ii) have the lowest asymptotic cost for electronic structure Hamiltonians in second quantization in the plane wave basis~\cite{SHC21}; and (iii) are advantageous for simulating general $\kappa$-local Hamiltonians~\cite{CSTWZ19} (although our work achieves further improvements regarding this last point). In addition, product formulas have also been widely used in classical simulations of quantum systems and in areas beyond quantum computing~\cite{bk:BC16}.

By definition, the cost of the product-formula approach is determined by both the repetition number of short-time steps and the complexity of implementing each step. To elaborate, first consider a simple example where we use the Lie-Trotter formula $e^{-itB}e^{-itA}$ to simulate a $2$-term Hamiltonian $H=A+B$. Then it holds that
\begin{equation}
    e^{-itB}e^{-itA}-e^{-it(A+B)}=
    \int_{0}^{t}\mathrm{d}\tau_1\int_{0}^{\tau_1}\mathrm{d}\tau_2\
	e^{-i(t-\tau_1)(A+B)}e^{-i\tau_1B}e^{i\tau_2B}\big[iB,iA\big]e^{-i\tau_2B}e^{-i\tau_1A},
\end{equation}
which implies the \emph{Trotter error} bound
\begin{equation}
\label{eq:pf1bound}
    \norm{e^{-itB}e^{-itA}-e^{-it(A+B)}}\leq\frac{t^2}{2}\norm{[A,B]}.
\end{equation}
This approximation error is small for a sufficiently short-time \emph{Trotter step}. For a longer simulation, we apply $r$ repetitions of the same step with time $t/r$, obtaining
\begin{equation}
    \norm{\left(e^{-i\frac{t}{r}B}e^{-i\frac{t}{r}A}\right)^r-e^{-it(A+B)}}\leq r\norm{e^{-i\frac{t}{r}B}e^{-i\frac{t}{r}A}-e^{-i\frac{t}{r}(A+B)}}\leq\frac{t^2}{2r}\norm{[A,B]}.
\end{equation}
To achieve an error at most $\epsilon$, it thus suffices to take
\begin{equation}
    r=\left\lceil\frac{\norm{[A,B]}t^2}{2\epsilon}\right\rceil.
\end{equation}
We now have a total of $2r$ elementary exponentials, where $r$ is the number of Trotter steps determined by the error bound \eq{pf1bound}.

In general, we can simulate the Hamiltonian $H=\sum_{\gamma=1}^\Gamma H_\gamma$ using a $p$th-order product formula $S_p(t)$, where $p$ is a positive integer that can be arbitrarily large. Similar to above, we need to bound the cost of implementing Trotter steps, as well as the repetition number $r$ which is in turn determined by the analysis of Trotter error. We introduce the following Trotter error bound with a commutator scaling established by~\cite{CSTWZ19}.
Although Trotter error analysis can be further tightened using additional assumptions of the quantum simulation problem~\cite{ChenBrandao21,Layden22,An2021timedependent,Burgarth22,Kocia22,Zhao21,Sahinoglu2021,Hatomura22,HHZ19}, those improvements are not relevant to our results and will not be further discussed here.
\begin{lemma}[Trotter error with commutator scaling]
\label{lem:comm_bound}
    Let $H=\sum_{\gamma=1}^\Gamma H_\gamma$ be a Hamiltonian in the LCH form and $S_p(t)$ be a $p$th-order product formula with respect to this decomposition. We have
    \begin{equation}
        \norm{S_p(t)-e^{-itH}}
        =\mathcal{O}\left(\sum_{\gamma_1,\gamma_2,\ldots,\gamma_{p+1}=1}^\Gamma\norm{\big[H_{\gamma_{p+1}},\cdots\big[H_{\gamma_2},H_{\gamma_1}\big]\big]}t^{p+1}\right).
    \end{equation}
    Furthermore, if $H$ is a fermionic Hamiltonian in second quantization and $H_\gamma$ are number preserving,
    \begin{equation}
        \norm{S_p(t)-e^{-itH}}_{\mathcal{W}_\eta}
        =\mathcal{O}\left(\sum_{\gamma_1,\gamma_2,\ldots,\gamma_{p+1}=1}^\Gamma\norm{\big[H_{\gamma_{p+1}},\cdots\big[H_{\gamma_2},H_{\gamma_1}\big]\big]}_{\mathcal{W}_\eta}t^{p+1}\right).
    \end{equation}
\end{lemma}

We denote the sum of nested commutators as
\begin{equation}
    \acommtilde:=\sum_{\gamma_1,\gamma_2,\ldots,\gamma_{p+1}=1}^\Gamma\norm{\big[H_{\gamma_{p+1}},\cdots\big[H_{\gamma_2},H_{\gamma_1}\big]\big]}.\label{eq:alpha-comm}
\end{equation}
Then to simulate for time $t$ with accuracy $\epsilon$, it suffices to take $r=\mathcal{O}\left(\acommtilde^{1/p}t^{1+1/p}/\epsilon^{1/p}\right)$, which simplifies to
\begin{equation}
    r=\frac{\acommtilde^{o(1)}t^{1+o(1)}}{\epsilon^{o(1)}}
\end{equation}
by choosing $p$ sufficiently large. For the fermionic case, we can simply replace the spectral norm $\norm{\cdot}$ by its restriction to the Hamming weight-$\eta$ subspace $\norm{\cdot}_{\mathcal{W}_\eta}$.

A sequential implementation of Trotter steps has an asymptotic cost of $\mathcal{O}(\Gamma)$, giving total gate complexity $\mathcal{O}(r\Gamma)$. The purpose of this work is to identify conditions under which the cost of Trotter steps does not explicitly scale with $\Gamma$. We will present three methods, one based on the block-encoding technique (\sec{block}), one based on an average-cost simulation (\sec{avgcost}) and one based on a recursive low-rank decomposition (\sec{rank}). We review the basic notion of block encoding as well as the related qubitization algorithm in the next subsection.

\subsection{Block encoding and qubitization}
\label{sec:prelim_qubitization}
Block encodings, together with the related qubitization algorithm, provide a versatile framework for simulating Hamiltonians in the LCU form and beyond. Such techniques enable quantum simulation with a complexity linear in the evolution time and logarithmic in the inverse accuracy~\cite{LC17}, and can be extended to solve problems other than simulation~\cite{GSLW19,Martyn21} (although these goals are sometimes also achievable via product formulas as explained in \sec{intro_algs}). In addition, these techniques have been utilized by recent work to design classical algorithms for simulating quantum systems~\cite{deQuantizeQSVT22}.

To explain the idea of block encoding and qubitization, we will use the notion of isometries. Specifically, consider two isometries $G_0:\mathcal{G}_0\rightarrow\mathcal{H}$, $G_1:\mathcal{G}_1\rightarrow\mathcal{H}$ and a unitary $U:\mathcal{H}\rightarrow\mathcal{H}$. We say an operator $B:\mathcal{G}_0\rightarrow\mathcal{G}_1$ is \emph{block encoded} by $G_0$, $G_1$, and $U$ if
\begin{equation}
    B=G_1^\dagger UG_0.
\end{equation}
Of course, this unitary dilation is mathematically feasible if and only if $\norm{B}\leq 1$~\cite[2.7.P2]{horn2012matrix}. 
However, there are many scenarios where additional normalization factors will be introduced when such block encodings are realized by quantum circuits. If we choose two bases with respect to the orthogonal decompositions $\mathcal{H}=\im(G_0G_0^\dagger)\obot\im(I-G_0G_0^\dagger)=\im(G_1G_1^\dagger)\obot\im(I-G_1G_1^\dagger)$ with $\obot$ being the orthogonal direct sum,
then $U$ has a matrix representation that looks like
\begin{equation}
    U=
    \begin{bmatrix}
    B & \cdot\\
    \cdot & \cdot
    \end{bmatrix},
\end{equation}
where $B$ is exhibited as the top-left matrix block; hence the name \emph{block encoding}.

For quantum simulation, we fix  $\mathcal{G}_0=\mathcal{G}_1=\mathcal{G}$ to be the target system space and we let $G_1^\dagger UG_0$ be Hermitian to represent physical Hamiltonians. 
Then, we have the following qubitization algorithm to perform quantum simulation of block-encoded operators~\cite{LC16}.
\begin{lemma}[Hamiltonian simulation by qubitization]
\label{lem:qubitization}
    Let $G_0,G_1:\mathcal{G}\rightarrow\mathcal{H}$ be isometries and $U:\mathcal{H}\rightarrow\mathcal{H}$ be a unitary such that $G_1^\dagger UG_0$ is Hermitian. Given a target evolution time $t$ and accuracy $\epsilon$, there exists a unitary $V_{\varphi}:\mathbb{C}^2\otimes\mathbb{C}^2\otimes\mathcal{H}\rightarrow\mathbb{C}^2\otimes\mathbb{C}^2\otimes\mathcal{H}$ parameterized by angles $\varphi_1,\ldots,\varphi_r$ such that
    \begin{equation}
    \label{eq:block_error}
        \norm{\left(\bra{+}\otimes \frac{\bra{0}\otimes G_0^\dagger+\bra{1}\otimes G_1^\dagger}{\sqrt{2}}\right)V_{\varphi}\left(\ket{+}\otimes \frac{\ket{0}\otimes G_0+\ket{1}\otimes G_1}{\sqrt{2}}\right)-e^{-itG_1^\dagger UG_0}}\leq\epsilon.
    \end{equation}
    The number of steps $r$ is an even integer with the asymptotic scaling
    \begin{equation}
    \label{eq:qubitization_complexity}
        r=\mathcal{O}\left(t+\log\left(\frac{1}{\epsilon}\right)\right),
    \end{equation}
    and $V_\varphi$ is explicitly defined as
    \begin{equation}
    \label{eq:qubitization_circuit}
    \begin{aligned}
        V_\varphi&:=\prod_{k=0}^{r/2-1}\left(V_{\varphi_{2k+2}}^\dagger V_{\varphi_{2k+1}}\right)\\
        V_{\varphi_j}&:=\left(e^{-i\varphi_jZ/2}\otimes I\right)\left(\ketbra{+}{+}\otimes I+\ketbra{-}{-}\otimes V\right)\left(e^{i\varphi_jZ/2}\otimes I\right)\\
        V&:=\left(X\otimes I\right)\left(\ketbra{0}{0}\otimes U+\ketbra{1}{1}\otimes U^\dagger\right)\left(I-2
        \frac{\ket{0}\otimes G_0+\ket{1}\otimes G_1}{\sqrt{2}}\frac{\bra{0}\otimes G_0^\dagger+\bra{1}\otimes G_1^\dagger}{\sqrt{2}}\right).
    \end{aligned}
    \end{equation}
\end{lemma}
\noindent Note that by our definition of $\mathcal{O}$, the complexity expression \eq{qubitization_complexity} is valid for $t$ sufficiently large and $\epsilon$ sufficiently small. This expression may be modified if other parameter regimes are of interest.

To simulate an $n$-qubit Hamiltonian $H=\sum_{\gamma=1}^\Gamma \beta_\gamma U_\gamma$ in the LCU form, we define
\begin{equation}
\label{eq:lcu}
\begin{aligned}
    &G:\mathbb{C}^{2^n}\rightarrow\mathbb{C}^\Gamma\otimes\mathbb{C}^{2^n},\qquad\quad
    &&G=\frac{1}{\sqrt{\norm{\beta}_1}}\sum_{\gamma=1}^\Gamma \sqrt{\beta_\gamma}\ket{\gamma}\otimes I,\\
    &U:\mathbb{C}^\Gamma\otimes\mathbb{C}^{2^n}\rightarrow\mathbb{C}^\Gamma\otimes\mathbb{C}^{2^n},
    &&U=\sum_{\gamma=1}^\Gamma\ketbra{\gamma}{\gamma}\otimes U_\gamma,
\end{aligned}
\end{equation}
which implies
\begin{equation}
    G^\dagger UG:\mathbb{C}^{2^n}\rightarrow\mathbb{C}^{2^n},\qquad\quad
    G^\dagger UG=\frac{H}{\norm{\beta}_1}.
\end{equation}
We thus see that the target Hamiltonian is block encoded by $G$ (a state \emph{preparation subroutine}) and $U$ (an operator \emph{selection subroutine}), but with a normalization factor that depends on the vector $1$-norm $\norm{\beta}_1$, which slows down quantum simulation. To approximate $e^{-itH}$, we then need to invoke \lem{qubitization} with an effective evolution time of $\norm{\beta}_1t$. This implies that the number of steps will now scale as
\begin{equation}
    r=\mathcal{O}\left(\norm{\beta}_1t+\log\left(\frac{1}{\epsilon}\right)\right).
\end{equation}
The preparation of a $\Gamma$-dimensional state naively requires $\mathcal{O}(\Gamma)$ gates to implement~\cite{ShendeBullockMarkov06} and the selection of $\Gamma$ terms has cost $\mathcal{O}(\Gamma)$~\cite[Appendix G.4]{CMNRS18}, assuming each controlled-$U_\gamma$ has constant cost. This gives a total complexity of $\mathcal{O}(r\Gamma)$ for qubitization.

Having introduced the qubitization algorithm, we now make a few remarks about its circuit implementation.

First, qubitization approximately simulates the time evolution with a certain probability, and both the approximation error and the success probability can be analyzed using the condition \eq{block_error}. Indeed, an argument based on the triangle inequality shows that the qubitization algorithm succeeds with probability at least $\left(1-\epsilon\right)^2\geq1-2\epsilon$, and the postmeasurement state has an error at most $\sqrt{1+\epsilon}-\sqrt{1-\epsilon}\leq\sqrt{2}\epsilon$~\cite[Lemma G.4 and Appendix H.1]{CMNRS18}. It therefore suffices to adjust the value of $\epsilon$ to meet the desired error and probability requirements.

Second, in our above LCU example, we have defined $G$ and $U$ that faithfully block encode $H/\norm{\beta}_1$. More realistically, we may consider state preparation and operator selection circuits that have lower cost but introduce error to the block encoding. Specifically, suppose $\widetilde{G}^\dagger\widetilde{U}\widetilde{G}=\widetilde{H}/\norm{\widetilde{\beta}}_1$, then we are effectively block encoding an erroneous Hamiltonian with error growing at most linearly in time: $\norm{e^{-it\norm{\widetilde{\beta}}_1\widetilde{G}^\dagger\widetilde{U}\widetilde{G}}-e^{-itH}}\leq t\norm{H-\widetilde{H}}$. And we can suppress this error accordingly by increasing the accuracy of block encoding.

We will focus on the complexity of the state preparation and operator selection subroutines for the remainder of our paper. There is also an additional cost in qubitization to implement the operations in between these subroutines, but that cost is typically logarithmic in $\Gamma$ and only makes a mild contribution to the total gate complexity. In particular, one can check that this holds for our block encodings of power-law Hamiltonians to be described in \sec{block} and \sec{avgcost}. So there is no loss of rigor to ignore this subdominant contribution in our analysis.

Finally, we point out that a host of techniques have been recently developed to improve the circuit implementation of block encoding and qubitization over the naive approach. For instance, if the target Hamiltonian consists of tensor product of Pauli operators, then the selection subroutine $U=U^\dagger$ is both unitary and Hermitian, so we can reduce $1$ ancilla from the qubitization circuit in \lem{qubitization}. Even when this Hermitian condition is not satisfied, one may still improve the implementation of the bidirectional control $\ketbra{0}{0}\otimes U+\ketbra{1}{1}\otimes U^\dagger$ using explicit structures of $U$. Also, the qubitization algorithm we have introduced in \lem{qubitization} is mainly for dynamical simulation. When qubitization is used as a subroutine in quantum phase estimation, we only need to implement the operator $V$ in \eq{qubitization_circuit} as opposed to the full circuit~\cite{Poulin2018,Berry2018}.\footnote{This simplification follows from the following spectral property of $V$: there exists an orthonormal basis under which $V$ is exhibited as a direct sum of $1$-by-$1$ and $2$-by-$2$ matrices, whose eigenvalues correspond to eigenvalues of the target Hamiltonian (which is enough for phase estimation). We can then visualize $V$ as a collection of single-qubit unitaries acting on orthogonal subspaces; hence the name ``qubitization''.}\
There are also some flexibilities about block encoding the LCU Hamiltonian beyond the standard approach of \eq{lcu}, such as allowing garbage registers~\cite{LKS18}, performing unary encodings~\cite{BCCKS14}, or using even more advanced circuit compilation tricks~\cite{Steudtner20}.
Additionally, given a block encoding $G_0,G_1:\mathcal{G}\rightarrow\mathcal{H}$ and $U:\mathcal{H}\rightarrow\mathcal{H}$, there always exists unitary $W:\mathcal{H}\rightarrow\mathcal{H}$ such that $WG_1=G_0$. 
Any such $W$ satisfies $G_1^\dagger UG_0=G_0^\dagger(WU)G_0$ and hence gives an alternative block encoding with only one isometry, simplifying the circuit implementation.

But perhaps the most significant improvement comes from the following observation: terms in the LCU Hamiltonian are sometimes indexed by vectors with a product structure, i.e.,
\begin{equation}
    U_\gamma=U_{\gamma_1}\cdots U_{\gamma_s},\qquad
    \gamma_j=1,\ldots,\Gamma^{1/s}.
\end{equation}
In this case, the selection subroutine takes a product form
\begin{equation}
\label{eq:sel_product}
    \sum_{\gamma=1}^\Gamma\ketbra{\gamma}{\gamma}\otimes U_\gamma=
    \sum_{\gamma_1,\ldots,\gamma_s=1}^{\Gamma^{1/s}}\ketbra{\gamma_1\cdots\gamma_s}{\gamma_1\cdots\gamma_s}\otimes U_{\gamma_1}\cdots U_{\gamma_s}
\end{equation}
and can be implemented with complexity $\mathcal{O}(s\Gamma^{1/s})$ as opposed to $\mathcal{O}(\Gamma)$. Meanwhile, we can sometimes exploit the structure of Hamiltonian coefficients to also perform the preparation subroutine faster than the naive approach. This then yields a step of qubitization with complexity sublinearly in the number of terms $\Gamma$.

However, the above discussion ignored the scaling of the normalization factor $\norm{\beta}_1$, which determines the number of repetition steps and is still the bottleneck of qubitization. In fact, the reduction of the $1$-norm has been the central technical problem studied by many recent work of quantum simulation~\cite{Lee21,vonBurg21,Zhao2021exploiting,Loaiza22}. We show in \sec{block} that this $1$-norm can be significantly reduced using product formulas ``almost for free'', which leads to new methods for simulating power-law Hamiltonians that improve the state of the art.

\subsection{Volume of diagonal unitaries}
\label{sec:prelim_volume}
In this section, we review techniques for computing the volume of a parameterized object in a high-dimensional Euclidean space. In particular, we compute the volume of diagonal unitaries with phase angles taking a continuum range of values between $-\theta_{\max}$ and $\theta_{\max}$, which will be used later in \sec{lowerbound} to prove our circuit lower bound. These calculations can be seen as generalizations of line and surface integrals defined for low-dimensional spaces.

Let $\mathcal{M}\subset\mathbb{R}^m$ be a subset and $f:\mathcal{M}\rightarrow\mathbb{R}^q\ (m\leq q)$ be a mapping. Here, $f=[f_1\ \cdots\ f_q]$ consists of $q$ scalar functions and we can think of it as a parameterized description of a high-dimensional geometrical object, where $\mathcal{M}$ is the set of all possible parameters. Then, given a subset of parameters $\mathcal{A}\subset\mathcal{M}$, we may compute the volume of $f(\mathcal{A})$ as
\begin{equation}
    \vol\left(f(\mathcal{A})\right)
    =\int_{\mathcal{A}}\mathrm{d}\theta \sqrt{\mathrm{det}(G_f)},
\end{equation}
where $G_f$ is the Gramian whose element in the $x$th row and $y$th column is given by the inner product
\begin{equation}
    \begin{bmatrix}
    \frac{\partial f_1}{\partial \theta_x} & \cdots & \frac{\partial f_q}{\partial \theta_x}
    \end{bmatrix}
    \begin{bmatrix}
    \frac{\partial f_1}{\partial \theta_y} \\ \vdots \\ \frac{\partial f_q}{\partial \theta_y}
    \end{bmatrix}.
\end{equation}
For the above formula to hold mathematically, we require that $\mathcal{M}$ is open, $\mathcal{A}$ is Lebesgue measurable, and $f$ is continuously differentiable and injective on $\mathcal{A}$. We refer the reader to standard analysis textbooks for a detailed explanation of this formula~\cite{GM10,GM11,Courant}.

Now, consider the set of diagonal unitaries
\begin{equation}
    \mathcal{D}_{\theta_{\max}}:=\left\{\sum_{x=0}^{m-1}e^{i\theta_x}\ketbra{x}{x},\ |\theta_{x}|\leq\theta_{\max}<\pi\right\}.
\end{equation}
We may view them as objects in the Euclidean space $\mathbb{R}^{2m}$ by choosing the parameterization
\begin{equation}
    f(\theta_0,\ldots,\theta_{m-1})=\begin{bmatrix}
    \cos\theta_0 & \sin\theta_0 & \cdots & \cos\theta_{m-1} & \sin\theta_{m-1}
    \end{bmatrix}.
\end{equation}
Then, the Gramian matrix is simply the identity matrix
\begin{equation}
\begin{aligned}
    G_f=
    \left[\begin{smallmatrix}
    -\sin\theta_0 & \cos\theta_0 & 0 & 0 & \cdots & 0 & 0\\
    0 & 0 & -\sin\theta_1 & \cos\theta_1 & \cdots & 0 & 0\\
    \vdots & \vdots & \vdots & \vdots & \vdots & \vdots & \vdots\\
    0 & 0 & 0 & 0 & \cdots & -\sin\theta_{m-1} & \cos\theta_{m-1}
    \end{smallmatrix}\right]
    \left[\begin{smallmatrix}
    -\sin\theta_0 & 0 & \cdots & 0\\
    \cos\theta_0 & 0 & \cdots & 0\\
    0 & -\sin\theta_1 & \cdots & 0\\
    0 & \cos\theta_1 & \cdots & 0\\
    \vdots & \vdots & \vdots & \vdots\\
    0 & 0 & \cdots & -\sin\theta_{m-1}\\
    0 & 0 & \cdots & \cos\theta_{m-1}\\
    \end{smallmatrix}\right]
    =I,
\end{aligned}
\end{equation}
and as a result we obtain the volume formula
\begin{equation}
    \vol\left(\mathcal{D}_{\theta_{\max}}\right)=\left(2\theta_{\max}\right)^m.
\end{equation}

Alternatively, we may also consider the mapping
\begin{equation}
    \varphi:\ \theta=[\theta_0\ \cdots\ \theta_{m-1}]
    \mapsto\sum_{x=0}^{m-1}e^{i\theta_x}\ketbra{x}{x}
\end{equation}
that parameterizes all diagonal unitaries with phase angles $-\pi\leq\theta_x<\pi$. This is a bijection, and we may define the pushforward measure on the set of diagonal unitaries using the Lebesgue measure $\lambda$ as
\begin{equation}
    \varphi_{*}(\lambda)(\mathcal{B})=\lambda(\varphi^{-1}(\mathcal{B})).
\end{equation}
The measure $\varphi_{*}(\lambda)$ can be verified to be invariant under multiplication and, up to a constant factor, is the unique Haar measure for the group of diagonal unitaries. This then gives the volume $\varphi_{*}(\lambda)\left(\mathcal{D}_{\theta_{\max}}\right)=\left(2\theta_{\max}\right)^m$ same as our previous formula.

\section{Faster Trotter steps using block encoding}
\label{sec:block}
We now introduce our first method to perform faster Trotter steps based on block encoding. The structure of this section is as follows. We describe how to reduce the $1$-norm of Hamiltonian terms using product formulas in \sec{block_norm}, removing the bottleneck of the qubitization algorithm. We then give a circuit implementation of the preparation and selection subroutines in \sec{block_prepsel}, assuming that the Hamiltonian coefficients are efficiently computable. This gives simulations with complexity nearly linear in the spacetime volume for power-law exponent $\alpha\geq 2$. Readers who wish to see the overall algorithm may skip ahead to \sec{block_summary}.

\subsection{Reducing the \texorpdfstring{$1$}{1}-norm of Hamiltonian coefficients}
\label{sec:block_norm}
Consider an $n$-qubit Hamiltonian $H=\sum_{1\leq j<k\leq n}H_{j,k}$, where $H_{j,k}$ are $2$-local terms acting nontrivially only on sites $j$ and $k$ with norm
$\norm{H_{j,k}}\leq 1/|j-k|^\alpha$. 
Our aim is to implement Trotter steps using block encodings and qubitization. To this end, we rewrite the Hamiltonian in the Pauli basis:
\begin{equation}
\label{eq:pauli_expansion}
\begin{aligned}
    H&=\sum_{1\leq j<k\leq n}\beta^{(xx)}_{j,k}X_jX_k
    +\sum_{1\leq j<k\leq n}\beta^{(xy)}_{j,k}X_jY_k
    +\sum_{1\leq j<k\leq n}\beta^{(xz)}_{j,k}X_jZ_k\\
    &\qquad+\sum_{1\leq j<k\leq n}\beta^{(yx)}_{j,k}Y_jX_k
    +\sum_{1\leq j<k\leq n}\beta^{(yy)}_{j,k}Y_jY_k
    +\sum_{1\leq j<k\leq n}\beta^{(yz)}_{j,k}Y_jZ_k\\
    &\qquad+\sum_{1\leq j<k\leq n}\beta^{(zx)}_{j,k}Z_jX_k
    +\sum_{1\leq j<k\leq n}\beta^{(zy)}_{j,k}Z_jY_k
    +\sum_{1\leq j<k\leq n}\beta^{(zz)}_{j,k}Z_jZ_k\\
    &\qquad+\sum_{1\leq j\leq n}\beta^{(x)}_{j}X_j
    +\sum_{1\leq j\leq n}\beta^{(y)}_{j}Y_j
    +\sum_{1\leq j\leq n}\beta^{(z)}_{j}Z_j
    +\beta^{(i)}I,
\end{aligned}
\end{equation}
where the coefficients have the scaling~\cite[Theorem H.1]{CSTWZ19}
\begin{equation}
    \beta^{(\sigma,\sigma')}_{j,k}=\mathcal{O}\left(\frac{1}{|j-k|^\alpha}\right),\quad
    \beta^{(\sigma)}_j=
    \begin{cases}
    \mathcal{O}(1),\quad&\alpha>1,\\
    \mathcal{O}(\log n),&\alpha=1,\\
    \mathcal{O}(n^{1-\alpha}),&0<\alpha<1,
    \end{cases}\quad
    \beta^{(i)}=
    \begin{cases}
    \mathcal{O}(n),\quad&\alpha>1,\\
    \mathcal{O}(n\log n),&\alpha=1,\\
    \mathcal{O}(n^{2-\alpha}),&0<\alpha<1,
    \end{cases}
\end{equation}
for $\sigma\neq\sigma'\in\{x,y,z\}$. 
Since $\beta^{(i)}I$ commutes with all the other terms and its evolution only introduces a global phase, we may separate out this term without introducing error. We then use product formulas to make a coarse-grained decomposition of the remaining evolution into exponentials of $12$ groups of terms. 
The commutator norm \eq{alpha-comm} corresponding to this decomposition has the scaling~\cite[Theorem H.2]{CSTWZ19}
\begin{equation}
    \acommtilde^{o(1)}=
    \begin{cases}
    n^{o(1)},\quad&\alpha\geq 1,\\
    n^{1-\alpha+o(1)},&0<\alpha<1,
    \end{cases}
\end{equation}
which implies
\begin{equation}
\label{eq:powerlaw_r}
    r=\begin{cases}
    \frac{n^{o(1)}t^{1+o(1)}}{\epsilon^{o(1)}},\quad&\alpha\geq 1,\\[5pt]
    \frac{n^{1-\alpha+o(1)}t^{1+o(1)}}{\epsilon^{o(1)}},&0<\alpha< 1.
    \end{cases}
\end{equation}
In particular, the Trotter error bound corresponding to this coarse-grained decomposition is never larger than that of the fine-grained decomposition (in which all $\Theta(n^2)$ terms are split). 

Note that all on-site terms from the same group pairwise commute and can be exponentiated with complexity $\mathcal{O}(n)$. So without loss of generality, we will only consider Hamiltonians of the form\footnote{Note that $H$ is indeed Hermitian as $X_j$ and $Y_k$ commute for $j<k$.}
\begin{equation}
    H=\sum_{1\leq j<k\leq n}\beta_{j,k}X_jY_k.
\end{equation}
The coefficients of this Hamiltonian have vector $1$-norm scaling like~\cite[Lemma H.1]{CSTWZ19}
\begin{equation}
\label{eq:naive_1norm}
    \norm{\beta}_1=\sum_{1\leq j<k\leq n}\abs{\beta_{j,k}}=
    \begin{cases}
    \mathcal{O}(n),\quad&\alpha>1,\\
    \mathcal{O}(n\log n),&\alpha=1,\\
    \mathcal{O}(n^{2-\alpha}),&0<\alpha<1.
    \end{cases}
\end{equation}
For $\alpha>1$, even if we use an improved qubitization circuit with product structure like in \eq{sel_product}, we still need a cost roughly
\begin{equation}
    \underbrace{n}_{\text{block encoding}}\cdot\quad\underbrace{n\left(\frac{t}{r}\right)^{o(1)}}_{\text{effective time}}
    =\underbrace{n^2\left(\frac{t}{r}\right)^{o(1)}}_{\text{Trotter step complexity}}
\end{equation}
to simulate for an almost constant time, to implement a single Trotter step. Thus, block encoding does not seem to offer a significant benefit over the sequential approach. In what follows, we show that product formulas can be used to reduce the $1$-norm of Hamiltonian coefficients, which resolves this technical obstacle and, when combined with block encoding, provide a fast implementation of Trotter steps.

\begin{figure}[t]
    \centering
    \includegraphics[width=0.4\textwidth]{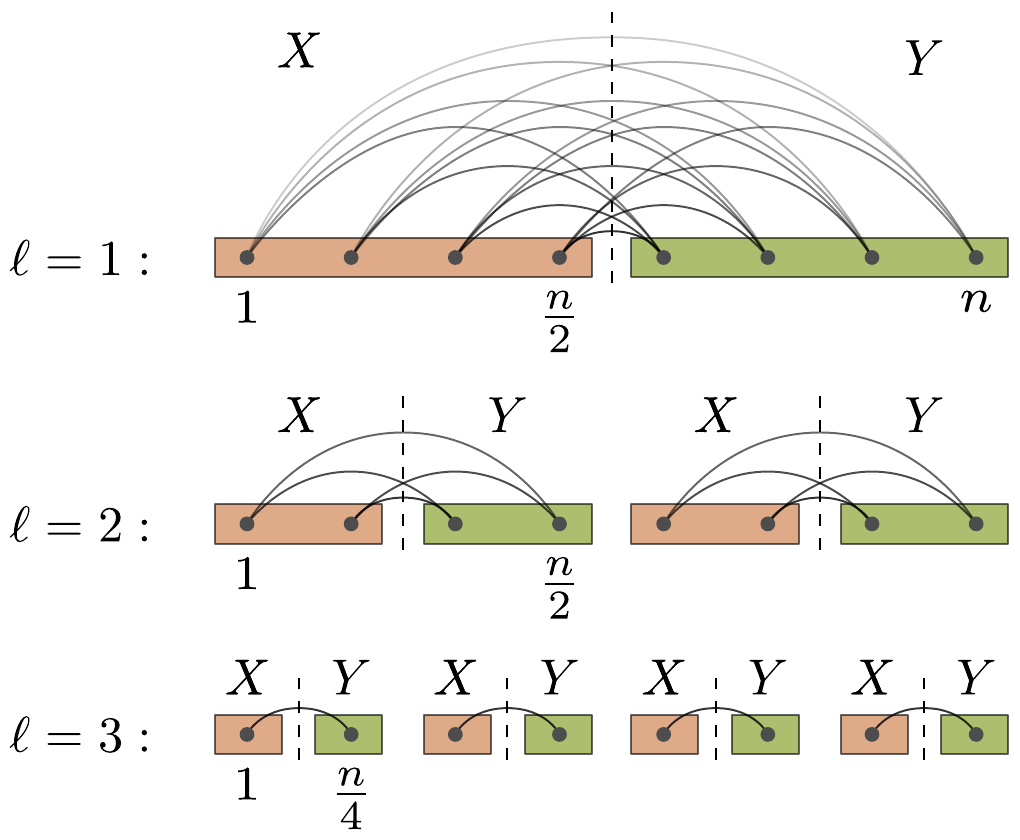}
\caption{Illustration of the recursive decomposition of $\sum_{j<k}\beta_{j,k} X_jY_k$ with three layers. The black dots represent qubits and the curves correspond to the interactions between them. In each layer, we cut the system into equal halves (dashed lines) and consider only the interactions across the cuts. The colors indicate whether the interaction consists of Pauli-$X$ (shaded orange) or Pauli-$Y$ (green).}
\label{fig:recursion}
\end{figure}

To simplify the discussion, we assume that the system size $n$ is a power of $2$. 
We also use
\begin{equation}
\begin{aligned}
    H_{[j,k]}&:=\sum_{j\leq u<v\leq k}\beta_{u,v}X_uY_v\
    \text{($1\leq j< k\leq n$)},\\
    H_{[j,k]:[l,m]}&:=\sum_{\substack{j\leq u\leq k\\l\leq v\leq m}}\beta_{u,v}X_uY_v\
    \text{($1\leq j\leq k<l\leq m\leq n$)}
\end{aligned}
\end{equation}
to represent terms within a specific interval and across two disjoint intervals of sites.
Then, we define a decomposition via the recurrence relation
\begin{equation}
\label{eq:block_recursion}
    H_{[j,k]}=H_{[j,\floor{\frac{j+k}{2}}]:[\floor{\frac{j+k}{2}}+1,k]}
    +H_{[j,\floor{\frac{j+k}{2}}]}
    +H_{[\floor{\frac{j+k}{2}}+1,k]},
\end{equation}
which unwraps to
\begin{equation}
\label{eq:decomp_block}
\begin{aligned}
    H&=H_{[1,n]}\\
    &=H_{[1,\frac{n}{2}]:[\frac{n}{2}+1,n]}
    +H_{[1,\frac{n}{2}]}
    +H_{[\frac{n}{2}+1,n]}\\
    &=H_{[1,\frac{n}{2}]:[\frac{n}{2}+1,n]}
    +H_{[1,\frac{n}{4}]:[\frac{n}{4}+1,\frac{n}{2}]}
    +H_{[\frac{n}{2}+1,\frac{3n}{4}]:[\frac{3n}{4}+1,n]}
    +H_{[1,\frac{n}{4}]}
    +H_{[\frac{n}{4}+1,\frac{n}{2}]}
    +H_{[\frac{n}{2}+1,\frac{3n}{4}]}
    +H_{[\frac{3n}{4}+1,n]}\\
    &=\cdots\\
    &=\sum_{\ell=1}^{\log n-1}\sum_{b=0}^{2^{\ell-1}-1}H_{[2b\frac{n}{2^\ell}+1,(2b+1)\frac{n}{2^\ell}]:[(2b+1)\frac{n}{2^\ell}+1,2(b+1)\frac{n}{2^\ell}]}.
\end{aligned}
\end{equation}
Here, the variable $\ell$ labels the layer of the decomposition and $b$ indexes the pairs of neighboring intervals within layer $\ell$. See \fig{recursion} for an illustration of this recursion for $\ell=1,2,3$. Correspondingly, we apply product formulas to decompose the entire evolution into exponentials of each individual $H_{[2b\frac{n}{2^\ell}+1,(2b+1)\frac{n}{2^\ell}]:[(2b+1)\frac{n}{2^\ell}+1,2(b+1)\frac{n}{2^\ell}]}$. Similar to the above analysis, this coarse-grained decomposition gives a Trotter error bound no larger than that of the fine-grained decomposition and implies the scaling \eq{powerlaw_r} as well.

To proceed, we take a closer look at this recursive decomposition and collect some of its features below.
\begin{enumerate}[leftmargin=*,label=(\roman*)]
    \item There are $\log n-1=\mathcal{O}(\log n)$ layers in the decomposition, all indexed by $\ell$.
    \item For a fixed layer $\ell$, there are $2^\ell$ consecutive intervals each of length $\frac{n}{2^\ell}$.
    \item Within each layer $\ell$, even and odd intervals are further grouped into $2^\ell-1$ pairs (indexed by $b$). Terms across intervals from the same pair are denoted by $H_{[2b\frac{n}{2^\ell}+1,(2b+1)\frac{n}{2^\ell}]:[(2b+1)\frac{n}{2^\ell}+1,2(b+1)\frac{n}{2^\ell}]}$. The total number of pairs of intervals is $\mathcal{O}(n)$ by the master theorem \lem{master}.
    \item When $\ell=1,\ldots,\log n-1$ and $b=0,\ldots,2^{\ell-1}-1$, $H_{[2b\frac{n}{2^\ell}+1,(2b+1)\frac{n}{2^\ell}]:[(2b+1)\frac{n}{2^\ell}+1,2(b+1)\frac{n}{2^\ell}]}$ provide a partition of all the terms in the original Hamiltonian $H$.
\end{enumerate}
We may use these features to further simplify our discussion. For instance, since we only have $2$-local terms of the same Pauli-type acting across disjoint intervals, we can simultaneously change the basis and consider only Pauli-$Z$ interactions without loss of generality:
\begin{equation}
    H_{[2b\frac{n}{2^\ell}+1,(2b+1)\frac{n}{2^\ell}]:[(2b+1)\frac{n}{2^\ell}+1,2(b+1)\frac{n}{2^\ell}]}
    =\sum_{\substack{2b\frac{n}{2^\ell}+1\leq u\leq (2b+1)\frac{n}{2^\ell}\\
    (2b+1)\frac{n}{2^\ell}+1\leq v\leq2(b+1)\frac{n}{2^\ell}}}
    \beta_{u,v}Z_uZ_v.
\end{equation}
Also, due to the nature of the recursive decomposition, it suffices to focus on implementing a specific $H_{[2b\frac{n}{2^\ell}+1,(2b+1)\frac{n}{2^\ell}]:[(2b+1)\frac{n}{2^\ell}+1,2(b+1)\frac{n}{2^\ell}]}$: we will have similar complexities $\cost\left(\cdot\right)$ for all pairs of intervals from different layers, so the total complexity
\begin{equation}
\label{eq:block_total_cost}
    \cost_{\text{rec}}(n)=
    \sum_{\ell=1}^{\log n-1}2^{\ell-1}\cost\left(\frac{n}{2^{\ell-1}}\right)
\end{equation}
can be immediately bounded by the master theorem \lem{master}.
For simplicity, we choose $\ell=1$, $b=0$ and we study the complexity $\cost(n)$ of exponentiating $H_{[1,\frac{n}{2}]:[\frac{n}{2}+1,n]}$.

We note that prior work has utilized above features of the recursive decomposition to develop measurement schedules for local quantum observables~\cite{Bonet-Monroig20}. For the purpose of quantum simulation, we will however need an additional key observation.
\begin{enumerate}[leftmargin=*,label=(\roman*),start=5]
    \item Coefficients of $H_{[1,\frac{n}{2}]:[\frac{n}{2}+1,n]}$ have the vector $1$-norm scaling:
    \begin{equation}
    \label{eq:decomposed_1norm}
    \norm{\beta}_1=\sum_{\substack{1\leq u\leq\frac{n}{2}\\
    \frac{n}{2}+1\leq v\leq n}}
    \abs{\beta_{u,v}}=
    \begin{cases}
    \mathcal{O}(1),\quad&\alpha>2,\\
    \mathcal{O}(\log n),&\alpha=2,\\
    \mathcal{O}(n^{2-\alpha}),&0<\alpha<2.
    \end{cases}
    \end{equation}
\end{enumerate}
Comparing \eq{decomposed_1norm} with \eq{naive_1norm}, the coefficients of $H_{[1,\frac{n}{2}]:[\frac{n}{2}+1,n]}$ have a significantly smaller 1-norm than that of the full Hamiltonian $H$.  
In particular, the 1-norm in \eq{decomposed_1norm} is asymptotically constant for $\alpha>2$, a factor of $n$ smaller than that of the original Hamiltonian. 
Therefore, exponentiating $H_{[1,\frac{n}{2}]:[\frac{n}{2}+1,n]}$ would also take a factor of $n$ fewer quantum gates than exponentiating $H$. Meanwhile, the gate complexity of block encoding remains asymptotically the same:
we will show in \sec{block_prepsel} that block encoding can be implemented for the pair of intervals $[1,\frac{n}{2}]$ and $[\frac{n}{2}+1,n]$ with $\cost(n)=\mathcal{O}(n\polylog(nt/\epsilon))$, which gives the total cost by applying the master theorem to \eq{block_total_cost}. This leads to the desired faster Trotter steps for simulating power-law Hamiltonians.

\subsection{Preparation and selection subroutines}
\label{sec:block_prepsel}
In the previous subsection, we have reduced the $1$-norm of power-law Hamiltonians by applying a recursive decomposition using product formulas. 
For completeness, we now give an explicit circuit for block encoding and apply qubitization to simulate the decomposed Hamiltonian.

To be precise, our goal is to simulate the following Hamiltonian acting across the intervals $[1,\frac{n}{2}]$ and $[\frac{n}{2}+1,n]$:
\begin{equation}
    H_{[1,\frac{n}{2}]:[\frac{n}{2}+1,n]}=\sum_{\substack{1\leq u\leq\frac{n}{2}\\\frac{n}{2}+1\leq v\leq n}}\beta_{u,v}Z_uZ_v,
\end{equation}
where the coefficients $|\beta_{u,v}|\leq 1/|u-v|^\alpha$. Additionally, we assume that the coefficients are \emph{logarithmically computable}, meaning the oracle
\begin{equation}
    O_{\beta}\ket{u,v,0}=\ket{u,v,\beta_{u,v}}
\end{equation}
can be implemented with cost $\mathcal{O}\left(\polylog(nt/\epsilon)\right)$.\footnote{Alternatively, one may treat $O_\beta$ as a black box and consider the query complexity.}
Note that by making this assumption, we have implicitly assumed certain structural properties of the Hamiltonian coefficients. 
For example, when $\beta_{u,v} = 1/\abs{u-v}^\alpha$ decays exactly as a power law, the oracle $O_\beta$ can be implemented with cost $\mathcal{O}\left(\polylog(nt/\epsilon)\right)$, fulfilling this requirement.
This can be achieved by performing elementary arithmetics such as subtraction, multiplication, and division on the binary representation of $u$ and $v$ (which has length $\mathcal{O}\left(\log(nt/\epsilon)\right)$).
However, when Hamiltonian coefficients are arbitrarily given, we prove a lower bound in \append{lowerbound2} showing that  at least $\sim n^2$ gates are required to implement $O_\beta$ in the circuit model.

In what follows, we analyze the number of queries to $O_\beta$ as well as additional interquery gates. 
We aim to block encode $H_{[1,\frac{n}{2}]:[\frac{n}{2}+1,n]}$ with its $1$-norm scaling close to \eq{decomposed_1norm}. As suggested in \sec{prelim_qubitization}, we may simply choose the selection subroutine to have the product structure
\begin{equation}
    \sum_{\substack{1\leq u\leq\frac{n}{2}\\\frac{n}{2}+1\leq v\leq n}}
    \ketbra{uv}{uv}\otimes Z_uZ_v.
\end{equation}
However, some extra efforts are needed to define the preparation subroutine. In particular, the following black-box state preparation subroutine
\begin{equation}
\begin{aligned}
    \frac{1}{n/2}\sum_{\substack{1\leq u\leq\frac{n}{2}\\\frac{n}{2}+1\leq v\leq n}}\ket{u}\ket{v}
    &\stackrel{O_\beta}{\mapsto}\frac{1}{n/2}\sum_{\substack{1\leq u\leq\frac{n}{2}\\\frac{n}{2}+1\leq v\leq n}}\ket{u}\ket{v}\ket{\beta_{u,v}}\\
    &\mapsto\frac{1}{n/2}\sum_{\substack{1\leq u\leq\frac{n}{2}\\\frac{n}{2}+1\leq v\leq n}}\ket{u}\ket{v}\ket{\beta_{u,v}}\left(\sqrt{|\beta_{u,v}|}\ket{0}+\sqrt{1-|\beta_{u,v}|}\ket{1}\right)\\
    &\xmapsto{O_\beta^\dagger}\frac{1}{n/2}\sum_{\substack{1\leq u\leq\frac{n}{2}\\\frac{n}{2}+1\leq v\leq n}}\left(\sqrt{|\beta_{u,v}|}\ket{u}\ket{v}\ket{0}+\sqrt{1-|\beta_{u,v}|}\ket{\psi_\bot}\ket{1}\right)
\end{aligned}
\end{equation}
does not work since it enlarges the $1$-norm of block encoding by a factor of $\frac{n}{2}$, preventing us from achieving the scaling in \eq{decomposed_1norm}.

The issue with the naive preparation subroutine is that we have prepared uniform superposition states at the beginning which ignored the power-law decaying pattern of $\beta_{u,v}$. 
To address this issue, we group the coefficients into multiple boxes. We take the number of boxes to be logarithmic in the system size $n$ so we can prepare superposition states over these boxes efficiently. Meanwhile, we ensure that the coefficients within each box are approximately uniform so we can again prepare the corresponding superposition state efficiently. This idea was previously used for the efficient block encoding of the plane-wave-basis electronic structure Hamiltonian in first quantization~\cite{Babbush2019}. Here, we describe a variant of this technique to efficiently block encode power-law Hamiltonians.

To simplify the notation, we shift the lattice and consider the rectangle region $(-\frac{n}{2},-1]\times[1,\frac{n}{2})$. We then divide it into boxes
\begin{equation}
    \mathcal{B}_{\mu,\nu}:=\left\{(u,v)|-2^{\mu+1}<u\leq-2^\mu,2^\nu\leq v<2^{\nu+1}\right\}
\end{equation}
for $\mu,\nu=0,1,\ldots,\log n-2$. With the restricted $\max$-norm
\begin{equation}
    \norm{\beta}_{\max,\mathcal{B}_{\mu,\nu}}=\max_{\substack{-2^{\mu+1}<u\leq-2^\mu\\2^\nu\leq v<2^{\nu+1}}}\left|\beta_{u,v}\right|,
\end{equation}
we will describe a state preparation subroutine whose complexity depends on the ratio between $\norm{\beta}_1$ and
\begin{equation}
    \norm{\beta}_{1,\text{box}}:=
    \sum_{\mu,\nu=0}^{\log n-2}2^{\mu+\nu}\norm{\beta}_{\max,\mathcal{B}_{\mu,\nu}}.
\end{equation}
Thus, our approach will be efficient if the coefficients are almost uniformly distributed within each box. The boundary terms 
can be handled separately without changing the asymptotic complexity scaling.

We start by preparing the state
\begin{equation}
\label{eq:nest_box}
    \frac{1}{\sqrt{\norm{\beta}_{1,\text{box}}}}\sum_{\mu,\nu=0}^{\log n-2}\sqrt{2^{\mu+\nu}\norm{\beta}_{\max,\mathcal{B}_{\mu,\nu}}}\ket{e_\mu}\ket{e_\nu},
\end{equation}
where we use unary encoding to represent $\mu$ and $\nu$. This state can be prepared using $\mathcal{O}(\log^2n)$ gates. Conditioned on the unary value of $\mu$ and $\nu$, we prepare the uniform superposition
\begin{equation}
    \frac{1}{\sqrt{\norm{\beta}_{1,\text{box}}}}\sum_{\mu,\nu=0}^{\log n-2}\sqrt{2^{\mu+\nu}\norm{\beta}_{\max,\mathcal{B}_{\mu,\nu}}}\ket{e_\mu}\ket{e_\nu}
    \frac{1}{\sqrt{2^\mu}}\sum_{-2^{\mu+1}<u\leq-2^\mu}\ket{u}
    \frac{1}{\sqrt{2^\nu}}\sum_{2^\nu\leq v<2^{\nu+1}}\ket{v},
\end{equation}
where we represent $u$ and $v$ in binary. We can achieve this by first transforming $\ket{e_\mu}$ and $\ket{e_\nu}$ into run-length unary representation, and performing a sequence of $\mathcal{O}(\log n)$ controlled Hadamard gates to generate the superposition. We now introduce an auxiliary uniform superposition state
\begin{equation}
    \frac{1}{\sqrt{\norm{\beta}_{1,\text{box}}}}\sum_{\mu,\nu=0}^{\log n-2}\sqrt{2^{\mu+\nu}\norm{\beta}_{\max,\mathcal{B}_{\mu,\nu}}}\ket{e_\mu}\ket{e_\nu}
    \frac{1}{\sqrt{2^{\mu+\nu}}}\sum_{\substack{-2^{\mu+1}<u\leq-2^\mu\\2^\nu\leq v<2^{\nu+1}}}\ket{u}\ket{v}
    \frac{1}{\sqrt{\Xi}}\sum_{\xi=0}^{\Xi-1}\ket{\xi}
\end{equation}
and test the inequality
\begin{equation}
    \frac{\xi}{\Xi}<\frac{|\beta_{u,v}|}{\norm{\beta}_{\max,\mathcal{B}_{\mu,\nu}}}.
\end{equation}
This can be performed with one query to the coefficient oracle $O_\beta$ and mild uses of other gates as follows.
We first apply $O_\beta$ to load the coefficients $\beta_{u,v}$. We also load the values of $\norm{\beta}_{\max,\mathcal{B}_{\mu,\nu}}$ via the transformation
\begin{equation}
    \ket{e_\mu,e_\nu,0}\mapsto\ket{e_\mu,e_\nu,\norm{\beta}_{\max,\mathcal{B}_{\mu,\nu}}}.
\end{equation}
Note that there are only $\mathcal{O}(\log^2n)$ different values of $\norm{\beta}_{\max,\mathcal{B}_{\mu,\nu}}$, each of which can be represented using at most $\mathcal{O}(\log(nt/\epsilon))$ bits, so we can efficiently perform this transformation. The inequality test can then be performed with a complexity scaling with the maximum size of the inputs. Assuming $\Xi$ is sufficiently large, the state after the inequality test (omitting the failure part and garbage registers) becomes
\begin{equation}
\label{eq:preamp}
\begin{aligned}
    &\ \frac{1}{\sqrt{\norm{\beta}_{1,\text{box}}}}\sum_{\mu,\nu=0}^{\log n-2}\sqrt{2^{\mu+\nu}\norm{\beta}_{\max,\mathcal{B}_{\mu,\nu}}}\ket{e_\mu}\ket{e_\nu}
    \frac{1}{\sqrt{2^{\mu+\nu}}}\sum_{\substack{-2^{\mu+1}<u\leq-2^\mu\\2^\nu\leq v<2^{\nu+1}}}\sqrt{\frac{|\beta_{u,v}|}{\norm{\beta}_{\max,\mathcal{B}_{\mu,\nu}}}}\ket{u}\ket{v}\\
    =&\ \frac{1}{\sqrt{\norm{\beta}_{1,\text{box}}}}\sum_{\mu,\nu=0}^{\log n-2}
    \sum_{\substack{-2^{\mu+1}<u\leq-2^\mu\\2^\nu\leq v<2^{\nu+1}}}\sqrt{|\beta_{u,v}|}\ket{e_\mu}\ket{e_\nu}\ket{u}\ket{v}.
\end{aligned}
\end{equation}
In practice, we choose a finite value of $\Xi$ which results in an erroneous block encoding that can be analyzed as in our second remark in \sec{prelim_qubitization}. To achieve an overall error of at most $\epsilon$, it suffices to set $\Xi=\poly(nt/\epsilon)$. Under this nested-boxes representation, the selection subroutine becomes
\begin{equation}
\label{eq:sel}
    \sum_{\mu,\nu=0}^{\log n-2}
    \sum_{\substack{-2^{\mu+1}<u\leq-2^\mu\\2^\nu\leq v<2^{\nu+1}}}
    \ketbra{e_\mu e_\nu}{e_\mu e_\nu}\otimes\ketbra{uv}{uv}\otimes Z_{\mu,u}Z_{\nu,v}.
\end{equation}
Again, this operation has a product structure and can be implemented with gate complexity $\mathcal{O}\left(n\log n\right)$ per \sec{prelim_qubitization}.

It is clear that we have prepared a state proportional to the desired state for block encoding when all $\beta_{u,v}>0$. If $\beta_{u,v}<0$, we simply introduce a minus sign in the implementation of the selection subroutine. The success probability is given by
\begin{equation}
    \frac{\sum\limits_{\mu,\nu=0}^{\log n-2}
    \sum\limits_{\substack{-2^{\mu+1}<u\leq-2^\mu\\2^\nu\leq v<2^{\nu+1}}}|\beta_{u,v}|}{\norm{\beta}_{1,\text{box}}}
    =\frac{\norm{\beta}_1}{\norm{\beta}_{1,\text{box}}}.
\end{equation}
To boost this success probability to close to $1$, we can perform $\mathcal{O}(\sqrt{\norm{\beta}_{1,\text{box}}/\norm{\beta}_1})$ steps of amplitude amplification. In other words, the complexity of state preparation will indeed depend on the ratio between $\norm{\beta}_{1,\text{box}}$ and $\norm{\beta}_1$ as previously claimed. 

If the distribution of Hamiltonian coefficients exactly matches a power law, we have
\begin{equation}
\label{eq:strict_powerlaw}
    \norm{\beta}_1=\sum_{\substack{-\frac{n}{2}<u\leq-1\\
    1\leq v< \frac{n}{2}}}
    \abs{\beta_{u,v}}=
    \begin{cases}
    \Theta(1),\quad&\alpha>2,\\
    \Theta(\log n),&\alpha=2,\\
    \Theta(n^{2-\alpha}),&0<\alpha<2.
    \end{cases}
\end{equation}
Furthermore, $\norm{\beta}_1$ differs from $\norm{\beta}_{1,\text{box}}$ by at most a constant factor. This is because
\begin{equation}
\label{eq:min_sum_max}
\begin{aligned}
    \sum_{\mu,\nu=0}^{\log n-2}2^{\mu+\nu}\min_{\substack{-2^{\mu+1}<u\leq-2^\mu\\2^\nu\leq v<2^{\nu+1}}}|\beta_{u,v}|
    &\leq\sum\limits_{\mu,\nu=0}^{\log n-2}
    \sum\limits_{\substack{-2^{\mu+1}<u\leq-2^\mu\\2^\nu\leq v<2^{\nu+1}}}|\beta_{u,v}|
    =\norm{\beta}_1\\
    &\leq\norm{\beta}_{1,\text{box}}
    =\sum_{\mu,\nu=0}^{\log n-2}2^{\mu+\nu}\max_{\substack{-2^{\mu+1}<u\leq-2^\mu\\2^\nu\leq v<2^{\nu+1}}}|\beta_{u,v}|,
\end{aligned}
\end{equation}
where the first and last quantities differ termwise by at most a factor of $2^{\alpha}=\mathcal{O}(1)$.
Therefore, the number of amplitude amplification steps is constant.
The cost of the preparation subroutine is then dominated by the query to the oracle $O_\beta$, which costs $\mathcal{O}\left(\polylog(nt/\epsilon)\right)$ to implement by our assumption.

We now use qubitization (\lem{qubitization}) to simulate $H_{[1,\frac{n}{2}]:[\frac{n}{2}+1,n]}$ for time $t/r$ with accuracy $\mathcal{O}\left(\epsilon/(rn)\right)$, where
\begin{equation}
    r=\begin{cases}
    \frac{n^{o(1)}t^{1+o(1)}}{\epsilon^{o(1)}},\quad&\alpha\geq1,\\[5pt]
    \frac{n^{1-\alpha+o(1)}t^{1+o(1)}}{\epsilon^{o(1)}},&0<\alpha<1,
    \end{cases}
\end{equation}
with the selection and preparation subroutine defined above. We bound the gate complexity $\cost(n)$ as follows:
\begin{equation}
\begin{aligned}
    \cost(n)&=
    \underbrace{\mathcal{O}\left(\norm{\beta}_1\frac{t}{r}+\log\left(\frac{rn}{\epsilon}\right)\right)}_{\text{number of qubitization steps}}
    \cdot\left(\underbrace{\mathcal{O}(n\log n)}_{\text{selection}}
    +\underbrace{\mathcal{O}\left(\sqrt{\frac{\norm{\beta}_{1,\text{box}}}{\norm{\beta}_1}}n\polylog(nt/\epsilon)\right)}_{\text{preparation}}\right)\\
    &=\begin{cases}
        \mathcal{O}\left(\left(\frac{t}{r}+1\right)n\polylog\left(\frac{nt}{\epsilon}\right)\right),\quad&\alpha\geq2,\\
        \mathcal{O}\left(\left(n^{2-\alpha}\frac{t}{r}+1\right)n\polylog\left(\frac{nt}{\epsilon}\right)\right),&0<\alpha<2.
    \end{cases}
\end{aligned}
\end{equation}
By using the same evolution and target accuracy with different values of $n$ and invoking the master theorem (\lem{master}), we obtain the gate complexity of implementing one Trotter step
\begin{equation}
    \cost_{\text{rec}}(n)=
    \sum_{\ell=1}^{\log n-1}2^{\ell-1}\cost\left(\frac{n}{2^{\ell-1}}\right)=
    \begin{cases}
        \mathcal{O}\left(\left(\frac{t}{r}+1\right)n\polylog\left(\frac{nt}{\epsilon}\right)\right),\quad&\alpha\geq2,\\
        \mathcal{O}\left(\left(n^{2-\alpha}\frac{t}{r}+1\right)n\polylog\left(\frac{nt}{\epsilon}\right)\right),&0<\alpha<2.
    \end{cases}\label{eq:sum-over-layers}
\end{equation}
This then gives the complexity of the entire quantum simulation:
\begin{equation}
    \begin{cases}
        nt\left(\frac{nt}{\epsilon}\right)^{o(1)},\quad&\alpha\geq2,\\[5pt]
        \mathcal{O}\left(n^{3-\alpha}t\polylog\left(\frac{nt}{\epsilon}\right)\right)+nt\left(\frac{nt}{\epsilon}\right)^{o(1)},&1\leq\alpha<2,\\[5pt]
        \mathcal{O}\left(n^{3-\alpha}t\polylog\left(\frac{nt}{\epsilon}\right)\right)+n^{2-\alpha}t\left(\frac{nt}{\epsilon}\right)^{o(1)},&0<\alpha<1.\\
    \end{cases}
\end{equation}

In general, the complexity of block encoding will depend on the closeness of distribution of Hamiltonian coefficients to a power-law distribution. To quantify this, we let $\lambda_{\text{block}}$ be the maximum ratio between $\norm{\beta}_{1,\text{box}}$ and $\norm{\beta}_1$, maximized over all pairs of intervals in the recursive decomposition \eq{decomp_block}. More explicitly,
\begin{equation}
\label{eq:lambda_block}
\begin{aligned}
    \lambda_{\text{block}}&:=\max_{\sigma\neq\sigma'\in\{x,y,z\}}\max_{\ell=1,\ldots,\log n-1}\max_{b=0,\ldots,2^{\ell-1}-1}\frac{\norm{\beta^{(\sigma,\sigma')}}_{1,\text{box},\mathcal{I}_{\ell,b}}}{\norm{\beta^{(\sigma,\sigma')}}_{1,\mathcal{I}_{\ell,b}}},
\end{aligned}
\end{equation}
where $\norm{\cdot}_{1,\mathcal{I}_{\ell,b}}$ and $\norm{\cdot}_{1,\text{box},\mathcal{I}_{\ell,b}}$ are the above norms defined with respect to the region
\begin{equation}
    \mathcal{I}_{\ell,b}:=\left\{(u,v)|\ 2b\frac{n}{2^\ell}+1\leq u\leq(2b+1)\frac{n}{2^\ell},(2b+1)\frac{n}{2^\ell}+1\leq v\leq2(b+1)\frac{n}{2^\ell}\right\}.
\end{equation}
Then, the complexity of quantum simulation should be revised to
\begin{equation}
    \begin{cases}
        \left(\sqrt{\lambda_{\text{block}}}+n\right)t\left(\frac{nt}{\epsilon}\right)^{o(1)},\quad&\alpha\geq2,\\[5pt]
        \left(\sqrt{\lambda_{\text{block}}}+n\right)\left(\mathcal{O}\left(n^{2-\alpha}t\polylog\left(\frac{nt}{\epsilon}\right)\right)+t\left(\frac{nt}{\epsilon}\right)^{o(1)}\right),&1\leq\alpha<2,\\[5pt]
        \left(\sqrt{\lambda_{\text{block}}}+n\right)\left(\mathcal{O}\left(n^{2-\alpha}t\polylog\left(\frac{nt}{\epsilon}\right)\right)+n^{1-\alpha}t\left(\frac{nt}{\epsilon}\right)^{o(1)}\right),&0<\alpha<1.\\
    \end{cases}
\end{equation}
It is clear that when the distribution of coefficients is close to a power law, \eq{min_sum_max} implies that $\lambda_{\text{block}}=\mathcal{O}(1)$
so we have recovered the cost scaling claimed earlier. When coefficients deviate significantly from a power law distribution, our gate complexity will enlarge by a factor of $\sqrt{\lambda_{\text{block}}}$ due to the use of amplitude amplification. However, we always have
\begin{equation*}
    \frac{\norm{\beta}_{1,\text{box}}}{\norm{\beta}_1}
    =\frac{\sum\limits_{\mu,\nu=0}^{\log n-2}2^{\mu+\nu}\max\limits_{\substack{-2^{\mu+1}<u\leq-2^\mu\\2^\nu\leq v<2^{\nu+1}}}|\beta_{u,v}|}{\sum\limits_{\mu,\nu=0}^{\log n-2}
    \sum\limits_{\substack{-2^{\mu+1}<u\leq-2^\mu\\2^\nu\leq v<2^{\nu+1}}}|\beta_{u,v}|}
    \leq\frac{n^2\sum\limits_{\mu,\nu=0}^{\log n-2}\max\limits_{\substack{-2^{\mu+1}<u\leq-2^\mu\\2^\nu\leq v<2^{\nu+1}}}|\beta_{u,v}|}{\sum\limits_{\mu,\nu=0}^{\log n-2}\max\limits_{\substack{-2^{\mu+1}<u\leq-2^\mu\\2^\nu\leq v<2^{\nu+1}}}|\beta_{u,v}|}
    =n^2,
\end{equation*}
which implies $\sqrt{\lambda_{\text{block}}}\leq n$. So the preparation subroutine still costs less than the selection subroutine, even when we perform the amplitude amplification. Thus our asymptotic gate complexity remains the same.

\subsection{Summary of the algorithm}
\label{sec:block_summary}
We now summarize the block-encoding method for implementing faster Trotter steps:
\begin{enumerate}
    \item Construct a preamplified preparation subroutine for \eq{preamp}.
    \item Perform $\mathcal{O}(\sqrt{\lambda_{\text{block}}})$ steps of amplitude amplification to construct the actual preparation subroutine with $\lambda_{\text{block}}$ defined in \eq{lambda_block}.
    \item Define the selection subroutine according to \eq{sel}.
    \item Use qubitization (\lem{qubitization}) to simulate for time $t/r$ with accuracy $\mathcal{O}\left(\epsilon/(rn)\right)$, with $r$ scaling like \eq{powerlaw_r}.
    \item Perform qubitization for all combinations of Pauli operators $\sigma,\sigma'=x,y,z$, layers indexed by $\ell=1,\ldots,\log n-1$ and pairs of intervals indexed by $b=0,\ldots,2^{\ell-1}-1$.
    \item Handle the remaining cases involving the identity matrix to implement a single Trotter step.
    \item Repeat $r$ Trotter steps to simulate the entire evolution.
\end{enumerate}

\begin{theorem}[Faster Trotter steps using block encoding]
\label{thm:block}
Consider $2$-local Hamiltonians $H=\sum_{\sigma,\sigma'\in\{i,x,y,z\}}\sum_{1\leq j<k\leq n}\beta^{(\sigma,\sigma')}_{j,k}P_j^{(\sigma)}P_k^{(\sigma')}$, where $\abs{\beta^{(\sigma,\sigma')}_{j,k}}\leq 1/|j-k|^\alpha$ for some constant $\alpha>0$ and $P^{(\sigma)}$ ($\sigma=i,x,y,z$) are the identity and Pauli matrices. Let $t>0$ be the simulation time and $\epsilon>0$ be the target accuracy. Assume that the coefficient oracle
\begin{equation}
    O_{\beta,\sigma,\sigma'}\ket{j,k,0}=\ket{j,k,\beta_{j,k}^{(\sigma,\sigma')}}
\end{equation}
can be implemented with gate complexity $\mathcal{O}\left(\polylog(nt/\epsilon)\right)$.
Then $H$ can be simulated using the algorithm of \sec{block_summary} with $\mathcal{O}\left(\log(nt/\epsilon)\right)$ ancilla qubits and gate complexity
\begin{equation}
    \begin{cases}
        nt\left(\frac{nt}{\epsilon}\right)^{o(1)},\quad&\alpha\geq2,\\[5pt]
        \mathcal{O}\left(n^{3-\alpha}t\polylog\left(\frac{nt}{\epsilon}\right)\right)+nt\left(\frac{nt}{\epsilon}\right)^{o(1)},&1\leq\alpha<2,\\[5pt]
        \mathcal{O}\left(n^{3-\alpha}t\polylog\left(\frac{nt}{\epsilon}\right)\right)+n^{2-\alpha}t\left(\frac{nt}{\epsilon}\right)^{o(1)},&0<\alpha<1.\\
    \end{cases}
\end{equation}
\end{theorem}

\section{Faster Trotter steps using average-cost simulation}
\label{sec:avgcost}
In the previous section, we have described a block-encoding-based method to simulate power-law Hamiltonians with efficiently computable coefficients. The complexity of our method is almost linear in the spacetime volume $(nt)^{1+o(1)}$ for $\alpha\geq2$ and is close to $n^{3-\alpha+o(1)}t^{1+o(1)}$ for $\alpha<2$, both of which improve the best results from previous work. 

In this section, we obtain a further improvement by performing an average-cost quantum simulation of commuting terms. We explain the basic idea of this technique in \sec{avgcost_comm}, with further details on the circuit implementation presented in \sec{avgcost_prepsel}. Readers may skip ahead to \sec{avgcost_summary} for a summary of the entire algorithm.

\subsection{Simulating commuting terms with average combination cost}
\label{sec:avgcost_comm}
We now show that gate complexities can be further reduced, using the simple fact that commuting Hamiltonian terms can be simulated with an \emph{average-cost linear combination}. To elaborate, consider a Hamiltonian $H=\sum_{\gamma=1}^\Gamma H_{\gamma}$ with Hermitian $H_{\gamma}$ and suppose we have block encodings $G_{\gamma,1}^\dagger U_\gamma G_{\gamma,0}=H_\gamma/\beta_\gamma$,
where $\beta_\gamma>0$, and $G_{\gamma,0},G_{\gamma,1}:\mathcal{G}\rightarrow\mathcal{H}$ and $U_\gamma:\mathcal{H}\rightarrow\mathcal{H}$ can be implemented with cost $c_\gamma>0$. Then, $H$ can be block encoded by defining
\begin{equation}
\begin{aligned}
    G_0&:\mathcal{G}\rightarrow\mathbb{C}^\Gamma\otimes\mathcal{H},\qquad\qquad &&G_0=\frac{1}{\sqrt{\norm{\beta}_1}}\sum_{\gamma=1}^\Gamma\sqrt{\beta_\gamma}\ket{\gamma}\otimes G_{\gamma,0}\\
    G_1&:\mathcal{G}\rightarrow\mathbb{C}^\Gamma\otimes\mathcal{H},\qquad\qquad &&G_1=\frac{1}{\sqrt{\norm{\beta}_1}}\sum_{\gamma=1}^\Gamma\sqrt{\beta_\gamma}\ket{\gamma}\otimes G_{\gamma,1}\\
    U&:\mathbb{C}^\Gamma\otimes\mathcal{H}\rightarrow\mathbb{C}^\Gamma\otimes\mathcal{H},
    &&U=\sum_{\gamma=1}^\Gamma\ketbra{\gamma}{\gamma}\otimes U_\gamma,\\
    G_1^\dagger UG_0&:\mathcal{G}\rightarrow\mathcal{G},
    &&G_1^\dagger UG_0=\frac{H}{\norm{\beta}_1}.
\end{aligned}
\end{equation}
This block encoding has a normalization factor of $\norm{\beta}_1$ and can be implemented with cost $\sum_\gamma c_\gamma$ (plus some additional cost for preparing the ancilla state $\frac{1}{\sqrt{\norm{\beta}_1}}\sum_{\gamma=1}^\Gamma\sqrt{\beta_\gamma}\ket{\gamma}$). Invoking \lem{qubitization}, we can simulate $H$ for time $t$ with accuracy $\epsilon$ with a cost scaling like
\begin{equation}
    \mathcal{O}\left(\left(\norm{\beta}_1t+\log\left(\frac{1}{\epsilon}\right)\right)\cdot\left(\sum_{\gamma=1}^\Gamma c_\gamma\right)\right),
\end{equation}
which reduces to $\sim t\left(\sum_\gamma\beta_\gamma\right)\left(\sum_\gamma c_\gamma\right)$ ignoring the error scaling. This is a worst-case combination because we are paying the same total cost $\sum_\gamma c_\gamma$ for each of the $\sim t\sum_\gamma\beta_\gamma$ qubitization steps.

This worst-case cost scaling may sometimes be avoided by recursively performing simulation in the interaction picture~\cite{LW18,Low18}. Roughly speaking, the recursive interaction-picture approach has a cost scaling like $\sim t\left(\sum_\gamma\beta_\gamma c_\gamma\right)\log^{2\Gamma-1}\left(t\norm{\beta}_{\max}\right)$. Thus we achieve the desired average-case combination cost, but also pick up a factor that scales exponentially with the number of terms $\Gamma$, which prevents the approach from being useful in many cases.

Instead, we make the following simple yet important observation about simulating commuting terms using block encodings and qubitization.
\begin{lemma}[Simulating commuting terms with average combination cost]
Let $G_{\gamma,0},G_{\gamma,1}:\mathcal{G}\rightarrow\mathcal{H}$ be isometries and $U_\gamma:\mathcal{H}\rightarrow\mathcal{H}$ be unitaries such that $G_{\gamma,1}^\dagger U_\gamma G_{\gamma,0}=H_\gamma/\beta_\gamma$ are Hermitian with $\beta_\gamma>0$. Assume that $H_\gamma$ pairwise commute. Given a target evolution time $t$ and accuracy $\epsilon$, there exist unitaries $V_{\varphi_\gamma}:\mathbb{C}^2\otimes\mathbb{C}^2\otimes\mathcal{H}\rightarrow\mathbb{C}^2\otimes\mathbb{C}^2\otimes\mathcal{H}$ parameterized by angles $\varphi_{\gamma,1},\ldots,\varphi_{\gamma,r_\gamma}$ such that
\begin{equation}
    \norm{\prod_{\gamma=1}^\Gamma\left(\left(\bra{+}\otimes\frac{\bra{0}\otimes G_{\gamma,0}^\dagger+\bra{1}\otimes G_{\gamma,1}^\dagger}{\sqrt{2}} \right)V_{\varphi_\gamma}\left(\ket{+}\otimes \frac{\ket{0}\otimes G_{\gamma,0}+\ket{1}\otimes G_{\gamma,1}}{\sqrt{2}}\right)\right)
    -e^{-itH}}\leq\epsilon.
\end{equation}
The number of steps $r_\gamma$ are even integers with the asymptotic scaling
\begin{equation}
    r_\gamma=\mathcal{O}\left(\beta_\gamma t+\log\left(\frac{\Gamma}{\epsilon}\right)\right),
\end{equation}
and $V_{\varphi_\gamma}$ are obtained by applying \lem{qubitization} to simulate $G_{\gamma,1}^\dagger U_\gamma G_{\gamma,0}$ for time $\beta_\gamma t$ with accuracy $\epsilon/\Gamma$.
\end{lemma}
\noindent In essence, we are just using the first-order Lie-Trotter formula $e^{-itH}=e^{-i tH_\Gamma}\cdots e^{-itH_1}$ with each exponential further simulated by the qubitization algorithm. Because Hamiltonian terms pairwise commute, there is no Trotter error introduced in this decomposition. As a result, we can simulate the target Hamiltonian with an average-case combination cost $\sim t\left(\sum_\gamma\beta_\gamma c_\gamma\right)$, without the unwanted exponential factor from the interaction-picture approach.\footnote{Note that we are using the same qubitization algorithm to simulate all the Hamiltonian terms, so there is a term-independent constant prefactor for all the gate complexities. This is why we have the summation rule $\sum_{\gamma=1}^\Gamma c_\gamma\mathcal{O}(\beta_\gamma t+\log(\Gamma/\epsilon))=\mathcal{O}\left(\left(\sum_{\gamma=1}^\Gamma\beta_\gamma c_\gamma\right)t+\sum_{\gamma=1}^\Gamma c_\gamma\log(\Gamma/\epsilon)\right)$. This rule will be used without declaration in the remainder of the paper.}

\begin{figure}[t]
	\centering
\includegraphics[width=0.7\textwidth]{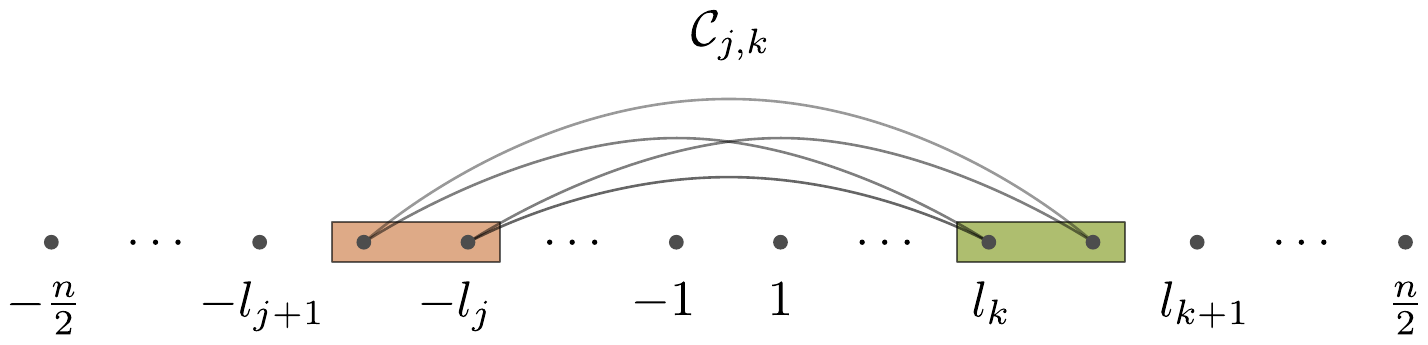}
\caption{Illustration of the interaction pairs in $\mathcal C_{j,k}$ defined in \eq{Cjk}. For convenience, we relabel the sites to $-n/2,\dots,-1$ and $1,\dots,n/2$.}
\label{fig:avg}
\end{figure}

For power-law Hamiltonians with $\alpha<2$, recall from \sec{block_norm} that we can without loss of generality consider
\begin{equation}
    H_{(-\frac{n}{2},-1]:[1,\frac{n}{2})}=\sum_{\substack{-\frac{n}{2}< u\leq-1\\1\leq v< \frac{n}{2}}}\beta_{u,v}Z_uZ_v,
\end{equation}
where we have dropped the subdominant terms and shifted the intervals for notational convenience.
Such a Hamiltonian term is directly block enocded and simulated by qubitization in \sec{block_prepsel}. We now show how that result can be further improved using the average-cost simulation technique. To this end, we divide each interval into $m$ subintervals at $\pm l_1,\ldots,\pm l_m$ and define
\begin{equation}
    \mathcal{C}_{j,k}:=
    \left\{(u,v)|-l_{j+1}< u\leq-l_{j},l_k\leq v<l_{k+1}\right\}.\label{eq:Cjk}
\end{equation}
See \fig{avg} for an illustration of the corresponding decomposition of Hamiltonian. For terms corresponding to $\mathcal{C}_{j,k}$, we have that the $1$-norm is asymptotically bounded by
\begin{equation}
    \norm{\beta}_{1,\mathcal{C}_{j,k}}=
    \mathcal{O}\left(\frac{\left(l_{j+1}-l_j\right)\left(l_{k+1}-l_k\right)}{\left(l_j+l_k\right)^\alpha}\right).
\end{equation}
We will describe in \sec{avgcost_prepsel} how to block encode the Hamiltonian terms within $\mathcal{C}_{j,k}$ with $1$-norm scaling exactly as above. Specifically, we show that up to polylogarithmic factors the preparation and selection subroutines have a total cost of
\begin{equation}
\label{eq:avgcost_block}
    \widetilde{\mathcal{O}}\left(1+\left(l_{j+1}-l_j\right)+\left(l_{k+1}-l_k\right)\right).
\end{equation}
In our above analysis, we have omitted a factor of $\lambda_{\text{avg}}$ due to the use of amplitude amplification. Just like \sec{block_prepsel}, this factor is close to $1$ when the distribution of Hamiltonian coefficients are close to a power-law distribution. We assume this is the case to simplify the following discussion, and present the full complexity expression in \sec{avgcost_prepsel}.

We use the uniform division $l_j=\Theta\left(j\frac{n}{m}\right)$ here for simplicity, although other divisions may lead to circuits with lower cost. Choosing the evolution time $t/r$ and ignoring the $\polylog(nt/\epsilon)$ scaling, we estimate the cost of simulating $H_{(-\frac{n}{2},-1]:[1,\frac{n}{2})}$ (for $\alpha<2$) as
\begin{equation}
\label{eq:avg_cost}
\begin{aligned}
    \widetilde{\mathcal{O}}\left(\sum_{j,k=1}^{m}\left(\frac{\left(\frac{n}{m}\right)^2\frac{t}{r}}{\left(j\frac{n}{m}+k\frac{n}{m}\right)^\alpha}+1\right)\left(1+\frac{n}{m}\right)\right)
    &=\widetilde{\mathcal{O}}\left(\left(\sum_{j,k=1}^{m}\frac{\left(\frac{n}{m}\right)^{2-\alpha}\frac{t}{r}}{(j+k)^\alpha}+m^2\right)\left(1+\frac{n}{m}\right)\right)\\
    &=\widetilde{\mathcal{O}}\left(\left(\left(\sum_{j=1}^{m}\frac{1}{j^{\alpha/2}}\right)^2\left(\frac{n}{m}\right)^{2-\alpha}\frac{t}{r}+m^2\right)\left(1+\frac{n}{m}\right)\right)\\
    &=\widetilde{\mathcal{O}}\left(\left(\left(m^{1-\frac{\alpha}{2}}\right)^2\left(\frac{n}{m}\right)^{2-\alpha}\frac{t}{r}+m^2\right)\left(1+\frac{n}{m}\right)\right)\\
    &=\widetilde{\mathcal{O}}\left(\left(n^{2-\alpha}\frac{t}{r}+m^2\right)\left(1+\frac{n}{m}\right)\right).
\end{aligned}
\end{equation}
We balance the two scalings in the first parentheses to optimize the gate complexity,
which implies 
\begin{equation}
    m=
    \begin{cases}
        \Theta\left(n^{1-\frac{\alpha}{2}}\right),\quad&1\leq\alpha<2,\\[5pt]
        \Theta\left(n^{\frac{1}{2}}\right),\quad&0<\alpha<1.
    \end{cases}
\end{equation}
For all values of $\alpha<2$, one can verify that $m=\Omega(1)$ and $m=\mathcal{O}(n)$ (in fact $m=\mathcal{O}(\sqrt{n})$), so this choice of $m$ is indeed valid. We give a slightly better (yet more complicated) choice of $m$ in \sec{avgcost_prepsel}.

The remaining analysis proceeds similarly as in \sec{block_prepsel}. We use qubitization (\lem{qubitization}) to simulate all $m^2$ pairs of subintervals in $H_{[1,\frac{n}{2}]:[\frac{n}{2}+1,n]}$ for time $t/r$ with accuracy $\mathcal{O}\left(\epsilon/(rm^2n)\right)$, where
\begin{equation}
    r=\begin{cases}
    \frac{n^{o(1)}t^{1+o(1)}}{\epsilon^{o(1)}},\quad&\alpha\geq1,\\[5pt]
    \frac{n^{1-\alpha+o(1)}t^{1+o(1)}}{\epsilon^{o(1)}},&0<\alpha<1,
    \end{cases}
\end{equation}
with an average-case combination cost as discussed above. We have the cost function
\begin{equation}
\begin{aligned}
    \begin{cases}
        \mathcal{O}\left(n^{2-\frac{\alpha}{2}}\polylog\left(\frac{nt}{\epsilon}\right)\right),\quad&1\leq\alpha<2,\\[5pt]
        \mathcal{O}\left(n^{\frac{3}{2}}\polylog\left(\frac{nt}{\epsilon}\right)\right),\quad&0<\alpha<1.
    \end{cases}
\end{aligned}
\end{equation}
By using the same evolution time and target accuracy with different values of $n$ and invoking the master theorem (\lem{master}), we obtain the gate complexity of implementing one Trotter step
\begin{equation}
    \cost_{\text{rec}}(n)=
    \sum_{l=1}^{\log n-1}2^{l-1}\cost\left(\frac{n}{2^{l-1}}\right)
    =\begin{cases}
        \mathcal{O}\left(n^{2-\frac{\alpha}{2}}\polylog\left(\frac{nt}{\epsilon}\right)\right),\quad&1\leq\alpha<2,\\[5pt]
        \mathcal{O}\left(n^{\frac{3}{2}}\polylog\left(\frac{nt}{\epsilon}\right)\right),\quad&0<\alpha<1.
    \end{cases}
\end{equation}
This then gives the complexity of the entire quantum simulation:
\begin{equation}
    \begin{cases}
        \frac{n^{2-\alpha/2+o(1)}t^{1+o(1)}}{\epsilon^{o(1)}},\quad&1\leq\alpha<2,\\[5pt]
        \frac{n^{5/2-\alpha+o(1)}t^{1+o(1)}}{\epsilon^{o(1)}},&0<\alpha<1.
    \end{cases}
\end{equation}

\subsection{Preparation and selection subroutines}
\label{sec:avgcost_prepsel}
We now describe a circuit that achieves the average-case combination cost for block encoding power-law Hamiltonians claimed in the previous subsection.

Specifically, let $l_j,l_{j+1},l_k,l_{k+1}$ be arbitrary integers such that $1\leq l_j<l_{j+1}\leq\frac{n}{2}$ and $1\leq l_k<l_{k+1}\leq\frac{n}{2}$. Our goal is to block encode the Hamiltonian
\begin{equation}
    H_{(-l_{j+1},-l_j]:[l_k,l_{k+1})}
    =\sum_{\substack{-l_{j+1}<u\leq-l_j\\l_k\leq v<l_{k+1}}}\beta_{u,v}Z_uZ_v,
\end{equation}
where $|\beta_{u,v}|\leq 1/|u-v|^\alpha$ and we have shifted the intervals for notational convenience. Additionally, we assume that the coefficients are \emph{logarithmically computable},\footnote{As before, one may instead treat $O_\beta$ as a black box and consider the query complexity.}\ meaning the oracle
\begin{equation}
    O_\beta\ket{u,v,0}=\ket{u,v,\beta_{u,v}}
\end{equation}
can be implemented with cost $\mathcal{O}\left(\polylog(nt/\epsilon)\right)$. By making this oracle assumption, we have implicitly assumed certain underlying structure of the Hamiltonian coefficients: in \append{lowerbound2} we show that one needs $\sim n^2$ gates to implement $O_\beta$ in the circuit model when structural properties of coefficients are unavailable. The selection subroutine can simply be chosen as
\begin{equation}
\label{eq:avg_sel}
    \sum_{-l_{j+1}<u\leq-l_j}\sum_{l_k\leq v<l_{k+1}}\ketbra{u,v}{u,v}\otimes Z_uZ_v.
\end{equation}
Due to the product structure of this operation, we can implement it with gate complexity $\mathcal{O}\left(\frac{n}{m}\right)$.

In what follows, we analyze the preparation subroutine. Here, we only consider the uniform division $l_j=\Theta\left(j\frac{n}{m}\right)$, as this is enough to justify the gate complexity claimed in \sec{avgcost_comm}. We will use the naive black-box state preparation technique, because coefficients within the divided subintervals are already close to uniform. We start by preparing the uniform superposition
\begin{equation}
    \frac{1}{\sqrt{l_{j+1}-l_{j}}}\sum_{-l_{j+1}<u\leq-l_j}\ket{u}
    \frac{1}{\sqrt{l_{k+1}-l_{k}}}\sum_{l_k\leq v<l_{k+1}}\ket{v}.
\end{equation}
Then we invoke the black-box state preparation subroutine
\begin{equation}
\label{eq:avg_preamp}
\begin{aligned}
    &\ \frac{1}{\sqrt{\left(l_{j+1}-l_{j}\right)\left(l_{k+1}-l_{k}\right)}}\sum_{\substack{-l_{j+1}<u\leq-l_j\\l_k\leq v<l_{k+1}}}\ket{u,v}\\
    \xmapsto{O_\beta}&\ \frac{1}{\sqrt{\left(l_{j+1}-l_{j}\right)\left(l_{k+1}-l_{k}\right)}}\sum_{\substack{-l_{j+1}<u\leq-l_j\\l_k\leq v<l_{k+1}}}\ket{u,v}\ket{\beta_{u,v}}\\
    \mapsto&\ \frac{1}{\sqrt{\left(l_{j+1}-l_{j}\right)\left(l_{k+1}-l_{k}\right)}}\sum_{\substack{-l_{j+1}<u\leq-l_j\\l_k\leq v<l_{k+1}}}\ket{u,v}\ket{\beta_{u,v}}\left(\sqrt{\frac{|\beta_{u,v}|}{\norm{\beta}_{\max}}}\ket{0}+\sqrt{\frac{\norm{\beta}_{\max}-|\beta_{u,v}|}{\norm{\beta}_{\max}}}\ket{1}\right),
\end{aligned}
\end{equation}
where the last step can be realized using inequality test in a way similar to \sec{block_prepsel}.
There is no need to uncompute the ancilla register $\ket{\beta_{u,v}}$, as the uncomputation is automatically performed in the qubitization algorithm. It is clear that we have prepared a state proportional to the desired state for block encoding when all $\beta_{u,v}>0$. If some $\beta_{u,v}<0$, we simply introduce a minus sign in the implementation of the selection subroutine. The success probability is given by
\begin{equation}
    \frac{\sum_{\substack{-l_{j+1}<u\leq-l_j\\l_k\leq v<l_{k+1}}}|\beta_{u,v}|}{\left(l_{j+1}-l_{j}\right)\left(l_{k+1}-l_{k}\right)\norm{\beta}_{\max}}=\frac{\norm{\beta}_1}{\left(l_{j+1}-l_{j}\right)\left(l_{k+1}-l_{k}\right)\norm{\beta}_{\max}}.
\end{equation}
To boost this probability to close to $1$, we can perform $\mathcal{O}\left(\sqrt{\left(l_{j+1}-l_{j}\right)\left(l_{k+1}-l_{k}\right)\norm{\beta}_{\max}/\norm{\beta}_1}\right)$ steps of amplitude amplification.

If the distribution of Hamiltonian coefficients exactly matches a power law, we have
\begin{equation}
    \norm{\beta}_1=\sum_{\substack{-l_{j+1}<u\leq-l_j\\l_k\leq v<l_{k+1}}}|\beta_{u,v}|\geq\frac{\left(l_{j+1}-l_{j}\right)\left(l_{k+1}-l_{k}\right)}{\left(l_{j+1}+l_{k+1}\right)^\alpha},\qquad
    \norm{\beta}_{\max}\leq\frac{1}{\left(l_{j}+l_{k}\right)^\alpha},
\end{equation}
so the number of amplification steps scales like
\begin{equation}
    \mathcal{O}\left(\sqrt{\frac{\left(l_{j+1}-l_{j}\right)\left(l_{k+1}-l_{k}\right)\norm{\beta}_{\max}}{\norm{\beta}_1}}\right)=\mathcal{O}\left(\sqrt{\frac{\left(l_{j+1}+l_{k+1}\right)^\alpha}{\left(l_{j}+l_{k}\right)^\alpha}}\right)=\mathcal{O}\left(1\right),
\end{equation}
which justifies the previous claim in \eq{avgcost_block}. In general, the complexity of block encoding will depend on the closeness of the distribution of Hamiltonian coefficients to a power-law distribution. To quantify this, we let $\lambda_{\text{avg}}$ be the maximum ratio between $\left(l_{j+1}-l_{j}\right)\left(l_{k+1}-l_{k}\right)\norm{\beta}_{\max}$ and $\norm{\beta}_1$, maximized over all pairs of intervals in the decomposition \eq{decomp_block} with a further uniform division. More explicitly,
\begin{equation}
\label{eq:lambda_avg}
\begin{aligned}
    \lambda_{\text{avg}}&:=\max_{\sigma\neq\sigma'\in\{x,y,z\}}\max_{\ell=1,\ldots,\log n-1}\max_{b=0,\ldots,2^{\ell-1}-1}\\
    &\qquad\max_{j,k=1,\ldots,m}\frac{\left(l_{\ell,b,j+1}-l_{\ell,b,j}\right)\left(l_{\ell,b,k+1}-l_{\ell,b,k}\right)\norm{\beta^{(\sigma,\sigma')}}_{\max,\mathcal{I}_{\ell,b,j,k}}}{\norm{\beta^{(\sigma,\sigma')}}_{1,\mathcal{I}_{\ell,b,j,k}}},
\end{aligned}
\end{equation}
where $l_{\ell,b,j}$ are the uniform division points, and $\norm{\cdot}_{\max,\mathcal{I}_{\ell,b,j,k}}$ and $\norm{\cdot}_{1,\mathcal{I}_{\ell,b,j,k}}$ are the $\max$- and $1$-norm restricted to the region
\begin{equation}
\begin{aligned}
    \mathcal{I}_{\ell,b,j,k}:=\Big\{(u,v)|\ &(2b+1)\frac{n}{2^\ell}-l_{\ell,b,j+1}< u\leq(2b+1)\frac{n}{2^\ell}-l_{\ell,b,j},\\
    &(2b+1)\frac{n}{2^\ell}+l_{\ell,b,k}\leq v\leq(2b+1)\frac{n}{2^\ell}+l_{\ell,b,k+1}\Big\}.
\end{aligned}
\end{equation}

With this definition, the cost of simulating $H_{(-\frac{n}{2},-1]:[1,\frac{n}{2})}$ for time $t/r$ should be revised to
\begin{equation}
    \cost(n)=\widetilde{\mathcal{O}}\left(\left(n^{2-\alpha}\frac{t}{r}+m^2\right)\left(\sqrt{\lambda_{\text{avg}}}+\frac{n}{m}\right)\right).
\end{equation}
However, we always have
\begin{equation*}
    \frac{\left(l_{j+1}-l_{j}\right)\left(l_{k+1}-l_{k}\right)\norm{\beta}_{\max}}{\norm{\beta}_1}
    \leq\frac{\left(\frac{n}{m}\right)^2\norm{\beta}_{\max}}{\norm{\beta}_{\max}}
    =\left(\frac{n}{m}\right)^2,
\end{equation*}
which implies $\sqrt{\lambda_{\text{avg}}}\leq n/m$. So the preparation subroutine still costs less than the selection subroutine, even when we perform the amplitude amplification.
We balance the first term by choosing
\begin{equation}
    m=\max\left\{\Theta\left(n^{1-\alpha/2}\left(\frac{t}{r}\right)^{1/2}\right),1\right\},
\end{equation}
which implies through the master theorem (\lem{master}) that the total simulation has complexity
\begin{equation}
    \widetilde{\mathcal{O}}\left(\left(n^{2-\alpha}t+r\right)\min\left\{n^{\alpha/2}\left(\frac{r}{t}\right)^{1/2},n\right\}\right).
\end{equation}
Inserting the scaling of $r$ from \eq{powerlaw_r}, we obtain the cost scaling
\begin{equation}
    \begin{cases}
        \min\bigg\{n^{2-\frac{\alpha}{2}}t\left(\frac{nt}{\epsilon}\right)^{o(1)},
        \ \mathcal{O}\left(n^{3-\alpha}t\polylog\left(\frac{nt}{\epsilon}\right)\right)
        +nt\left(\frac{nt}{\epsilon}\right)^{o(1)}\bigg\},
        \quad&1\leq\alpha<2,\\[0.35cm]
        \min\bigg\{n^{\frac{5}{2}-\alpha}t\left(\frac{nt}{\epsilon}\right)^{o(1)},
        \ \mathcal{O}\left(n^{3-\alpha}t\polylog\left(\frac{nt}{\epsilon}\right)\right)
        +n^{2-\alpha}t\left(\frac{nt}{\epsilon}\right)^{o(1)}\bigg\},
        &0<\alpha<1.
    \end{cases}
\end{equation}

\subsection{Summary of the algorithm}
\label{sec:avgcost_summary}
We now summarize the average-cost simulation method for implementing faster Trotter steps:
\begin{enumerate}
    \item Construct a preamplified preparation subroutine for \eq{avg_preamp}.
    \item Perform $\mathcal{O}(\sqrt{\lambda_{\text{avg}}})$ steps of amplitude amplification to construct the actual preparation subroutine with $\lambda_{\text{avg}}$ defined in \eq{lambda_avg}.
    \item Define the selection subroutine according to \eq{avg_sel}.
    \item Use qubitization (\lem{qubitization}) to simulate for time $t/r$ with accuracy $\mathcal{O}\left(\epsilon/(rn^2)\right)$, with $r$ scaling like \eq{powerlaw_r}.
    \item Perform qubitization for all combinations of Pauli operators $\sigma,\sigma'=x,y,z$, layers indexed by $\ell=1,\ldots,\log n-1$, intervals indexed by $b=0,\ldots,2^{\ell-1}-1$ and pairs of subintervals indexed by $j,k=1,\ldots,m$.
    \item Handle the remaining cases involving the identity matrix to implement a single Trotter step.
    \item Repeat $r$ Trotter steps to simulate the entire evolution.
\end{enumerate}

\begin{theorem}[Faster Trotter steps using average-cost simulation]
\label{thm:avgcost}
Consider $2$-local Hamiltonians $H=\sum_{\sigma,\sigma'\in\{i,x,y,z\}}\sum_{1\leq j<k\leq n}\beta^{(\sigma,\sigma')}_{j,k}P_j^{(\sigma)}P_k^{(\sigma')}$, where $\abs{\beta^{(\sigma,\sigma')}_{j,k}}\leq 1/|j-k|^\alpha$ for some constant $\alpha>0$ and $P^{(\sigma)}$ ($\sigma=i,x,y,z$) are the identity and Pauli matrices. Let $t>0$ be the simulation time and $\epsilon>0$ be the target accuracy. Assume that the coefficient oracle
\begin{equation}
    O_{\beta,\sigma,\sigma'}\ket{j,k,0}=\ket{j,k,\beta_{j,k}^{(\sigma,\sigma')}}
\end{equation}
can be implemented with gate complexity $\mathcal{O}\left(\polylog(nt/\epsilon)\right)$. Then $H$ can be simulated using the algorithm of \sec{avgcost_summary} with $\mathcal{O}\left(\log(nt/\epsilon)\right)$ ancilla qubits and gate complexity
\begin{equation}
    \begin{cases}
        \min\bigg\{n^{2-\frac{\alpha}{2}}t\left(\frac{nt}{\epsilon}\right)^{o(1)},
        \ \mathcal{O}\left(n^{3-\alpha}t\polylog\left(\frac{nt}{\epsilon}\right)\right)
        +nt\left(\frac{nt}{\epsilon}\right)^{o(1)}\bigg\},
        \quad&1\leq\alpha<2,\\[0.35cm]
        \min\bigg\{n^{\frac{5}{2}-\alpha}t\left(\frac{nt}{\epsilon}\right)^{o(1)},
        \ \mathcal{O}\left(n^{3-\alpha}t\polylog\left(\frac{nt}{\epsilon}\right)\right)
        +n^{2-\alpha}t\left(\frac{nt}{\epsilon}\right)^{o(1)}\bigg\},
        &0<\alpha<1.
    \end{cases}
\end{equation}
\end{theorem}

\section{Faster Trotter steps using low-rank decomposition}
\label{sec:rank}
In the previous sections, we have shown how the $1$-norm of Hamiltonian coefficients can be reduced via a recursive decomposition using product formulas, which results in faster circuit implementation of Trotter steps. We further improve our result by simulating commuting terms with an average-case combination cost. Assuming Hamiltonian coefficients are efficiently computable, these techniques together enable simulations of power-law systems with complexity nearly linear in the spacetime volume for $\alpha\geq2$, whereas the cost becomes $n^{2-\alpha/2+o(1)}t^{1+o(1)}$ for $1\leq\alpha<2$ and $n^{5/2-\alpha+o(1)}t^{1+o(1)}$ for $0<\alpha<1$.

In this section, we describe a method for implementing a Trotter step through a recursive decomposition of the Hamiltonian using its hierarchical low-rank structure \cite{Hackbusch1999,Grasedyck2003}. 
This low-rank structure was previously used in~\cite{NKL22} to block encode kernel matrices. Here, we directly use the recursive decomposition in \sec{rank_decomp} to construct circuits without block encoding. The overall algorithm and its complexity are then summarized in \sec{rank_summary}.

\subsection{Recursive low-rank decomposition}
\label{sec:rank_decomp}

The initial steps of our method are the same as that of \sec{block_norm}. In particular, we expand the power-law Hamiltonian in the Pauli basis as in \eq{pauli_expansion}, and use product formulas to perform a coarse-grained decomposition. Without loss of generality, we may focus on $2$-local terms
\begin{equation}
    H=\sum_{1\leq j<k\leq n}\beta_{j,k}X_jY_k,
\end{equation}
as the on-site terms can be implemented with subdominant cost and the remaining $2$-local terms can be handled similarly by a change of basis. As before, we assume that the system size $n$ is a power of $2$ and use
\begin{equation}
\begin{aligned}
    H_{[j,k]}&:=\sum_{j\leq u<v\leq k}\beta_{u,v}X_uY_v\
    \text{($1\leq j< k\leq n$)},\\
    H_{[j,k]:[l,m]}&:=\sum_{\substack{j\leq u\leq k\\l\leq v\leq m}}\beta_{u,v}X_uY_v\
    \text{($1\leq j\leq k<l\leq m\leq n$)}
\end{aligned}
\end{equation}
to represent terms within a specific interval and across two disjoint intervals of sites.

\begin{figure}[t]
	\centering
\includegraphics[width=0.95\textwidth]{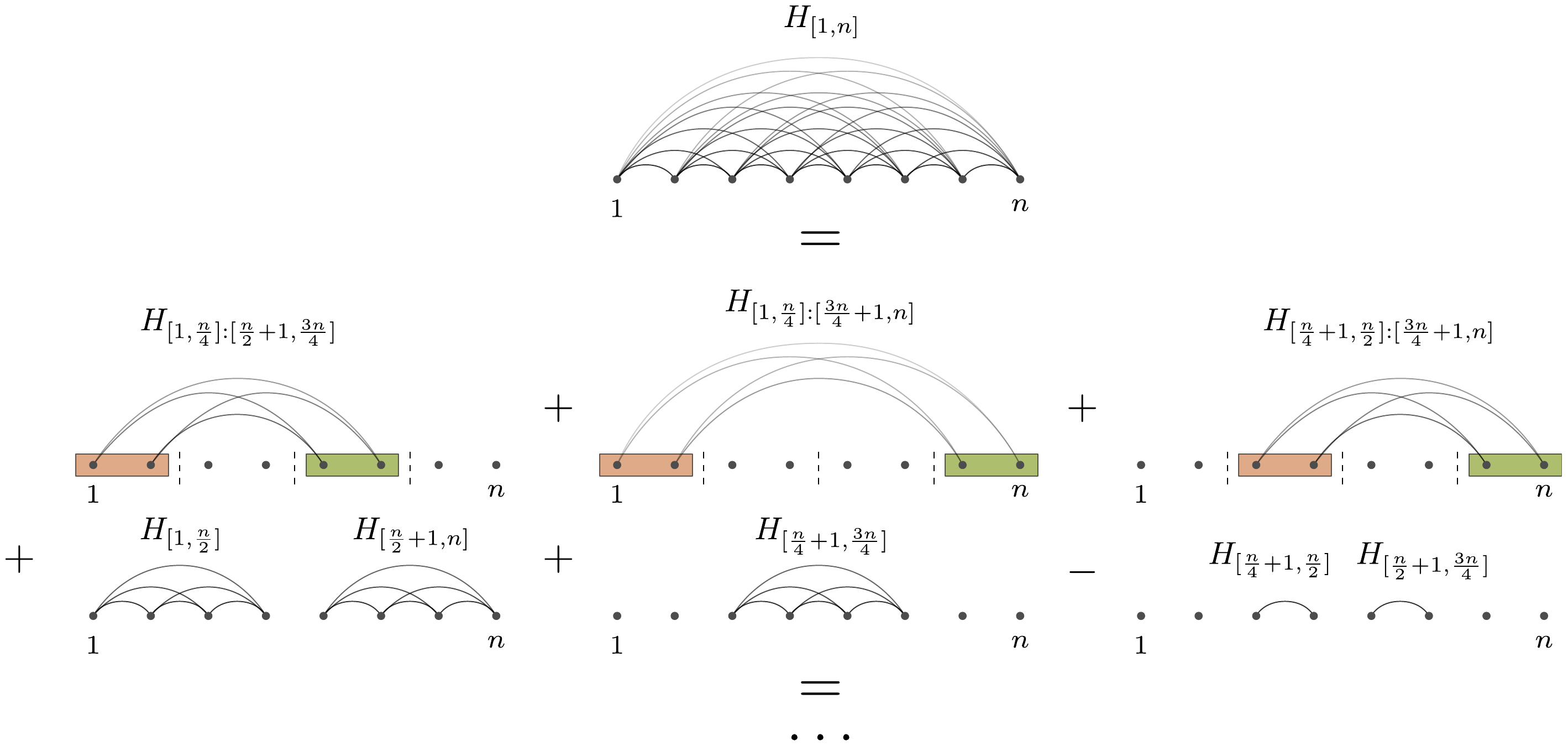}
\caption{Illustration of one iteration [\eq{low-rank-basic-decomp}] in the low-rank decomposition. We repeatedly apply \eq{low-rank-basic-decomp} to the remaining terms in the second row to arrive at the final decomposition \eq{decomp_rank}.}
\label{fig:rank}
\end{figure}

However, we now use a decomposition different from that of \sec{block_norm}. Specifically, we use the recurrence relation
\begin{equation}
\begin{aligned}
    H_{[j,k]}&= H_{[j,j+\delta-1]:[j+2\delta,j+3\delta-1]}
    +H_{[j,j+\delta-1]:[j+3\delta,k]}
    +H_{[j+\delta,j+2\delta-1]:[j+3\delta,k]}\\
    &\quad+H_{[j,j+2\delta-1]}
    +H_{[j+\delta,j+3\delta-1]}
    +H_{[j+2\delta,k]}\\
    &\quad-H_{[j+\delta,j+2\delta-1]}
    -H_{[j+2\delta,j+3\delta-1]}\label{eq:low-rank-basic-decomp}
\end{aligned}
\end{equation}
with $\delta=\floor{(k-j+1)/4}$, which unwraps to layer $h=\Theta(\log n)$ as
\begin{equation}
\label{eq:decomp_rank}
\begin{aligned}
    H&=H_{[1,n]}\\
    &=H_{[1,\frac{n}{4}]:[\frac{n}{2}+1,\frac{3n}{4}]}
    +H_{[1,\frac{n}{4}]:[\frac{3n}{4}+1,n]}
    +H_{[\frac{n}{4}+1,\frac{n}{2}]:[\frac{3n}{4}+1,n]}\\
    &\quad+H_{[1,\frac{n}{2}]}+H_{[\frac{n}{4}+1,\frac{3n}{4}]}+H_{[\frac{n}{2}+1,n]}\\
    &\quad-H_{[\frac{n}{4}+1,\frac{n}{2}]}-H_{[\frac{n}{2}+1,\frac{3n}{4}]}\\
    &=\cdots\\
    &=\sum_{\ell=2}^{h}
    \sum_{b=0}^{2^{\ell-1}-2}\bigg(
    H_{[1+2b\frac{n}{2^{\ell}},(2b+1)\frac{n}{2^{\ell}}]:[1+(2b+2)\frac{n}{2^{\ell}},(2b+3)\frac{n}{2^{\ell}}]}\\
    &\qquad\qquad\qquad+H_{[1+2b\frac{n}{2^{\ell}},(2b+1)\frac{n}{2^{\ell}}]:[1+(2b+3)\frac{n}{2^{\ell}},(2b+4)\frac{n}{2^{\ell}}]}\\
    &\qquad\qquad\qquad+H_{[1+(2b+1)\frac{n}{2^{\ell}},(2b+2)\frac{n}{2^{\ell}}]:[1+(2b+3)\frac{n}{2^{\ell}},(2b+4)\frac{n}{2^{\ell}}]}\bigg)\\
    &\quad+\sum_{b=0}^{2^h-1}H_{[1+b\frac{n}{2^\ell},(b+1)\frac{n}{2^\ell}]}
    +\sum_{b=0}^{2^h-2}H_{[1+b\frac{n}{2^\ell},(b+1)\frac{n}{2^\ell}]:[1+(b+1)\frac{n}{2^\ell},(b+2)\frac{n}{2^\ell}]}.
\end{aligned}
\end{equation}
See \fig{rank} for an illustration of this decomposition at the first nontrivial layer $\ell=2$.

We observe the following features of the decomposition that are helpful to describe our circuit implementation:
\begin{enumerate}[leftmargin=*,label=(\roman*)]
    \item There are $h-1=\Theta(\log n)$ layers in the decomposition, all indexed by $\ell$.
    \item For a fixed layer $\ell$, there are $2^{\ell}$ consecutive intervals each of length $\frac{n}{2^\ell}$.
    \item Within each layer $\ell$, intervals are further divided into $2^{\ell-1}$ blocks (indexed by $b$). Within each pair of consecutive blocks, we only keep the three terms that act on intervals with distance at least $\frac{n}{2^\ell}$. They are $H_{[1+2b\frac{n}{2^{\ell}},(2b+1)\frac{n}{2^{\ell}}]:[1+(2b+2)\frac{n}{2^{\ell}},(2b+3)\frac{n}{2^{\ell}}]}$, $H_{[1+2b\frac{n}{2^{\ell}},(2b+1)\frac{n}{2^{\ell}}]:[1+(2b+3)\frac{n}{2^{\ell}},(2b+4)\frac{n}{2^{\ell}}]}$, and $H_{[1+(2b+1)\frac{n}{2^{\ell}},(2b+2)\frac{n}{2^{\ell}}]:[1+(2b+3)\frac{n}{2^{\ell}},(2b+4)\frac{n}{2^{\ell}}]}$. The total number of pairs of intervals is $\mathcal{O}(n)$ by the master theorem \lem{master}.
    \item When $\ell=2,\ldots,h$ and $b=0,\ldots,2^{\ell-1}-2$, the decomposition provides a partition of all the terms in the Hamiltonian with distance at least $\frac{n}{2^h}$.
\end{enumerate}
As in \sec{block_norm}, we may use these features to simplify our analysis. For instance, since we only have $2$-local terms with the same type of Pauli operators acting across disjoint intervals, we can simultaneously change the basis and consider only Pauli-$Z$ interactions, e.g.,
\begin{equation}
    H_{[1+2b\frac{n}{2^{\ell}},(2b+1)\frac{n}{2^{\ell}}]:[1+(2b+2)\frac{n}{2^{\ell}},(2b+3)\frac{n}{2^{\ell}}]}
    =\sum_{\substack{1+2b\frac{n}{2^{\ell}}\leq u\leq(2b+1)\frac{n}{2^{\ell}}\\1+(2b+2)\frac{n}{2^{\ell}}\leq v\leq(2b+3)\frac{n}{2^{\ell}}}}
    \beta_{u,v}Z_uZ_v.
\end{equation}
Also, due to the nature of the recursive decomposition, it suffices to focus on the implementation of a specific term such as $H_{[1+2b\frac{n}{2^{\ell}},(2b+1)\frac{n}{2^{\ell}}]:[1+(2b+2)\frac{n}{2^{\ell}},(2b+3)\frac{n}{2^{\ell}}]}$. We will show momentarily that all the decomposed terms can be implemented with similar complexities $\cost(\cdot)$, while the remaining terms such as $H_{[1+b\frac{n}{2^\ell},(b+1)\frac{n}{2^\ell}]}$ and $H_{[1+b\frac{n}{2^\ell},(b+1)\frac{n}{2^\ell}]:[1+(b+1)\frac{n}{2^\ell},(b+2)\frac{n}{2^\ell}]}$ act on constant-size intervals and can be handled by a sequential implementation using product formulas. This means the total complexity can be bounded as
\begin{equation}
    \cost_{\text{rec}}(n)=\sum_{\ell=2}^h 3\left(2^{\ell-1}-1\right)\cost\left(\frac{n}{2^{\ell-1}}\right)+\mathcal{O}(n),
\end{equation}
which again reduces to the study of $\cost(\cdot)$ because of the master theorem (\lem{master}).
For simplicity, we choose $\ell=2$, $b=0$ and we study the complexity $\cost(\frac{n}{2})$ of simulating $H_{[1,\frac{n}{4}]:[\frac{n}{2}+1,\frac{3n}{4}]}$.

Our main motivation to consider this decomposition is made clear through the following rank assumption of $H_{[1,\frac{n}{4}]:[\frac{n}{2}+1,\frac{3n}{4}]}$:
\begin{enumerate}[leftmargin=*,label=(\roman*),start=5]
    \item Coefficients of $H_{[1,\frac{n}{4}]:[\frac{n}{2}+1,\frac{3n}{4}]}$, when organized into an $\frac{n}{4}$-by-$\frac{n}{4}$ matrix, have rank at most $\rho$, i.e., $\rank\left([\beta_{u,v}]\right)\leq\rho$.
\end{enumerate}
Because of this low-rank property, the coefficient matrix admits the thin singular value decomposition~\cite[Theorem 7.3.2]{horn2012matrix}
\begin{equation}
    \beta_{u,v}=\sum_{s=1}^\rho \mu_{u,s}\sigma_{s}\nu_{v,s}
\end{equation}
for $u=1\ldots,\frac{n}{4}$ and $v=\frac{n}{2}+1,\ldots,\frac{3n}{4}$, where $\mu$ and $\nu$, when viewed as $\frac{n}{4}$-by-$\rho$ matrices, have real orthonormal columns, and $\sigma_s$ are singular values bounded by the induced $1$-norm of $\beta$ and $\beta^\top$: $0\leq\sigma_s\leq\sqrt{\vertiii{\beta}_1\vertiii{\beta^\top}_1}$~\cite[5.6.P21]{horn2012matrix}. Correspondingly, the exponential of $H_{[1,\frac{n}{4}]:[\frac{n}{2}+1,\frac{3n}{4}]}$ can be rewritten as
\begin{equation}
    e^{-itH_{[1,\frac{n}{4}]:[\frac{n}{2}+1,\frac{3n}{4}]}}
    =e^{-it\sum_{s=1}^\rho\sigma_{s} \left(\sum_{u=1}^{\frac{n}{4}}\mu_{u,s}Z_u\right)\left(\sum_{v=\frac{n}{2}+1}^{\frac{3n}{r}}\nu_{v,s}Z_v\right)}.
\end{equation}

With respect to the computational basis, our target exponential has the action
\begin{equation}
\label{eq:rank_phase}
    e^{-itH_{[1,\frac{n}{4}]:[\frac{n}{2}+1,\frac{3n}{4}]}}\ket{z_n,\ldots,z_1}
    =e^{-it\sum_{s=1}^\rho\sigma_{s} \left(\sum_{u=1}^{\frac{n}{4}}\mu_{u,s}(-1)^{z_u}\right)\left(\sum_{v=\frac{n}{2}+1}^{\frac{3n}{r}}\nu_{v,s}(-1)^{z_v}\right)}\ket{z_n,\ldots,z_1}.
\end{equation}
This suggests a circuit implementation as follows. We first introduce an ancilla register of size $\mathcal{O}(\log(nt/\epsilon))$, and compute the following function in superposition
\begin{equation}
    \begin{bmatrix}
    z_n & \cdots & z_1
    \end{bmatrix}
    \mapsto \sum_{s=1}^\rho\sigma_{s} \left(\sum_{u=1}^{\frac{n}{4}}\mu_{u,s}(-1)^{z_u}\right)\left(\sum_{v=\frac{n}{2}+1}^{\frac{3n}{r}}\nu_{v,s}(-1)^{z_v}\right).
\end{equation}
Note that the function values are real numbers of size at most $\mathcal{O}(\poly(n))$ and can thus be approximately stored in the $\mathcal{O}(\log(nt/\epsilon))$-qubit ancilla register. We then use a sequence of controlled rotations to introduce the phase. We may now uncompute the ancilla register by reverting the circuit. This implements the desired exponential $e^{-itH_{[1,\frac{n}{4}]:[\frac{n}{2}+1,\frac{3n}{4}]}}$.

Our above circuit has a gate complexity of $\cost\left(\frac{n}{2}\right)=\mathcal{O}\left(n\rho\polylog(nt/\epsilon)\right)$ for simulating one pair of intervals in the recursive decomposition. To implement the Trotter step, we redefine $\rho$ to be the maximum truncation rank of coefficient matrices, maximized over all pairs of intervals in the decomposition \eq{decomp_block}. Explicitly,
\begin{equation}
\label{eq:rho_rank}
    \rho:=\max_{\sigma\neq\sigma'\in\{x,y,z\}}\max_{\ell=2,\ldots,h}\max_{b=0,\ldots,2^{l-1}-2}\rho_{\ell,b}^{(\sigma,\sigma')},
\end{equation}
where $\rho_{\ell,b}$ is the largest truncation rank for the terms $H_{[1+2b\frac{n}{2^{\ell}},(2b+1)\frac{n}{2^{\ell}}]:[1+(2b+2)\frac{n}{2^{\ell}},(2b+3)\frac{n}{2^{\ell}}]}$,\linebreak $H_{[1+2b\frac{n}{2^{\ell}},(2b+1)\frac{n}{2^{\ell}}]:[1+(2b+3)\frac{n}{2^{\ell}},(2b+4)\frac{n}{2^{\ell}}]}$, and $H_{[1+(2b+1)\frac{n}{2^{\ell}},(2b+2)\frac{n}{2^{\ell}}]:[1+(2b+3)\frac{n}{2^{\ell}},(2b+4)\frac{n}{2^{\ell}}]}$. By the master theorem (\lem{master}), this implies the circuit for Trotter steps has the same asymptotic cost
\begin{equation}
    \cost_{\text{rec}}(n)=\sum_{\ell=2}^h 3\left(2^{\ell-1}-1\right)\cost\left(\frac{n}{2^{\ell-1}}\right)+\mathcal{O}(n)
    =\mathcal{O}\left(n\rho\polylog\left(\frac{nt}{\epsilon}\right)\right).
\end{equation}
See \append{master} for further details. By simulating for time $t/r$ in each step and repeating $r$ steps where
\begin{equation}
    r=\begin{cases}
    \frac{n^{o(1)}t^{1+o(1)}}{\epsilon^{o(1)}},\quad&\alpha\geq1,\\[5pt]
    \frac{n^{1-\alpha+o(1)}t^{1+o(1)}}{\epsilon^{o(1)}},&0<\alpha<1,
    \end{cases}
\end{equation}
we obtain the total gate complexity of the low-rank simulation method
\begin{equation}
    \begin{cases}
    \frac{\rho(nt)^{1+o(1)}}{\epsilon^{o(1)}},\quad&\alpha\geq1,\\[5pt]
    \frac{\rho n^{2-\alpha+o(1)}t^{1+o(1)}}{\epsilon^{o(1)}},&0<\alpha<1.
    \end{cases}
\end{equation}

Before ending this section, we discuss the important question of how the rank $\rho$ is determined in the above decomposition. For various classes of power-law interactions including the Coulomb interaction, there are rigorous analyses based on the multipole expansion showing that $\rho=\mathcal{O}(\log(nt/\epsilon))$ suffices to guarantee that the simulation is $\epsilon$-accurate,\footnote{Note that this logarithmic factor will be absorbed by $(nt/\epsilon)^{o(1)}$ in the final gate complexity.} which leads to the gate complexity claimed in \tab{result_summary}. In fact, we will soon examine an application of this method to simulating the real-space electronic structure Hamiltonian in \sec{app}. In practice, one may also run numerical simulations to empirically determine the rank value $\rho$. We refer the reader to~\cite{Greengard1987} and papers citing this work for detailed studies of the low-rank decomposition in the classical setting.

\subsection{Summary of the algorithm}
\label{sec:rank_summary}
We now summarize the low-rank method for implementing faster Trotter steps:
\begin{enumerate}
    \item Use diagonalization to implement the low-rank matrix exponential \eq{rank_phase}.
    \item Perform matrix exponentials for all combinations of Pauli operators $\sigma,\sigma'=x,y,z$, layers indexed by $\ell=2,\ldots,h$ and blocks indexed by $b=0,\ldots,2^{\ell-1}-2$, as well as the constant-size blocks.
    \item Handle the remaining cases involving the identity matrix to implement a single Trotter step.
    \item Repeat $r$ Trotter steps to simulate the entire evolution.
\end{enumerate}

\begin{theorem}[Faster Trotter steps using low-rank decomposition]
\label{thm:rank}
Consider $2$-local Hamiltonians $H=\sum_{\sigma,\sigma'\in\{i,x,y,z\}}\sum_{1\leq j<k\leq n}\beta^{(\sigma,\sigma')}_{j,k}P_j^{(\sigma)}P_k^{(\sigma')}$, where $\abs{\beta^{(\sigma,\sigma')}_{j,k}}\leq 1/|j-k|^\alpha$ for some constant $\alpha>0$ and $P^{(\sigma)}$ ($\sigma=i,x,y,z$) are the identity and Pauli matrices. Let $t>0$ be the simulation time and $\epsilon>0$ be the target accuracy. Then $H$ can be simulated using the algorithm of \sec{rank_summary} with $\mathcal{O}\left(\log(nt/\epsilon)\right)$ ancilla qubits and gate complexity
\begin{equation}
    \begin{cases}
        \rho nt\left(\frac{nt}{\epsilon}\right)^{o(1)},\quad&\alpha\geq1,\\[5pt]
        \rho n^{2-\alpha}t\left(\frac{nt}{\epsilon}\right)^{o(1)},&\alpha<1.
    \end{cases}
\end{equation}
Here, $1\leq\rho\leq n$ defined in \eq{rho_rank} is the maximum truncation rank of certain off-diagonal blocks of coefficient matrices ($\rho=\mathcal{O}\left(\log(nt/\epsilon)\right)$ if the coefficient distribution exactly matches a power law in one spatial dimension).
\end{theorem}

\section{Applications to real-space quantum simulation}
\label{sec:app}
Simulating electronic structure Hamiltonians is one of the most widely studied problems in quantum simulation~\cite{mcardle2020quantum,cao2019quantum,MottaEmerging2022}. An efficient solution of the electronic structure problem could lead to better understandings of catalysts and materials, which has applications in numerous subareas of physics and chemistry. Here, we consider mapping the electronic structure Hamiltonian on a grid and performing simulation in the second quantization in real space. 
Compared to general molecular basis Hamiltonians~\cite{Lee21}, the grid-based Hamiltonian contains much fewer terms with well-structured coefficients, which is useful for reducing the resource requirement of quantum simulation.
We will present an algorithm combining our method for Trotter step implementation with a tighter error analysis, which improves the best simulation results from previous work.

We consider the following class of Hamiltonians
\begin{equation}
\label{eq:second_quantized_ham}
    H=T+V:=\sum_{j,k}\tau_{j,k}A_j^\dagger A_k+\sum_{l,m}\nu_{l,m}N_l N_m,
\end{equation}
where $A_j^\dagger$, $A_k$ are the fermionic creation and annihilation operators, $N_l$ are the occupation-number operators, $\tau$, $\nu$ are coefficient matrices, and the summations are over $n$ spin orbitals. We can represent the real-space electronic structure Hamiltonians in the above form with specific choices of coefficients $\tau$ and $\nu$; we will come back to this point momentarily. Then, we simulate the Hamiltonian using product formulas by splitting $e^{-itH}$ into products of $e^{-itT}$ and $e^{-itV}$. The Trotter error corresponding to this splitting was studied in~\cite{SHC21}. There, they found that a $p$th-order product formula $S_p(t)$ has error scaling like
\begin{equation}
\label{eq:old_fermionic_bound1}
    \norm{S_p(t)-e^{-itH}}_{\mathcal{W}_\eta}
	=\mathcal{O}\left(\left(\norm{\tau}+\norm{\nu}_{\max}\eta\right)^{p-1}
		\norm{\tau}\norm{\nu}_{\max}\eta^2 t^{p+1}\right),
\end{equation}
where recall $\norm{\cdot}_{\max}$ is the max-norm denoting the largest matrix element in absolute value, $\norm{\cdot}$ is the operator norm and
\begin{equation}
\label{eq:fermionic_seminorm_def}
	\norm{X}_{\mathcal{W}_\eta}
	:=\max_{\ket{\psi_\eta},\ket{\phi_\eta}\in\mathcal{W}_\eta}\abs{\bra{\phi_\eta}X\ket{\psi_\eta}}
\end{equation}
is the restriction of the operator norm to the $\eta$-electron subspace. Furthermore, if the coefficient matrices $\tau$ and $\nu$ are $s$-sparse (with at most $s$ nonzero elements in each row and column), then it holds that
\begin{equation}
\label{eq:old_fermionic_bound2}
    \norm{S_p(t)-e^{-itH}}_{\mathcal{W}_\eta}
	=\mathcal{O}\left(\left(\norm{\tau}_{\max}+\norm{\nu}_{\max}\right)^{p-1}
		\norm{\tau}_{\max}\norm{\nu}_{\max}s^{p+1}\eta t^{p+1}\right).
\end{equation}

The new bound we prove is as follows:
\begin{equation}
    \norm{S_p(t)-e^{-itH}}_{\mathcal{W}_\eta}
	=\mathcal{O}\left(\left(\vertiii{\tau}_{1}+\vertiii{\nu}_{1,[\eta]}\right)^{p-1}
		\vertiii{\tau}_{1}\vertiii{\nu}_{1,[\eta]}\eta t^{p+1}\right),
\end{equation}
where we have used the induced $1$-norm and its restricted version
\begin{equation}
    \vertiii{\tau}_1:=\max_j\sum_k|\tau_{j,k}|,\qquad
    \vertiii{\nu}_{1,[\eta]}:=\max_{j}\max_{k_1<\cdots<k_\eta}\left(\abs{\nu_{j,k_1}}+\cdots+\abs{\nu_{j,k_\eta}}\right).
\end{equation}
This new bound is strictly better than \eq{old_fermionic_bound2} because for an $s$-sparse coefficient matrix $\beta$, it holds that $\vertiii{\beta}_{1,[\eta]}\leq\vertiii{\beta}_1\leq s\norm{\beta}_{\max}$ and the equalities are not always attainable. In fact, as we will see momentarily, the gap between the induced $1$-norm $\vertiii{\beta}_{1,[\eta]}$ and the max-norm $s\norm{\beta}_{\max}$ is quite significant for the Coulomb interaction which is relevant for the electronic structure simulation. Compared with \eq{old_fermionic_bound1}, our induced $1$-norm scaling $\vertiii{\nu}_{1,[\eta]}$ is again better than $\norm{\nu}_{\max}\eta$, but our dependence on $\tau$ is slightly worse since $\vertiii{\tau}_1\geq\norm{\tau}$ for a Hermitian $\tau$. Fortunately, this is not an issue with the real-space electronic structure Hamiltonian. To avoid diluting the main message of our work, we state our Trotter error bound below and leave its proof to \append{trotter}.

\begin{theorem}[Trotter error with fermionic induced $1$-norm scaling]
\label{thm:fermionic_induced_1norm}
Let $H=T+V:=\sum_{j,k}\tau_{j,k}A_j^\dagger A_k\linebreak+\sum_{l,m}\nu_{l,m}N_l N_m$ be an interacting-electronic Hamiltonian, and $S_p(t)$ be a $p$th-order product formula splitting the evolutions under $T$ and $V$. Then,
\begin{equation}
    \norm{S_p(t)-e^{-itH}}_{\mathcal{W}_\eta}
	=\mathcal{O}\left(\left(\vertiii{\tau}_{1}+\vertiii{\nu}_{1,[\eta]}\right)^{p-1}
		\vertiii{\tau}_{1}\vertiii{\nu}_{1,[\eta]}\eta t^{p+1}\right).
\end{equation}
Here, $\vertiii{\tau}_{1}$ and $\vertiii{\nu}_{1,[\eta]}$ are the (restricted) fermionic induced $1$-norm defined by
\begin{equation}
    \vertiii{\tau}_{1}=\max_j\sum_k\abs{\tau_{j,k}},\qquad
    \vertiii{\nu}_{1,[\eta]}=\max_{j}\max_{k_1<\cdots<k_\eta}\left(\abs{\nu_{j,k_1}}+\cdots+\abs{\nu_{j,k_\eta}}\right).
\end{equation}
\end{theorem}

We now apply the above theorem to the simulation of electronic structure Hamiltonians in real space. In this case, we have $H=T+U+V$, where $T$ represents the kinetic term, $U$ represents the external potential term introduced by the Born-Oppenheimer approximation, and $V$ represents the Coulomb potential term. We first consider the uniform electron gas without the external potential $U$. The Coulomb interaction takes the form~\cite[Eq.\ (K4)]{SBWRB21}
\begin{equation}
    V=\frac{n^{1/3}}{2\omega^{1/3}}\sum_{l,m}\frac{1}{\norm{l-m}}N_lN_m,
\end{equation}
where $n$ is the number of spin orbitals, $\omega$ is the volume of the computational cell, $l$ and $m$ are 3D vectors with each coordinate taking integer values between $-\frac{n^{1/3}-1}{2}$ and $\frac{n^{1/3}-1}{2}$. This term can be represented as $V=\sum_{l,m}\nu_{l,m}N_lN_m$ and it is easy to check that $\norm{\nu}_{\max}=\mathcal{O}\left(n^{1/3}/\omega^{1/3}\right)$.
On the other hand, we can estimate the (restricted) induced $1$-norm as follows. For a fixed value of $l$, the $\eta$ spin orbitals nearest to $l$, if occupied, will have the largest possible coefficients $1/\norm{l-m}$. These $\eta$ spin orbitals form a ball of radius $\mathcal{O}(\eta^{1/3})$, which is also inside of a cube of linear size $\mathcal{O}(\eta^{1/3})$. This implies that~\cite[Lemma H.1]{CSTWZ19}
\begin{equation}
    \vertiii{\nu}_{1,[\eta]}=\mathcal{O}\left(\frac{\eta^{2/3}n^{1/3}}{\omega^{1/3}}\right).
\end{equation}
We thus see that the scaling of $\vertiii{\nu}_{1,[\eta]}$ is strictly less than $\eta\norm{\nu}_{\max}$, giving a factor of $\eta^{1/3}$ improvement over the best previous result. Meanwhile, the scaling for the kinetic term is\footnote{The scaling we report here corresponds to the fermionic-Fourier-transform implementation~\cite[Eq.\ (K2)]{SBWRB21} (whose analysis in turn follows from the bounds~\cite[Eqs.\ (F11)-(F13)]{BWMMNC18} in second quantization). The finite-difference scheme of~\cite[Eq.\ (A7)]{BWMMNC18} seems to have a different scaling for the kinetic term, which should be further studied by future work.}
\begin{equation}
    \vertiii{\tau}_1=\mathcal{O}\left(\frac{n^{2/3}}{\omega^{2/3}}\right).
\end{equation}
This determines the asymptotic scaling of the number of Trotter steps as
\begin{equation}
    \left(\frac{\eta^{2/3}n^{1/3}}{\omega^{1/3}}
    +\frac{n^{2/3}}{\omega^{2/3}}\right)\frac{n^{o(1)}t^{1+o(1)}}{\epsilon^{o(1)}}.
\end{equation}

To implement each Trotter step, one could sequentially exponentiate all terms in the Hamiltonian with cost $\Theta(n^2)$. However, note that under the Jordan-Wigner encoding, the Coulomb potential term is translated to the spin Hamiltonian
\begin{equation}
    V=\frac{n^{1/3}}{2\omega^{1/3}}\sum_{l,m}\frac{1}{\norm{l-m}}\frac{I-Z_l}{2}\frac{I-Z_m}{2},
\end{equation}
with interaction coefficients decaying with distance according to power law. This can be handled by the recursive low-rank method described in \sec{rank} for the 1D case with straightforward generalizations in \append{generalize_dim} to higher spatial dimensions. One can show using the multipole expansion that the decomposed terms can be approximated using matrices with rank polylogarithmic in the input parameters~\cite{Greengard1987,LinTongLowRank}. So the complexity of implementing the Coulomb potential term is $\mathcal{O}\left(n\polylog(nt/\epsilon)\right)$.

For the kinetic part, we can either represent it using a finite difference scheme~\cite{BWMMNC18}, or using a diagonal form in the kinetic basis and fermionic-Fourier-transforming back~\cite{SBWRB21}. In the first representation, we have geometrically local terms and we can further split them using product formulas without increasing the asymptotic scaling of Trotter error. Alternatively, we can modify boundary terms (using product formulas) so that the coefficients are cyclic, and the kinetic operator can then be implemented using the fermionic Fourier transform \cite{BWMMNC18} (we can also use this fermionic-Fourier-transform implementation for the diagonal-form representation in the kinetic basis). As the first derivatives of the electronic wavefunction are not continuous at electron cusps~\cite{Myers1991KatoCusp}, we note that the error introduced by even a first-order difference scheme should not differ asymptotically from high-order difference schemes, or even an exact diagonalization of the kinetic part, in the limit of large $n$. 
In either case, we can exponentiate the kinetic part with an asymptotic cost of $\mathcal{O}\left(n\polylog(nt/\epsilon)\right)$ as well. This gives the total gate complexity
\begin{equation}
\label{eq:real_space_cost}
    \left(\frac{\eta^{2/3}n^{4/3}}{\omega^{1/3}}
    +\frac{n^{5/3}}{\omega^{2/3}}\right)\frac{n^{o(1)}t^{1+o(1)}}{\epsilon^{o(1)}}
\end{equation}
for the uniform electron gas, which simplifies to $\left(\eta^{1/3}n^{1/3}+\frac{n^{2/3}}{\eta^{2/3}}\right)n^{1+o(1)}$ assuming that the density of electrons $\eta/\omega=\mathcal{O}(1)$, the evolution time $t=\mathcal{O}(1)$ and the accuracy $\epsilon=\Theta(1)$ are all fixed.

In second quantization, electronic structure Hamiltonians can be simulated in real space by implementing sequential Trotter steps, which is ancilla-free with the asymptotic gate complexity $\left(\frac{n^{8/3}}{\eta^{2/3}}+n^{7/3}\eta^{2/3}\right)n^{o(1)}$. This complexity can be further improved by diagonalizing the Coulomb potential using the (classical) fast Fourier transform~\cite{LW18,SHC21}, although the ancilla space complexity increases significantly to $\mathcal{O}(n\log(n))$. By contrast, our recursive low-rank method combined with the tighter Trotter error bound achieves a strictly lower complexity while using only $\mathcal{O}\left(\log(n)\right)$ ancillas.
In first quantization, recent work~\cite{SBWRB21} gives an algorithm based on the interaction-picture simulation~\cite{LW18} with complexity $\widetilde{\mathcal{O}}\left(\eta^{8/3}n^{1/3}\right)$. So our result, while not dominating in all parameter regimes, is asymptotically better as long as $n=\eta^{7/3-o(1)}$. See \tab{electron_result_summary} for a more detailed comparison.

\begin{table}
	\begin{center}
        \renewcommand{\arraystretch}{1.5}
        \resizebox{1.\textwidth}{!}{%
		\begin{tabular}{c|cc|cc}
			\hline
   \multirow{2}{*}{Simulation algorithms} & \multicolumn{2}{c|}{Gate complexity} & \multicolumn{2}{c}{Space complexity}\\
   \cline{2-5}
			 & $n,\eta$ & $\eta=\Theta(n)$ & System & Ancilla\\
			\hline
                First-quantized qubitization~\cite{Babbush2019} & $\widetilde{\mathcal{O}}\left(n^{2/3}\eta^{4/3}+n^{1/3}\eta^{8/3}\right)$ & $\widetilde{\mathcal{O}}\left(n^3\right)$ & $\eta\log(n)$ & $\widetilde{\mathcal{O}}\left(\eta\right)$\\
			First-quantized interaction-picture~\cite{Babbush2019} & $\widetilde{\mathcal{O}}\left(n^{1/3}\eta^{8/3}\right)$ & $\widetilde{\mathcal{O}}\left(n^3\right)$ & $\eta\log(n)$ & $\widetilde{\mathcal{O}}\left(\eta\right)$\\
			Second-quantized interaction-picture~\cite{LW18} & $\widetilde{\mathcal{O}}\left(\frac{n^{8/3}}{\eta^{2/3}}\right)$ & $\widetilde{\mathcal{O}}\left(n^2\right)$ & $n$ & $\mathcal{O}\left(n\log(n)\right)$\\
                Second-quantized Trotterization~\cite{SHC21} (sequential) & $\left(\frac{n^{8/3}}{\eta^{2/3}}+n^{7/3}\eta^{2/3}\right)n^{o(1)}$ & $n^{3+o(1)}$ & $n$ & $0$\\
			Second-quantized Trotterization~\cite{SHC21} (fast-Fourier-transform) & $\left(\frac{n^{5/3}}{\eta^{2/3}}+n^{4/3}\eta^{2/3}\right)n^{o(1)}$ & $n^{2+o(1)}$ & $n$ & $\mathcal{O}\left(n\log(n)\right)$\\
			\hline
			Second-quantized Trotterization (recursive low-rank) & $\left(\frac{n^{5/3}}{\eta^{2/3}}+n^{4/3}\eta^{1/3}\right)n^{o(1)}$ & $n^{5/3+o(1)}$ & $n$ & $\mathcal{O}\left(\log(n)\right)$\\
			\hline
		\end{tabular}
        }
        \renewcommand{\arraystretch}{1}
	\end{center}
	\caption{Comparison of our result and previous results for simulating the uniform electron gas with $n$ spin orbitals and $\eta$ electrons. We use $\widetilde{\mathcal{O}}(\cdot)$ to suppress $\polylog(n)$ factors in the complexity scaling. The complexity of the recursive low-rank method follows from \thm{rank} (with its 3D extension discussed in \append{generalize_dim}), as well as the tighter Trotter error bound described in \thm{fermionic_induced_1norm}.}
	\label{tab:electron_result_summary}
\end{table}

For the general case, we have the external potential term 
\begin{equation}
    U=\sum_{m}\left(\sum_l\frac{\zeta_l}{\norm{\widetilde{r}_l-r_m}}\right)N_m
\end{equation}
introduced under the Born-Oppenheimer approximation, where $\zeta_l$ are nuclear charges, $\widetilde{r}_l$ are nuclear coordinates, and $r_m$ are the positions of electrons ($r_m=\frac{\omega^{1/3}}{n^{1/3}}m$). In this case, we will have a complexity similar to \eq{real_space_cost}, but with an additional contribution from the external potential that depends on
\begin{equation}
    \max_m\sum_l\frac{\zeta_l}{\norm{\widetilde{r}_l-r_m}}.
\end{equation}
We note that this is strictly better than the interaction-picture algorithm from previous work~\cite{SBWRB21} which instead depends on a larger quantity
\begin{equation}
    \left(\sum_l\zeta_l\right)\max_{l,m}\frac{1}{\norm{\widetilde{r}_l-r_m}}
    =\mathcal{O}\left(\frac{\eta n^{1/3}}{\omega^{1/3}}\right).
\end{equation}
A study of how much this improvement is for practical electronic structure problems will be left as a subject for future work.

\section{First circuit lower bound}
\label{sec:lowerbound}
We have presented three methods for implementing faster Trotter steps and identified applications to electronic structure simulation in second quantization in real space. It is worth mentioning that all our methods make use of structural properties of the Hamiltonian in an essential way. Without such guarantees, the coefficient oracle in the block-encoding and average-cost simulation described in \sec{block} and \sec{avgcost} would have an implementation cost that depends on the number of terms $\Theta(n^2)$, whereas the blocks in the recursive decomposition of \sec{rank} would no longer have low rank. Consequently, all our methods will reduce to the sequential circuit implementation of Trotter steps.

In this section, we construct a class of $2$-local Hamiltonians with only Pauli-$Z$ interactions with coefficients taking a continuum range of values. We show that one needs at least $\Omega(n^2)$ gates to evolve this class of Hamiltonians with accuracy $\epsilon=\Omega(1/\poly(n))$ for time $t=\Omega(\epsilon)$. This suggests that using structural properties of the Hamiltonian is both necessary and sufficient for performing faster Trotter steps. We prove a related circuit lower bound in \append{lowerbound2} to justify the necessity of structural properties to efficiently implement the coefficient oracle $O_\beta$ with approximate circuit synthesis.

The proof of our circuit lower bound is based on a gate-efficient reduction to the problem of approximately synthesizing diagonal unitaries in the Hamming weight-$2$ subspace, which we will describe in detail in \sec{lowerbound_hamming}. With this strategy in mind, we first examine the approximate synthesis of diagonal unitaries in \sec{lowerbound_diag}.

\subsection{Approximate synthesis of diagonal unitaries}
\label{sec:lowerbound_diag}
We now consider the gate complexity lower bound for synthesizing diagonal unitaries. The exact version of this problem was considered by Bullock and Markov~\cite{BM04}. Here, we generalize their lower bound to the approximate synthesis, by adapting a volume comparison technique of Knill~\cite{Knill95}.

To be specific, our goal is to implement $\mu$-qubit diagonal unitaries with accuracy $\delta$
\begin{equation}
    \ket{x_\mu,\ldots,x_2,x_1}\mapsto
    e^{i\theta_{x}}\ket{x_\mu,\ldots,x_2,x_1},
\end{equation}
where the phase angles take values up to $|\theta_x|\leq \theta_{\max}<\pi$. Recall that the set of all such unitaries is denoted by $\mathcal{D}_{\theta_{max}}$. Our circuit acts on a total number of $b\geq\mu$ qubits. We will first prove a lower bound for a $2$-qubit gate set $\mathcal{K}$ of finite size $|\mathcal{K}|$, and then bootstrap the result to obtain a lower bound for arbitrary $2$-qubit gates.

As aforementioned, this lower bound will be proved using a volume comparison argument: we will compare the volume of all the diagonal unitaries within $\delta$-distance to the set of constructable quantum circuits, against the volume of our target set of diagonal unitaries. A similar argument was used by Knill to prove a circuit lower bound for synthesizing the full unitary group $\mathcal{U}(2^\mu)$~\cite{Knill95}, but our proof here is significantly simplified as we restrict to the set of diagonal unitaries whose volume is easier to compute (see \sec{prelim_volume}).

We start by bounding the number of $b$-qubit circuits that can be constructed using $g$ gates from a finite two-qubit gate set $\mathcal{K}$. For each of the $g$ gates, we have $\binom{b}{2}$ choices for the locations in which the two-qubit gate acts and, fixing the locations, we have $|\mathcal{K}|$ choices for the specific gate. This gives the upper bound
\begin{equation}
    \left(\binom{b}{2}|\mathcal{K}|\right)^g
\end{equation}
for the number of distinct circuits we can construct.

Now, to synthesize diagonal unitaries with accuracy $\delta$, we need all of them to fall into the $\delta$-ball of some quantum circuits (in spectral norm distance). For each circuit $V$, the volume of diagonal unitaries within its $\delta$-ball can be estimated as follows. We take any two diagonal unitaries $U$ and $\widetilde{U}$ from the $\delta$-ball centered at $V$. By the triangle inequality,
\begin{equation}
    \norm{U-V},\norm{\widetilde{U}-V}\leq\delta\quad\Rightarrow\quad\norm{\widetilde{U}-U}\leq2\delta,
\end{equation}
which implies
\begin{equation}
    \abs{e^{i\widetilde{\theta}_x}-e^{i\theta_x}}\leq2\delta
\end{equation}
when restricting to the specific diagonal elements indexed by $x$.
Solving for $\widetilde{\theta}_x$ and $\theta_x$ when $\delta<1$, we have
\begin{equation}
    \abs{\widetilde{\theta}_x-\theta_x}\leq\arcsin\left(2\delta\sqrt{1-\delta^2}\right).
\end{equation}
This means $\max_x\max_{\widetilde{\theta}_x,\theta_x}\abs{\widetilde{\theta}_x-\theta_x}\leq\arcsin\left(2\delta\sqrt{1-\delta^2}\right)$, so the volume of such diagonal unitaries is upper bounded by $\left(\arcsin\left(2\delta\sqrt{1-\delta^2}\right)\right)^{2^\mu}$.
We thus have that the total volume of synthesizable diagonal unitaries is at most
\begin{equation}
\begin{aligned}
    &\ \vol\left(\left\{U|U\text{ is diagonal unitary},\exists\text{ circuit }V,\norm{U-V}\leq\delta\right\}\right)\\
    \leq&\ \left(\binom{b}{2}|\mathcal{K}|\right)^g\max_{\text{circuit }V}\vol\left(\left\{U|U\text{ is diagonal unitary},\norm{U-V}\leq\delta\right\}\right)\\
    \leq&\ \left(\binom{b}{2}|\mathcal{K}|\right)^g\vol\left(\mathcal{D}_{\frac{\arcsin\left(2\delta\sqrt{1-\delta^2}\right)}{2}}\right)
    =\left(\binom{b}{2}|\mathcal{K}|\right)^g\left(\arcsin\left(2\delta\sqrt{1-\delta^2}\right)\right)^{2^\mu}.
\end{aligned}
\end{equation}

On the other hand, we know that the total volume of the target diagonal unitaries is
\begin{equation}
\label{eq:vol_diag}
    \vol\left(\mathcal{D}_{\theta_{\max}}\right)
    =\left(2\theta_{\max}\right)^{2^\mu}.
\end{equation}
For our synthesis to succeed, we require that
\begin{equation}
    \left\{U|U\text{ is diagonal unitary},\exists\text{ circuit }V,\norm{U-V}\leq\delta\right\}\supseteq\mathcal{D}_{\theta_{\max}},
\end{equation}
which implies
\begin{equation}
    \left(\binom{b}{2}|\mathcal{K}|\right)^g
    \left(\arcsin\left(2\delta\sqrt{1-\delta^2}\right)\right)^{2^\mu}
    \geq\left(2\theta_{\max}\right)^{2^\mu},\qquad
    g\geq2^\mu\cdot\frac{\log\left(\frac{2\theta_{\max}}{\arcsin\left(2\delta\sqrt{1-\delta^2}\right)}\right)}{\log\big(\binom{b}{2}|\mathcal{K}|\big)}.
\end{equation}
Note that this lower bound is only useful when $\theta_{\max}=\Omega\left(\delta\right)$, for otherwise the right-hand side of the bound becomes negative. Intuitively, we would simply use the identity operator in the circuit synthesis if the target diagonal unitary itself is close to identity. On the other hand, the restriction $\theta_{\max}<\pi$ is not fundamental: if $\theta_{\max}\geq\pi$, we can simply reset it to a smaller value.

We have so far proved a circuit lower bound when the underlying gate set $\mathcal{K}$ is of finite size. If we are allowed to use arbitrary $2$-qubit gates, then the bound can be modified as follows. We first synthesize the circuit with respect to a fixed finite universal gate set $\mathcal{K}'$, say Clifford+T. This gives a new circuit approximating the original one to accuracy $\delta$ with a slightly larger gate complexity $cg\log(g/\delta)$ for some constant $c>0$. We then invoke the above bound with the number of gates $cg\log(g/\delta)$, accuracy $2\delta$ and the gate set $\mathcal{K}'=\text{Clifford+T}$ of constant size. This gives
\begin{equation}
    cg\log\left(\frac{g}{\delta}\right)= \Omega\left(2^\mu\cdot\frac{\log\left(\frac{\theta_{\max}}{\delta}\right)}{\log b}\right).
\end{equation}
Solving $g$ using the Lambert-W function~\cite[Lemma 59]{GSLW19}, we obtain the same asymptotic scaling
\begin{equation}
    g=\Omega\left(2^\mu\cdot\frac{\log\left(\frac{\theta_{\max}}{\delta}\right)}{\log b}\right).
\end{equation}
We summarize this bound in the following theorem.

\begin{theorem}[Approximate synthesis of diagonal unitaries]
\label{thm:lowerbound_diag}
Consider diagonal unitaries $\mathcal{D}_{\theta_{\max}}=\left\{\sum_xe^{i\theta_x}\ketbra{x_\mu,\ldots,x_1}{x_\mu,\ldots,x_1}\ |\ |\theta_x|\leq\theta_{\max}<\pi\right\}$ on $\mu$ qubits with phase angles at most $\theta_{\max}<\pi$. Given accuracy $0<\delta<1$, number of qubits $b\geq\mu$, and $2$-qubit gate set $\mathcal{K}$ of finite size $|\mathcal{K}|$,
\begin{equation}
\begin{aligned}
    &\min\left\{g\ |\ \forall U\in\mathcal{D}_{\max},\exists\text{\ circuit\ }V\text{\ on $b$ qubits with $g$ gates from set $\mathcal{K}$},\norm{U-V}\leq\delta\right\}\\
    &\geq2^\mu\cdot\frac{\log\left(\frac{2\theta_{\max}}{\arcsin\left(2\delta\sqrt{1-\delta^2}\right)}\right)}{\log\big(\binom{b}{2}|\mathcal{K}|\big)}.
\end{aligned}
\end{equation}
Under the same assumption (with $0<\delta<1/2$ and $\theta_{\max}=\Omega\left(\delta\right)$) but choosing $\mathcal{K}$ to be the set of arbitrary $2$-qubit gates,
\begin{equation}
\begin{aligned}
    &\min\left\{g\ |\ \forall U\in\mathcal{D}_{\max},\exists\text{\ circuit\ }V\text{\ on $b$ qubits with $g$ gates from set $\mathcal{K}$},\norm{U-V}\leq\delta\right\}\\
    &=\Omega\left(\frac{2^\mu}{\log b}\right).
\end{aligned}
\end{equation}
\end{theorem}

\subsection{Reduction from simulation in the Hamming weight-\texorpdfstring{$2$}{2} subspace}
\label{sec:lowerbound_hamming}
We now show a circuit lower bound for simulating a class of $2$-local Hamiltonians with only Pauli-$Z$ interactions. Specifically, we consider $H=\sum_{1\leq j<k\leq n}\beta_{j,k}Z_jZ_k$, where coefficients take a continuum range of values up to $|\beta_{j,k}|\leq t$. Suppose first that this evolution can be simulated to accuracy $\delta$ using circuits acting on $b\geq n$ qubits with gates chosen from a $2$-qubit gate set $\mathcal{K}$ of finite size $|\mathcal{K}|$. An analogous lower bound holds when the gate set contains arbitrary $2$-qubit gates.

The key technical ingredient of our proof is a gate-efficient reduction from the approximate synthesis of diagonal unitaries within the Hamming weight-$2$ subspace $\mathcal{W}_2$. To elaborate, we rewrite the evolution as
\begin{equation}
    e^{-iH}=e^{i\sum_{j<k}\beta_{j,k}I}
    e^{-i\sum_{j<k}\beta_{j,k}Z_j}
    e^{-i\sum_{j<k}\beta_{j,k}Z_k}
    e^{-i\sum_{j<k}4\beta_{j,k}\frac{I-Z_j}{2}\frac{I-Z_k}{2}}.
\end{equation}
It suffices for us to consider only the last exponential since all the remaining ones can be performed with gate complexity $\mathcal{O}(n)$ (with only logarithmic overhead when compiled with respect to a fixed finite universal gate set, say Clifford+T). We have
\begin{equation}
    \sum_{j<k}\beta_{j,k}\frac{I-Z_j}{2}\frac{I-Z_k}{2}\ket{e_u\oplus e_v}
    =\beta_{u,v}\ket{e_u\oplus e_v},
\end{equation}
for all $1\leq u<v\leq n$, where $e_u$ represents the $n$-bit string with $1$ at the $u$th position and $0$ elsewhere and $\oplus$ denotes the modulo-2 addition. Using a conversion mapping that will be introduced below, we can then relate the problem of simulating $2$-local Hamiltonians with coefficients $\beta_{j,k}$, to the synthesis of diagonal unitaries with phase angles $\beta_{j,k}$, within the Hamming weight-$2$ subspace.

\begin{figure}[t]
	\centering
\resizebox{1.\textwidth}{!}{%
\includegraphics{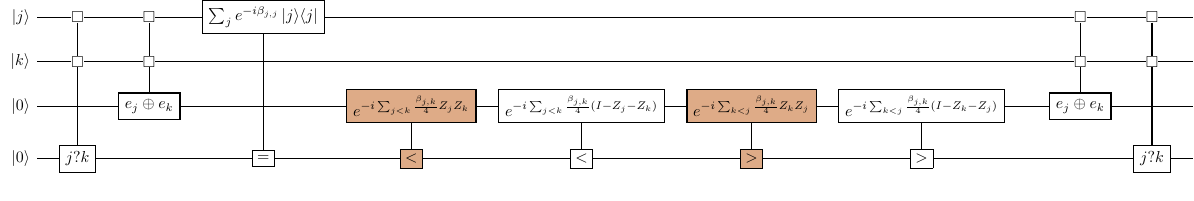}
}
\caption{Illustration of the reduction. This circuit implements diagonal unitaries on registers $\ket{j}$ and $\ket{k}$ by simulating $2$-local commuting Hamiltonians on an ancilla register (shaded orange). The reduction is gate efficient, so the asymptotic cost of quantum simulation is the same as that of synthesizing diagonal unitaries.}
\label{fig:reduction}
\end{figure}

To complete the reduction, we need to convert the usual computational basis states to the basis states of the Hamming weight-$2$ subspace $\mathcal{W}_2$. This can be done using an inequality test, followed by a binary-to-unary conversion. Specifically, suppose we have two quantum registers each of length $\log n$ holding the binary representation of $j$ and $k$ (assuming $n$ is a power of $2$). We then use an inequality test with cost $\mathcal{O}(\log n)$ to implement
\begin{equation}
    \ket{j}\ket{k}\mapsto\ket{j}\ket{k}\ket{j?k}=
    \begin{cases}
    \ket{j}\ket{k}\ket{j=k},\quad &j=k,\\
    \ket{j}\ket{k}\ket{j<k},\quad &j<k,\\
    \ket{j}\ket{k}\ket{j>k},\quad &j>k.\\
    \end{cases}
\end{equation}
We can introduce arbitrary phases for the first case using $\mathcal{O}(n)$ gates (which can again be compiled using Clifford+T with only logarithmic overhead). For the case where $j<k$, we apply the binary-to-unary conversion with cost $\mathcal{O}(n)$ to get
\begin{equation}
    \ket{j}\ket{k}\ket{j<k}\mapsto
    \ket{j}\ket{k}\ket{e_j\oplus e_k}\ket{j<k},
\end{equation}
after which we introduce the desired phases by simulating the corresponding $2$-local Hamiltonian with accruacy $\epsilon$. The last case $j>k$ can be handled similarly by swapping the role of $j$ and $k$. See \fig{reduction} for an illustration of this reduction.

To summarize, we can use the above circuit to implement arbitrary diagonal unitaries on the two quantum registers $\ket{j}\ket{k}$ of total length $\log n+\log n=\log(n^2)$ to accuracy $3\epsilon$ with phase angles taking values up to $|\beta_{j,k}|\leq t$. Our circuit uses two instances of controlled simulations of $2$-local Hamiltonians, together with $\mathcal{O}(n\polylog(n/\epsilon))$ additional Clifford+T gates, acting on a total number of $\mathcal{O}(b)$ qubits (recall $b\geq n$).
In other words, if the simulation uses $g$ gates from $\mathcal{K}$, then we can implement arbitrary diagonal unitaries using
\begin{equation}
    2g+\mathcal{O}\left(n\polylog(n/\epsilon)\right)
\end{equation}
gates from the gate set
\begin{equation}
    \mathcal{K}'=\text{controlled-}\mathcal{K}\cup\left(\mathrm{Clifford+T}\right).
\end{equation}
However, our above lower bound indicates that
\begin{equation}
    \Omega\left(\frac{n^2}{\log(b|\mathcal{K}'|)}\right)
\end{equation}
gates are necessary to implement diagonal unitaries acting on $b=\log(n^2)$ qubits with accuracy $3\epsilon$ for $t=\Omega(\epsilon)$. 
Thus, a similar asymptotic bound applies to the uncontrolled simulation as well by observing that $|\mathcal{K}'|=\Theta(|\mathcal{K}|)$.

\begin{corollary}[Simulating $2$-local commuting Hamiltonians]
\label{cor:lowerbound_ham}
Consider $2$-local Hamiltonians $H=\sum_{1\leq j<k\leq n}\beta_{j,k}Z_jZ_k$, where coefficients take values up to $|\beta_{j,k}|\leq t$. Given accuracy $0<\epsilon<1/3$, number of qubits $b\geq n$, and $2$-qubit gate set $\mathcal{K}$ of finite size $|\mathcal{K}|$, if $t=\Omega(\epsilon)$,
\begin{equation}
\begin{aligned}
    &\min\left\{g\ |\ \forall\text{ $2$-local Hamiltonian\ }H,\exists\text{\ circuit\ }V\text{\ on $b$ qubits with $g$ gates from $\mathcal{K}$},\norm{e^{-iH}-V}\leq\epsilon\right\}\\
    &=\Omega\left(\frac{n^2}{\log\big(b|\mathcal{K}|\big)}-n\polylog\left(\frac{n}{\epsilon}\right)\right).
\end{aligned}
\end{equation}
Under the same assumption but choosing $\mathcal{K}$ to be the set of arbitrary $2$-qubit gates,
\begin{equation}
\begin{aligned}
    &\min\left\{g\ |\ \forall\text{ $2$-local Hamiltonian\ }H,\exists\text{\ circuit\ }V\text{\ on $b$ qubits with $g$ gates from $\mathcal{K}$},\norm{e^{-iH}-V}\leq\epsilon\right\}\\
    &=\Omega\left(\frac{n^2}{\log b}-n\polylog\left(\frac{n}{\epsilon}\right)\right).
\end{aligned}
\end{equation}
\end{corollary}

To apply our bound for implementing Trotter steps, we first divide the evolution into $r$ steps and simulate each step using product formulas for time $t/r$ with accuracy $\epsilon/r$. Depending on the specific definition of the Hamiltonian, we may perform further decompositions like in \sec{block}, \sec{avgcost} and \sec{rank} and apply the bound only to commuting terms. We note that when the Hamiltonian coefficients have nonuniform magnitudes, the minimum effective evolution time can often be much smaller. For instance, the effective time for power-law interactions scale like $\Theta\left(\frac{t}{n^{\alpha}r}\right)$ as the smallest terms have size $\Theta\left(1/n^{\alpha}\right)$. To apply our bound, we would then require that $t=\Omega(n^\alpha\epsilon)$, which is a pessimistic estimate. Of course, one can replace the volume formula in \eq{vol_diag} with a better estimate that takes account of the nonuniformness of the Hamiltonian coefficients. However, we emphasize that this modification will not lead to a significant improvement to our bound as the dependence on the magnitude of coefficients is only logarithmic. We leave a detailed study of such improvement for future work.

\section{Discussion}
\label{sec:discussion}
In this work, we have developed circuit implementations of Trotter steps going beyond the sequential approach that cycles through all terms in the Hamiltonian. We have presented three methods, one based on the block-encoding technique, one based on an average-cost simulation and one based on a recursive low-rank decomposition. This allows us to simulate systems with power-law decaying interactions with significantly lower cost. As an application, we have given an algorithm for simulating electronic structure Hamiltonians in real space, with complexity $\left(\eta^{1/3}n^{1/3}+\frac{n^{2/3}}{\eta^{2/3}}\right)n^{1+o(1)}$ for the uniform electron gas and a similar cost when the external potential is included, improving over the best results from previous work. In achieving these speedups, we have made essential use of structural properties of the system. We further construct a class of $2$-local commuting Hamiltonians on $n$ qubits with continuum range of coefficients whose simulation requires $\Omega(n^2)$ gates. Our results thus suggest that the use of structural properties of the Hamiltonian is both necessary and sufficient to implement Trotter steps with lower asymptotic gate complexity.

Although we have employed different techniques to develop various faster simulation algorithms, the core idea behind all our improvements is the use of recursion. With suitable recursive decompositions, we have shown that the target Hamiltonian can be significantly simplified with lower $1$-norm and smaller rank value, both of which are useful for quantum simulation. Meanwhile, because of the master theorem (\lem{master}), recursions only introduce logarithmic factors to the overall gate complexity, thus simplifying the target problem ``almost for free''. We expect that recursions will find further applications in reducing the cost of simulating many-body Hamiltonians.

We have focused on simulating $2$-local spin models throughout our paper, and it would be interesting to study general $\kappa$-local models as well as fermionic systems.
For this purpose, previous work~\cite{MYMLMBC18} utilized different low-rank decompositions for $1$- and $2$-body fermionic operators which led to Trotter steps with lower cost. However, an issue not rigorously addressed in their work is that the decomposition does not respect the commutation relations of Hamiltonian terms and could lead to a Trotter error larger than the naive approach. Thus, some additional efforts are needed to justify the utility of those methods in quantum simulation~\cite{Assess22}. Note that this drawback can be slightly alleviated using the phase gradient construction~\cite{kivlichan2020improved} if the fault-tolerant resource cost is of interest. On the other hand, our techniques do apply to other $\kappa$-local models for constant $\kappa$ and to fermionic systems, as long as the Hamiltonian can be recursively decomposed as in \sec{block}, \sec{avgcost} and \sec{rank}. Preliminary studies of such generalizations were available in the context of measuring quantum observables~\cite{Bonet-Monroig20}, and we hope future work could develop more efficient Hamiltonian decompositions along this line.

We have demonstrated that advanced quantum simulation techniques, such as block encoding and qubitization, can be used to improve the performance of product formulas. Such techniques can be directly applied to implement Trotter steps; however, this would introduce a slow-down factor proportional to the $1$-norm of Hamiltonian coefficients, offering no benefit over the sequential circuit implementation. We have shown how to overcome this obstacle for power-law interactions by recursively decomposing the Hamiltonian using product formulas to reduce the norm and by implementing an average-case cost simulation. Our result is complementary to recent results on the hybridized simulation~\cite{Rajput2022hybridizedmethods,TongGaugeTheory21}, where quantum algorithms are ``hybridized'' with the interaction-picture method to achieve better performance. Our finding indicates that product formulas can be just as effective in hybridizing with other algorithms for quantum simulation.

In our low-rank method, we have performed a recursive decomposition of the target Hamiltonian where certain off-diagonal blocks of coefficients have rank value logarithmic in the input parameters. Such a decomposition was used in a different context by~\cite{NKL22} to block encode kernel matrices. Although their results are potentially useful for improving electronic structure simulation in first quantization, we are not aware of a simple circuit to realize this idea. In our problem, the decomposed Hamiltonian terms can be handled directly, so block encoding is no longer needed (it would be inefficient anyway due to the normalization factor issue mentioned above). We apply this method in \sec{app} to implement the Coulomb potential term of the electronic structure Hamiltonian in second quantization in real space, obtaining improvements over the best previous simulation results.
As can be seen from \tab{result_summary}, this method achieves the lowest gate complexity among all our algorithms as long as the low-rank condition is satisfied. This condition can be rigorously justified using the multipole expansion for all power-law models in 1D and many power-law models such as Coulomb interactions in higher spatial dimensions. For other Hamiltonians where the rank condition does not hold, faster Trotter steps are still possible using methods based on block encoding and average-cost simulation.

When no structural property is available in the target Hamiltonian, then intuitively the best method one can hope for is to exponentiate the terms one by one. We justify this intuition in \sec{lowerbound} with a circuit lower bound, showing that a class of $2$-local commuting Hamiltonians on $n$ sites requires at least $\Omega(n^2)$ gates to simulate. However, we emphasize that our lower bound is useful only when the effective evolution time is sufficiently long: if the effective time is short compared to the accuracy threshold, our bound becomes trivial. However, such failures are often indications that improved implementations of Trotter steps exist, as one can truncate the Hamiltonian and replace small-angle exponentials by identity operators. Previous work has studied various truncation schemes for product formulas as well as more advanced quantum algorithms~\cite{CSTWZ19,XSSS20,Clinton22,Meister2022tailoringterm,Berry2019qubitizationof,Ivanov22}. By exploiting the relative scaling between evolution time and system size, it is plausible to get faster quantum simulation algorithms with better dependence on the spacetime volume.

Although our lower bound is mostly used to argue the difficulty of implementing product formulas, it is applicable to other simulation algorithms as well. In particular, it applies to the sampling-based approach~\cite{Campbell18}, as one can always reproduce classical computations on a quantum computer, and the use of randomness can be represented unitarily via the Stinespring dilation. A similar argument then lower bounds both the classical and quantum resources used by the simulation. But given the difficulty of realizing a scalable quantum computer, it would be beneficial if certain computations can be offloaded classically, which is exactly a main goal of the sampling-based algorithm. Looking forward, it would be interesting to further study the tradeoffs between classical and quantum resources in designing quantum algorithms.

Several other related questions are worth further investigation. To simplify our analysis, we have used boxes with exponentially growing sizes to group each qubit index in the block-encoding approach, but this may be improved by using different box definitions. We have also chosen a uniform division of intervals in the realization of average-cost simulation, but improvements might be possible through a nonuniform division scheme. We have almost exclusively considered the number of arbitrary $2$-qubit gates as the cost metric for quantum simulation, and future work may consider other metrics such as circuit depth, locality~\cite{GeneralSwapNetwork19} and fault-tolerant gate count (although we expect some of our techniques are useful under those metrics as well). 
With these questions in mind, we hope our work will motivate continuing progress on finding the most efficient methods to solve instances of quantum simulation problems with practical interest.

\section*{Acknowledgements}
Y.S.\ thanks Ryan Babbush for helpful discussions during the initial stages of this work. We thank Wim van Dam for his comments on an earlier draft.
Y.T.\ acknowledges funding from the U.S.\ Department of Energy Office of Science, Office of Advanced Scientific Computing Research, (DE-NA0003525, and DE-SC0020290). Work supported by DE-SC0020290 is supported by the DOE QuantISED program through the theory consortium ``Intersections of QIS and Theoretical Particle Physics'' at Fermilab. The Institute for Quantum Information and Matter is an NSF Physics Frontiers Center.
M.C.T.\ is supported by the Defense Advanced Research Project Agency (DARPA) under Contract No.\ 134371-5113608 and the Quantum Algorithms and Machine Learning Grant from Nippon Telegraph and Telephone (NTT), No.\ AGMT DTD 9/24/20.

\appendix
\section{Second circuit lower bound}
\label{append:lowerbound2}
In \sec{lowerbound}, we proved a lower bound on the gate complexity of approximatly synthesizing diagonal unitaries, and bootstrapped it to get a lower bound on the simulation of $2$-local commuting Hamiltonians. In this appendix, we prove a related lower bound on the cost of synthesizing the coefficient oracle $O_\beta$ used in \sec{block} and \sec{avgcost} for block encoding. To this end, we first prove a lower bound for the approximate synthesis of discrete diagonal unitaries in \append{lowerbound2_discrete}, and then do a reduction from applications of the coefficient oracle in \append{lowerbound2_coeff}. Our result extends a previous bound of~\cite{LKS18} by allowing approximate circuit synthesis and use of arbitrary $2$-qubit gates.

\subsection{Approximate synthesis of discrete diagonal unitaries}
\label{append:lowerbound2_discrete}
We first consider the gate complexity of synthesizing discrete diagonal unitaries of the form
\begin{equation}
\label{eq:discrete_diag}
    \ket{x}\mapsto
    e^{i\frac{\beta_{x,m-1} 2^{m-1}+\cdots+\beta_{x,0}}{2^m}}\ket{x},
\end{equation}
with accuracy $\delta$. Here, $x$ takes $2^\mu$ possible values, and $\beta_{x}$ is an $m$-bit string parameterized by $x$ (so $\beta_{x,j}\in\{0,1\}$). We use a setup similar to that of \sec{lowerbound}. Specifically, our circuit acts on a total number of $b\geq\mu$ qubits. We will first prove a lower bound for a $2$-qubit gate set $\mathcal{K}$ of finite size $|\mathcal{K}|$, and then bootstrap the result to obtain a lower bound for arbitrary $2$-qubit gates. 

As before, the number of distinct $b$-qubit circuits that can be constructed using $g$ gates from gate set $\mathcal{K}$ is at most
\begin{equation}
    \left(\binom{b}{2}|\mathcal{K}|\right)^g.
\end{equation}
We now repeat our volume argument. To synthesize all discrete diagonal unitaries with accuracy $\delta$, we need all of them to fall into the spectral norm $\delta$-ball of some quantum circuits. Fixing a specific circuit $V$, we consider any diagonal unitaries $U$ and $\widetilde{U}$ from the $\delta$-ball centered at $V$. Because of the triangle inequality, we must have
\begin{equation}
    \norm{\widetilde{U}-U}\leq2\delta.
\end{equation}

However, the new observation is that our target set is discrete and finite. 
For $\delta=\mathcal{O}(1)$ and $\delta=\Omega\left(\frac{1}{2^m}\right)$, these unitaries have the $x$th diagonal entries
\begin{equation}
    \abs{e^{i\frac{\widetilde{\beta}_{x,m-1} 2^{m-1}+\cdots+\widetilde{\beta}_{x,0}}{2^m}}-
    e^{i\frac{\beta_{x,m-1} 2^{m-1}+\cdots+\beta_{x,0}}{2^m}}}\leq2\delta.
\end{equation}
Solving for the (integers) $\widetilde{\beta}_x$ and $\beta_x$, we have
\begin{equation}
    \abs{\widetilde{\beta}_x-\beta_x}\leq2^m\arcsin\left(2\delta\sqrt{1-\delta^2}\right),
\end{equation}
which implies that the number of such discrete diagonal unitaries is at most
\begin{equation}
    \mathcal{O}\left(\left(2^m\delta\right)^{2^\mu}\right).
\end{equation}
Using the counting measure $\#\{\cdot\}$, we thus have that the total number of synthesizable discrete diagonal unitaries is at most
\begin{equation}
\begin{aligned}
    &\ \#\left\{U|U\text{ is discrete diagonal unitary},\exists\text{ circuit }V,\norm{U-V}\leq\delta\right\}\\
    \leq&\ \left(\binom{b}{2}|\mathcal{K}|\right)^g\max_{\text{circuit }V}\#\left\{U|U\text{ is discrete diagonal unitary},\norm{U-V}\leq\delta\right\}\\
    =&\ \left(\binom{b}{2}|\mathcal{K}|\right)^g\mathcal{O}\left(\left(2^m\delta\right)^{2^\mu}\right)
    =\mathcal{O}\left(b^{2g}|\mathcal{K}|^g2^{m2^\mu}\delta^{2^\mu}\right).
\end{aligned}
\end{equation}

On the other hand, by a counting argument, we know that the total number of discrete diagonal unitaries is exactly
\begin{equation}
    \#\left\{U|U\text{ is discrete diagonal unitary}\right\}
    =\left(2^m\right)^{2^\mu}=2^{m2^\mu}.
\end{equation}
For our synthesis to succeed, we require that
\begin{equation}
    \left\{U\text{ is discrete diagonal unitary }|\ \exists\text{ circuit }V,\norm{U-V}\leq\delta\right\}\supseteq
    \left\{U\text{ is discrete diagonal unitary}\right\},
\end{equation}
which implies
\begin{equation}
    g=\Omega\left(2^\mu\frac{\log\left(\frac{1}{\delta}\right)}{\log\left(b|\mathcal{K}|\right)}\right).
\end{equation}
If $\mathcal{K}$ is the arbitrary $2$-qubit gate set, we can use the Lambert-W function in a similar way as in \sec{lowerbound_diag} to get
\begin{equation}
    g=\Omega\left(2^\mu\frac{\log\left(\frac{1}{\delta}\right)}{\log b}\right).
\end{equation}
We have thus proved:

\begin{theorem}[Approximate synthesis of discrete diagonal unitaries]
\label{thm:lowerbound_discrete}
Consider discrete diagonal unitaries $\left\{\sum_{x}e^{i\frac{\beta_{x,m-1} 2^{m-1}+\cdots+\beta_{x,0}}{2^m}}\ketbra{x}{x}\right\}$ on $\mu$ qubits with phase angles determined by $m$-bit strings $\beta_{x}$. Given accuracy $\delta=\mathcal{O}(1)$ and $\delta=\Omega\left(\frac{1}{2^m}\right)$, number of qubits $b\geq\mu$, and $2$-qubit gate set $\mathcal{K}$ of finite size $|\mathcal{K}|$,
\begin{equation}
\begin{aligned}
    &\min\left\{g\ |\ \forall\text{discrete diagonal unitary } U,\exists\text{\ $b$-qubit circuit\ }V\text{\ with $g$ gates from set $\mathcal{K}$},\norm{U-V}\leq\delta\right\}\\
    &=\Omega\left(\frac{2^\mu \log\left(\frac{1}{\delta}\right)}{\log\left(b|\mathcal{K}|\right)}\right).
\end{aligned}
\end{equation}
Under the same assumption but choosing $\mathcal{K}$ to be the set of arbitrary $2$-qubit gates,
\begin{equation}
\begin{aligned}
    &\min\left\{g\ |\ \forall\text{discrete diagonal unitary } U,\exists\text{\ $b$-qubit circuit\ }V\text{\ with $g$ gates from set $\mathcal{K}$},\norm{U-V}\leq\delta\right\}\\
    &=\Omega\left(\frac{2^\mu \log\left(\frac{1}{\delta}\right)}{\log b}\right).
\end{aligned}
\end{equation}
\end{theorem}

\subsection{Reduction from the coefficient oracle}
\label{append:lowerbound2_coeff}
Recall that in the block-encoding method, we have used the following oracle of Hamiltonian coefficients:
\begin{equation}
    O_{\beta}\ket{u,v,0}=\ket{u,v,\beta_{u,v}}.
\end{equation}
Here, $u$ and $v$ each takes $n$ values and $\beta_{u,v}$ are $m$-bit binary strings. To ensure the efficiency of our method, we have assumed that the gate complexity of implementing oracle $O_\beta$ is
$\mathcal{O}(\polylog(nt/\epsilon))$. Implicitly, this implies the existence of certain structural properties of $\beta_{u,v}$: we now show that if $\beta_{u,v}$ are allowed to depend arbitrarily on $u$ and $v$, then one needs at least $\sim n^2$ gates to implement this oracle, resulting in an inefficient block-encoding approach.

The key ingredient of the proof is a reduction from the application of coefficient oracle to the approximate synthesis of discrete diagonal unitaries. This is done as follows:
\begin{equation}
\begin{aligned}
    \ket{u,v,0}&\xmapsto{O_\beta}\ket{u,v,\beta_{u,v}}\\
    &\mapsto e^{i\frac{\widetilde{\beta}_{u,v}}{2^m}}\ket{u,v,\beta_{u,v}}\\
    &\xmapsto{O_\beta^\dagger}e^{i\frac{\widetilde{\beta}_{u,v}}{2^m}}\ket{u,v,0},
\end{aligned}
\end{equation}
where in the second step we introduce the approximate phase factors $\widetilde{\beta}_{u,v}$ by applying $\Theta\left(\log\left(1/\epsilon\right)\right)$ rotations $\ketbra{0}{0}+e^{\frac{i}{2^{\ell}}}\ketbra{1}{1}$ ($\ell=1,\ldots,\Theta\left(\log\left(1/\epsilon\right)\right)$). We have thus implemented an arbitrary discrete diagonal unitary on registers $\ket{u,v}$ of total length $\mu=2\log n$ using a circuit of two oracle calls (each with accuracy $\epsilon$) and $\Theta\left(\log\left(1/\epsilon\right)\right)$ single-qubit gates (which can be compiled with respect to a fixed gate set like Clifford+T with only logarithmic overhead).

In summary, suppose $O_\beta$ can be implemented to accuracy $\epsilon$ using a $b$-qubit circuit of $g$ gates chosen from a gate set $\mathcal{K}$ of finite size $|\mathcal{K}|$. Then, we can implement the entire reduction to accuracy $3\epsilon$ using a $b$-qubit circuit of $2g+\mathcal{O}\left(\polylog(1/\epsilon)\right)$ gates chosen from $\mathcal{K}'=\mathcal{K}\cup(\mathrm{Clifford+T})$. But our lower bound indicates that
\begin{equation}
\Omega\left(\frac{n^2 \log\left(\frac{1}{\epsilon}\right)}{\log\left(b|\mathcal{K}'|\right)}\right)
=\Omega\left(\frac{n^2 \log\left(\frac{1}{\epsilon}\right)}{\log\left(b|\mathcal{K}|\right)}\right)
\end{equation}
gates are required for the synthesis task. We have thus proved:

\begin{corollary}[Approximate synthesis of coefficient oracles]
\label{cor:lowerbound_coeff}
Consider the set of coefficient oracles $\left\{O_{\beta}\ |\ O_{\beta}\ket{u,v,0}=\ket{u,v,\beta_{u,v}}\right\}$ where $u$ and $v$ each takes $n$ values and $\beta_{u,v}$ are $m$-bit strings. Given accuracy $\epsilon=\mathcal{O}(1)$ and $\epsilon=\Omega\left(\frac{1}{2^m}\right)$, number of qubits $b\geq m+2\log n$, and $2$-qubit gate set $\mathcal{K}$ of finite size $|\mathcal{K}|$,
\begin{equation}
\begin{aligned}
    &\min\left\{g\ |\ \forall\text{oracle } O_\beta,\exists\text{\ circuit\ }V\text{\ on $b$ qubits with $g$ gates from set $\mathcal{K}$},\norm{O_\beta-V}\leq\epsilon\right\}\\
    &=\Omega\left(\frac{n^2\log\left(\frac{1}{\epsilon}\right)}{\log\left(b|\mathcal{K}|\right)}-\polylog\left(\frac{1}{\epsilon}\right)\right).
\end{aligned}
\end{equation}
Under the same assumption but choosing $\mathcal{K}$ to be the set of arbitrary $2$-qubit gates,
\begin{equation}
\begin{aligned}
    &\min\left\{g\ |\ \forall\text{oracle } O_\beta,\exists\text{\ circuit\ }V\text{\ on $b$ qubits with $g$ gates from set $\mathcal{K}$},\norm{O_\beta-V}\leq\epsilon\right\}\\
    &=\Omega\left(\frac{n^2\log\left(\frac{1}{\epsilon}\right)}{\log b}-\polylog\left(\frac{1}{\epsilon}\right)\right).
\end{aligned}
\end{equation}
\end{corollary}

We see that implementations of the coefficient oracle $O_\beta$ will generally cost $\sim n^2$ gates as well, even when approximate circuit synthesis is allowed. Thus, one needs structural properties of the Hamiltonian not only to realize faster Trotter steps but to implement faster coefficient oracles as well.

\section{Trotter error with fermionic induced \texorpdfstring{$1$}{1}-norm scaling}
\label{append:trotter}
In this appendix, we prove the fermonic induced $1$-norm scaling of Trotter error claimed in \thm{fermionic_induced_1norm}, which is in turn used in the simulation of real-space electronic structure Hamiltonians. Specifically, we consider the Hamiltonian 
\begin{equation}
    H=T+V:=\sum_{j,k}\tau_{j,k}A_j^\dagger A_k+\sum_{l,m}\nu_{l,m}N_l N_m,
\end{equation}
where $A_j^\dagger$, $A_k$ are the fermionic creation and annihilation operators, $N_l$ are the occupation-number operators, $\tau$, $\nu$ are coefficient matrices, and the summations are over $n$ spin orbitals. Let $S_p(t)$ be a $p$th-order product formula that splits $e^{-itH}$ into products of $e^{-itT}$ and $e^{-itV}$. Our goal is to bound the Trotter error $\norm{S_p(t)-e^{-iH}}_{\mathcal{W}_\eta}$ restricted to the $\eta$-electron subspace $\mathcal{W}_\eta$ (spanned by basis states with Hamming weight $\eta$).

To describe our proof, we introduce some additional notations and terminologies. For $\gamma=0,1$, we let
\begin{equation}
    H^{(\gamma)}=\sum_{j,k}\mu_{j,k}^{(\gamma)}H_{j,k}^{(\gamma)},\qquad
    \mu^{(0)} = \nu,\qquad \mu^{(1)} = \tau,\qquad H^{(0)}_{j,k} = N_j N_k,\qquad H^{(1)}_{j,k} = A^\dagger_j A_k.
\end{equation}
In other words, $H^{(0)}=V$, $H^{(1)}=T$. Then, we have the commutator scaling of Trotter error from \lem{comm_bound}:
\begin{equation}
\norm{S_p(t)-e^{-itH}}_{\mathcal{W}_\eta}=\mathcal{O}\left(\max_{\gamma\in\{0,1\}^{p+1}}\max_{\ket{\psi_\eta}\in\mathcal{W}_\eta} \abs{\bra{\psi_\eta} \left[H^{(\gamma_{p+1})},\cdots\left[H^{(\gamma_2)},H^{(\gamma_1)}\right]\right] \ket{\psi_\eta}}t^{p+1}\right),
\end{equation}
where we have taken the expectation value since nested commutators of Hermitian operators are necessarily anti-Hermitian. Note the following commutation rules for fermionic operators~\cite{helgaker2014molecular}
\begin{equation} \label{eq:commutator_t_2}
	\left[A_j^\dagger A_k,A_{j_x}^\dagger\right]=\delta_{k,j_x}A_j^\dagger,\quad
	\left[A_j^\dagger A_k,A_{k_y}\right]=-\delta_{k_y,j}A_k,\quad
	\left[A_j^\dagger A_k,N_{l_z}\right]
	=\delta_{k,l_z}A_j^\dagger A_{l_z}
	-\delta_{j,l_z}A_{l_z}^\dagger A_k.
\end{equation}
\begin{equation} \label{eq:commutator_v_2}
	\left[N_lN_m,A_{j_x}^\dagger\right]
	=\delta_{m,j_x}N_lA_{j_x}^\dagger+\delta_{l,j_x}A_{j_x}^\dagger N_m, \quad
	\left[N_lN_m,A_{k_x}\right]
	=-\delta_{m,k_x}N_lA_{k_x}-\delta_{l,k_x}A_{k_x} N_m,
\end{equation}
where $\delta_{j,k}=1$ if and only if $j=k$.
Thus, the nested commutator $\left[H^{(\gamma_{p+1})}_{j_{p+1}, k_{p+1}}, \ldots \left[H^{(\gamma_{2})}_{j_{2}, k_{2}}, H^{(\gamma_{1})}_{j_{1}, k_{1}}\right]\right]$ can be written as a linear combination of products of elementary fermionic operators. Following~\cite[Proposition 11]{SHC21}, we call each summand $P$ from the nested commutator a \emph{fermionic path} and write $P \rhd \left(H^{(\gamma_{p+1})}_{j_{p+1}, k_{p+1}}, \ldots, H^{(\gamma_{1})}_{j_{1}, k_{1}} \right)$. Then, we have the following bound
\begin{equation}
\begin{aligned}
    &\abs{\bra{\psi_\eta} \left[H^{(\gamma_{p+1})},\cdots\left[H^{(\gamma_2)},H^{(\gamma_1)}\right]\right] \ket{\psi_\eta}}\\
    \leq&\max_{\norm{c}_1=\eta}
    \sum_{j_{p+1}, k_{p+1}} \ldots \sum_{j_{1}, k_{1}}  \sum_{P \rhd \left(H^{(\gamma_{p+1})}_{j_{p+1}, k_{p+1}}, \ldots, H^{(\gamma_{1})}_{j_{1}, k_{1}} \right)}
    \abs{\mu^{(\gamma_{p+1})}_{j_{p+1}, k_{p+1}}} \ldots \abs{\mu^{(\gamma_{1})}_{j_{1}, k_{1}}} 
    \frac{\norm{P \ket{{c}}}+\norm{P^\dagger \ket{{c}}}}{2},
\end{aligned}
\end{equation}
where the maximization is taken over all $n$-bit strings $c$ with Hamming weight $\eta$, and $P\ket{c}$ ($P^\dagger\ket{c}$) either gives another basis state or gives the zero vector.

Previous work~\cite{SHC21} further proceeds to upper bound the coefficients by the max-norm $\norm{\tau}_{\max}$ and $\norm{\nu}_{\max}$ and counts the number of fermionic paths with nonzero contribution to the summation. This results in a bound with the max-norm scaling asymptotically worse than our result. To get our improved \thm{fermionic_induced_1norm}, we need a more careful treatment of the summation order as well as the action of fermionic operators on a fixed basis state.

We now present a proof by induction. The base case of $p=0$ is easy to check. For any $n$-bit string $c$ with Hamming weight $\eta$, we have
\begin{equation}
\begin{aligned}
    \sum_{j,k}\abs{\tau_{j,k}}\frac{\norm{A_j^\dagger A_k\ket{c}}+\norm{A_k^\dagger A_j\ket{c}}}{2}
    &=\mathcal{O}\left(\vertiii{\tau}_1\eta\right),\\
    \sum_{l,m}\abs{\nu_{l,m}}\frac{\norm{N_l N_m\ket{c}}+\norm{N_m N_l\ket{c}}}{2}
    &=\mathcal{O}\left(\vertiii{\nu}_{1,[\eta]}\eta\right).
\end{aligned}
\end{equation}
This is because in order to make a nonzero contribution, the right-most fermionic operators only have $\eta$ possible choices. This motivates us to choose the summation ordering so that the indices of operators on the left are summed over first:
\begin{equation}
\begin{aligned}
    &\sum_k\sum_j\abs{\tau_{j,k}}\norm{A_j^\dagger A_k\ket{c}},\qquad
    &&\sum_j\sum_k\abs{\tau_{j,k}}\norm{A_k^\dagger A_j\ket{c}},\\
    &\sum_m\sum_l\abs{\nu_{l,m}}\norm{N_lN_m\ket{c}},\qquad
    &&\sum_l\sum_m\abs{\nu_{l,m}}\norm{N_mN_l\ket{c}}.
\end{aligned}
\end{equation}
Fixing one choice for the right-most operator, the remaining quantity can be upper bounded by either the induced $1$-norm $\vertiii{\tau}_1$ or the restricted induced $1$-norm $\vertiii{\nu}_{1,[\eta]}$ respectively. This proves the desired bound in the base case.

For the inductive step, assume that given any fermionic path P $\rhd \left(H^{(\gamma_{p})}_{j_{p}, k_{p}}, \ldots, H^{(\gamma_{1})}_{j_{1}, k_{1}} \right)$, we can always reorder the summation to yield the desired bound.
Now take the next layer of commutator $[T,P]$ or $[V,P]$. By \eq{commutator_t_2} and \eq{commutator_v_2} and a separate induction, $P=\prod_{q=1}^{p+2}P_\upsilon$ is the product of at most $p+2$ elementary fermionic operators (which include $A_j^{\dagger},A_k,N_l$), and at any point in the product the number of creation operators is always smaller than the number of annihilation operators. Then, the commutation may be performed sequentially as
\begin{equation}
    \left[T,P\right]=\sum_{q=1}^{p+2}P_{p+2}\cdots P_{q+1}\left[T,P_{q}\right]P_{q-1}\cdots P_1,\quad
    \left[V,P\right]=\sum_{q=1}^{p+2}P_{p+2}\cdots P_{q+1}\left[V,P_{q}\right]P_{q-1}\cdots P_1.
\end{equation}
Depending on $P_q=A_{j_x}^\dagger$, $A_{k_y}$ or $N_{l_z}$, we divide the analysis into five subcases:
\begin{enumerate}[leftmargin=*,label=(\roman*)]
    \item $\left[A_j^\dagger A_k,A_{j_x}^\dagger\right]=\delta_{k,j_x}A_j^\dagger$: We keep the original sum over $j_x$ while introducing a new sum over $j$ \textbf{on the right}. The summation range of $j$ is set to be over \textbf{all $n$ spin orbitals}. The number of electrons increases by $1$ due to $A_j^\dagger$ with the new location determined by $j$ (as opposed to $j_x$). Asymptotically, this does not affect the summation of other existing indices and increases the norm by a factor of $\vertiii{\tau}_1$ through the index $j$.
    \item $\left[A_j^\dagger A_k,A_{k_y}\right]=-\delta_{k_y,j}A_k$: We keep the original sum over $k_y$ while introducing a new sum over $k$ \textbf{on the right}. The summation range of $k$ is set to be over \textbf{all $n$ spin orbitals}. The number of electrons decreases by $1$ due to $A_k$ with the new location determined by $k$ (as opposed to $k_y$). Asymptotically, this does not affect the summation of other existing indices and increases the norm by a factor of $\vertiii{\tau}_1$ through the index $k$.
    \item $\left[A_j^\dagger A_k,N_{l_z}\right]
	=\delta_{k,l_z}A_j^\dagger A_{l_z}
	-\delta_{j,l_z}A_{l_z}^\dagger A_k$: We keep the original sum over $l_z$ while introducing a new sum over $j$ ($k$) \textbf{on the right} respectively. The summation range of $j$ ($k$) is set to be over \textbf{all $n$ spin orbitals}. The number of electrons remains the same but the occupation changes with the locations determined by $j$ ($k$) and $l_z$. Asymptotically, this does not affect the summation of other existing indices and increases the norm by a factor of $\vertiii{\tau}_1$ through the index $j$ ($k$).
	\item $\left[N_lN_m,A_{j_x}^\dagger\right]
	=\delta_{m,j_x}N_lA_{j_x}^\dagger+\delta_{l,j_x}A_{j_x}^\dagger N_m=\delta_{m,j_x}N_lA_{j_x}^\dagger+\delta_{l,j_x}N_mA_{j_x}^\dagger-\delta_{l,j_x}\delta_{m,j_x}A_{j_x}^\dagger$: We keep the original sum over $j_x$ while introducing a new sum over $l$ ($m$) \textbf{on the right} respectively. The summation range of $l$ ($m$) is set to be over \textbf{all occupied spin orbitals}. Both the number and the locations of electrons remain the same. Asymptotically, this does not affect the summation of other existing indices and increases the norm by a factor of $\vertiii{\nu}_{1,[\eta]}$ through the index $l$ ($m$).
	\item $\left[N_lN_m,A_{k_x}\right]
	=-\delta_{m,k_x}N_lA_{k_x}-\delta_{l,k_x}A_{k_x} N_m=-\delta_{m,k_x}N_lA_{k_x}-\delta_{l,k_x}N_mA_{k_x}-\delta_{l,k_x}\delta_{m,k_x}A_{k_x}$: We keep the original sum over $k_x$ while introducing a new sum over $l$ ($m$) \textbf{on the right} respectively. The summation range of $l$ ($m$) is set to be over \textbf{all occupied spin orbitals}. Both the number and the locations of electrons remain the same. Asymptotically, this does not affect the summation of other existing indices and increases the norm by a factor of $\vertiii{\nu}_{1,[\eta]}$ through the index $l$ ($m$).
\end{enumerate}
Note that in the fourth and fifth cases, we have implicitly used the fact that the number of occupied modes is no larger than $\eta$. 
The proof of \thm{fermionic_induced_1norm} is now complete.

\section{Generalization}
\label{append:generalize}
\subsection{Higher spatial dimensionality}
\label{append:generalize_dim}
In the main body of the paper, we have presented faster methods to perform Trotter steps for simulating power-law interactions in one spatial dimension. Here, we briefly describe how to generalize those methods to higher spatial dimensionality.

Specifically, consider an $n$-qubit $d$-dimensional square lattice $\mathcal{L}\subseteq\mathbb{Z}^d$. Without loss of generality, we assume $n^{1/d}$ is a power of $2$ and take $\mathcal{L}=\{1,\ldots,n^{1/d}\}^d$. Then, the Hamiltonian has the form $H=\sum_{j,k\in\mathcal{L}}H_{j,k}$. For power-law interactions, we have the norm condition $\norm{H_{j,k}}\leq 1/\norm{j-k}^\alpha$ for constant $\alpha>0$, where $\norm{j-k}$ is simply the Euclidean distance between $d$-dimensional vectors $j$ and $k$. Just like in the 1D case, we can expand the Hamiltonian with respect to tensor products of Pauli operators and hereafter assume the Hamiltonian takes the form
\begin{equation}
    H=\sum_{\substack{j,k\in\mathcal{L}\\j\neq k}}\beta_{j,k}X_jY_k.
\end{equation}
The commutator norm \eq{alpha-comm} corresponding to such decompositions has the scaling~\cite[Theorem H.2]{CSTWZ19}
\begin{equation}
    \acommtilde^{o(1)}=
    \begin{cases}
    n^{o(1)},\quad&\alpha\geq d,\\
    n^{1-\frac{\alpha}{d}+o(1)},&0<\alpha<d,
    \end{cases}
\end{equation}
which implies
\begin{equation}
    r=\begin{cases}
    \frac{n^{o(1)}t^{1+o(1)}}{\epsilon^{o(1)}},\quad&\alpha\geq d,\\[5pt]
    \frac{n^{1-\frac{\alpha}{d}+o(1)}t^{1+o(1)}}{\epsilon^{o(1)}},&0<\alpha< d.
    \end{cases}
\end{equation}

Depending on the order $j_s<k_s$ or $j_s>k_s$ between each pair of the corresponding coordinates, we can further divide the Hamiltonian into $2^d$ subterms. The remaining degenerate cases where some pair of indices collide $j_s=k_s$ can be merged accordingly. We assume that
\begin{equation}
    H=\sum_{1\leq j_1<k_1\leq n^{1/d}}\cdots\sum_{1\leq j_d<k_d\leq n^{1/d}}\beta_{j,k}X_jY_k
\end{equation}
in the following discussion, and a similar analysis holds for other cases with some notation changes.

\begin{figure}[t]
\centering

\begin{subfigure}[t]{.5\linewidth}
\includegraphics[width = 0.9\textwidth]{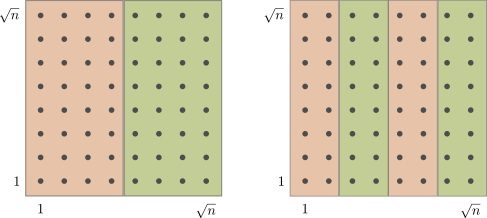}
\subcaption{}
\end{subfigure}
\begin{subfigure}[t]{.5\linewidth}
\includegraphics[width = 0.9\textwidth]{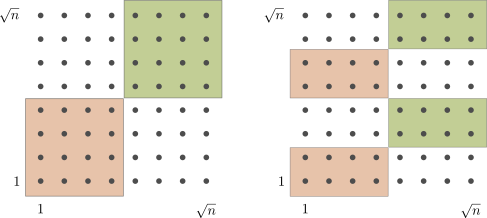}
\subcaption{}
\end{subfigure}
\caption{Illustration of the 2D recursive decomposition. Decomposed terms act across two disjoint cubes. The colors indicate types of the operator support following the convention in \fig{recursion}. Subfigure (a) represents terms in the first two layers of the decomposition of the first spatial dimension.
Subfigure (b) recursively decomposes the first layer from (a) in the second spatial dimension.}
\label{fig:2d_recursion}
\end{figure}

We now perform a recursive decomposition along each spatial dimension of the system. So for the first spatial dimension, we have the term
\begin{equation}
    \sum_{\substack{1\leq j_1<\frac{n^{1/d}}{2}\\\frac{n^{1/d}}{2}\leq k_1\leq n^{1/d}}}\cdots\sum_{1\leq j_d<k_d\leq n^{1/d}}\beta_{j,k}X_jY_k
\end{equation}
in the first layer and terms
\begin{equation}
    \sum_{\substack{1\leq j_1<\frac{n^{1/d}}{4}\\\frac{n^{1/d}}{4}\leq k_1<\frac{n^{1/d}}{2}}}\cdots\sum_{1\leq j_d<k_d\leq n^{1/d}}\beta_{j,k}X_jY_k,\qquad
    \sum_{\substack{\frac{n^{1/d}}{2}\leq j_1<\frac{3n^{1/d}}{4}\\\frac{3n^{1/d}}{4}\leq k_1\leq n^{1/d}}}\cdots\sum_{1\leq j_d<k_d\leq n^{1/d}}\beta_{j,k}X_jY_k
\end{equation}
in the second layer and so on. Because of the master theorem \lem{master}, we may restrict to a single term in the decomposition, say the one in the first layer. We then perform another recursive decomposition along the second spatial dimension, and once again restrict to only a specific term in the decomposition using \lem{master}. We iterate through all the $d$ spatial dimensions. See \fig{2d_recursion} for an illustration of this strategy in the 2D case.

For the block-encoding method in \sec{block}, the cost of circuit implementation depends on the $1$-norm of Hamiltonian coefficients, which determines the number of steps in the qubitization algorithm. Specifically, the cost of implementing decomposed Hamiltonian terms should now be revised to
\begin{equation}
    \cost(n)=
    \begin{cases}
        \mathcal{O}\left(\left(\frac{t}{r}+1\right)n\polylog\left(\frac{nt}{\epsilon}\right)\right),\quad&\alpha\geq2d,\\[5pt]
        \mathcal{O}\left(\left(n^{2-\frac{\alpha}{d}}\frac{t}{r}+1\right)n\polylog\left(\frac{nt}{\epsilon}\right)\right),&0<\alpha<2d,
    \end{cases}
\end{equation}
which implies that the entire simulation has a complexity of
\begin{equation}
    \begin{cases}
        nt\left(\frac{nt}{\epsilon}\right)^{o(1)},\quad&\alpha\geq2d,\\[5pt]
        \mathcal{O}\left(n^{3-\frac{\alpha}{d}}t\polylog\left(\frac{nt}{\epsilon}\right)\right)+nt\left(\frac{nt}{\epsilon}\right)^{o(1)},&d\leq\alpha<2d,\\[5pt]
        \mathcal{O}\left(n^{3-\frac{\alpha}{d}}t\polylog\left(\frac{nt}{\epsilon}\right)\right)+n^{2-\frac{\alpha}{d}}t\left(\frac{nt}{\epsilon}\right)^{o(1)},&0<\alpha<d.\\
    \end{cases}
\end{equation}

\begin{figure}[t]
	\centering
\includegraphics[width=0.9\textwidth]{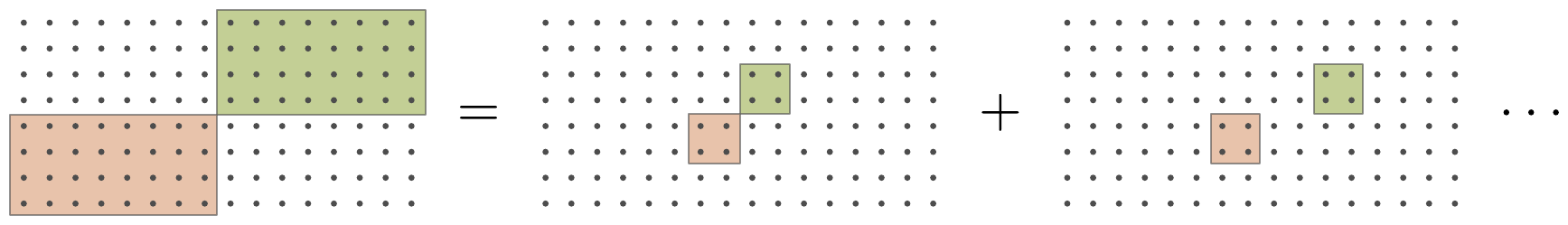}
\caption{Illustration of an average-cost block encoding for power-law Hamiltonians in 2D. }
\label{fig:avg_2D}
\end{figure}

This simulation can be further improved for $\alpha<2d$ using the average-cost simulation technique described in \sec{avgcost}. Specifically, for the two cubes sharing a common vertex, we further divide each spatial dimension into $m$ intervals of equal length; see \fig{avg_2D} for an illustration of this decomposition in 2D. Then, considering evolution time $t/r$, we have the modified cost function
\begin{equation}
    \cost(n)=\widetilde{\mathcal{O}}\left(\left(n^{2-\frac{\alpha}{d}}\frac{t}{r}+m^{2d}\right)\frac{n}{m^d}\right),
\end{equation}
which implies that the entire simulation has cost
\begin{equation}
    \begin{cases}
        \min\bigg\{n^{2-\frac{\alpha}{2d}}t\left(\frac{nt}{\epsilon}\right)^{o(1)},\ \mathcal{O}\left(n^{3-\frac{\alpha}{d}}t\polylog\left(\frac{nt}{\epsilon}\right)\right)
        +nt\left(\frac{nt}{\epsilon}\right)^{o(1)}\bigg\},
        \quad&d\leq\alpha<2d,\\[0.4cm]
        \min\bigg\{n^{\frac{5}{2}-\frac{\alpha}{d}}t\left(\frac{nt}{\epsilon}\right)^{o(1)},\ \mathcal{O}\left(n^{3-\frac{\alpha}{d}}t\polylog\left(\frac{nt}{\epsilon}\right)\right)
        +n^{2-\frac{\alpha}{d}}t\left(\frac{nt}{\epsilon}\right)^{o(1)}\bigg\},
        &0<\alpha<d.
    \end{cases}
\end{equation}

The generalization of the low-rank method in \sec{rank} is similar and straightforward, by using higher-dimensional cubes instead of 1D intervals to perform the recursive decomposition. Unlike the block-encoding method, here we do not have the issue of normalization factor; the circuit implementation is just a direct translation of classical computations. Moreover, for Coulomb interactions in higher spatial dimensions, one can show using the multipole expansion that the decomposed terms can also be approximated with a matrix with rank logarithmic in the input parameters. The resulting implementation has a complexity similar to that of the 1D case:
\begin{equation}
    \begin{cases}
        \rho nt\left(\frac{nt}{\epsilon}\right)^{o(1)},\quad&\alpha\geq d,\\[5pt]
        \rho n^{2-\frac{\alpha}{d}}t\left(\frac{nt}{\epsilon}\right)^{o(1)},&\alpha<d.
    \end{cases}
\end{equation}
We refer the reader to~\cite{Greengard1987} for a more detailed discussion of the decomposition.

\subsection{\texorpdfstring{$\kappa$}{k}-local Hamiltonians}
\label{append:generalize_local}
We have focused on the simulation of $2$-local Hamiltonians in the main body of the paper. In this subsection, we discuss how to extend our approach to handle more general $\kappa$-local Hamiltonians for constant $\kappa$.

Specifically, the Hamiltonian takes the form $H=\sum_{1\leq j_1<j_2<\cdots<j_\kappa\leq n}H_{j_1,j_2,\ldots,j_\kappa}$, where operator $H_{j_1,j_2,\ldots,j_\kappa}$ acts nontrivially only on qubits $j_1,\ldots,j_\kappa$. Similar to \eq{pauli_expansion}, we can rewrite the Hamiltonian using tensor products of Pauli operators and decompose the evolution accordingly using product formulas. This introduces a Trotter error no larger than the naive decomposition where all Hamiltonian terms are decomposed. Thus without loss of generality, we consider
\begin{equation}
    H=\sum_{1\leq j_1<j_2<\cdots<j_\kappa\leq n}\beta_{j_1,j_2,\ldots,j_\kappa}P_{j_1}^{(\sigma_{1})}P_{j_2}^{(\sigma_{2})}\cdots P_{j_\kappa}^{(\sigma_{\kappa})},
\end{equation}
where $P_{j_b}^{(\sigma_b)}$ is a Pauli operator on qubit $j$ labeled by $\sigma_b=x,y,z$.

To describe the recursive decomposition, we introduce the abbreviation
\begin{equation}
\begin{aligned}
    H_{[j,k]}&:=\sum_{j\leq u_1<\cdots<u_\kappa\leq k}\beta_{u_1,\ldots,u_\kappa}
    P_{u_1}^{(\sigma_{1})}\cdots P_{u_\kappa}^{(\sigma_{\kappa})},\\
    H_{[j_1,k_1]:[j_2,k_2]:\cdots:[j_{\kappa},k_\kappa],\kappa_1:\kappa_2:\cdots:\kappa_\kappa}&:=\sum_{j_1\leq u_1<\cdots<u_{\kappa_1}\leq k_1}\sum_{j_2\leq v_{1}<\cdots<v_{\kappa_2}\leq k_2}\cdots
    \sum_{j_{\kappa}\leq w_{1}<\cdots<w_{\kappa_\kappa}\leq k_\kappa}\\
    &\qquad \beta_{u_1,\ldots,u_{\kappa_1},v_1,\ldots,v_{\kappa_2},\ldots,w_1,\ldots,w_{\kappa_\kappa}}
    P_{u_1}^{(\sigma_{1})}\cdots P_{w_{\kappa_\kappa}}^{(\sigma_{\kappa})}
\end{aligned}
\end{equation}
to represent terms within a specific interval and across $\kappa$ disjoint intervals of sites, where $j_1\leq k_1<j_2\leq k_2<\ldots$ and $\kappa_1,\ldots,\kappa_\kappa$ are nonnegative integers summing to $\kappa_1+\cdots+\kappa_\kappa=\kappa$. Unlike the $2$-local case, we now need additional parameters $\kappa_1,\ldots,\kappa_\kappa$ to record the number of indices within each interval. With that, the decomposition used in the block-encoding method in \sec{block} should be modified to
\begin{equation}
\begin{aligned}
    H_{[1,n]}&=\sum_{\substack{\kappa_1+\cdots+\kappa_\kappa=\kappa\\\forall\kappa_b<\kappa}}H_{[1,\frac{n}{\kappa}]:[\frac{n}{\kappa}+1,2\frac{n}{\kappa}]:\cdots:[(\kappa-1)\frac{n}{\kappa},n],\kappa_1:\kappa_2:\cdots:\kappa_\kappa}\\
    &\quad+\sum_{b=0}^{\kappa-1}H_{[b\frac{n}{\kappa}+1,(b+1)\frac{n}{\kappa}]}.
\end{aligned}
\end{equation}
We have assumed that $\kappa$ is also a power of $2$, but the general case can be handled by redefining the boundary terms. Now, terms in the second line can be unwrapped under the same recursion, whereas terms in the first line correspond to degenerate cases with locality parameter $<\kappa$ (there are at most $\binom{2\kappa-1}{\kappa-1}=\mathcal{O}(1)$ cases) that can be handled by an outer induction on $\kappa$. We may modify the decompositions in \sec{avgcost} and \sec{rank} in a similar way. However, the Hamiltonian coefficients now have a more complicated tensor structure, so further studies are required to understand the complexity of implementing Trotter steps using such methods.

\subsection{General fermionic Hamiltonians}
\label{append:generalize_fermionic}
Fermionic Hamiltonians that are geometrically local can be mapped to spin models using a locality-preserving encoding such as~\cite{Verstraete_2005} and solved using methods from the previous subsection.
However, this strategy does not work for more general fermionic models, as the number of recursively decomposed Hamiltonian terms scale exponentially with the locality parameter $\kappa$, which is no longer constant under fermionic encodings such as the Jordan-Wigner or Bravyi-Kitaev encoding.
We now discuss how to potentially apply our techniques to simulating general fermionic Hamiltonians. For presentational purpose, we only examine the one-body operator in one spatial dimension:
\begin{equation}
    H=\sum_{j,k=1}^n\beta_{j,k}A_j^\dagger A_k,
\end{equation}
where $A_j^\dagger$ and $A_k$ are creation and annihilation operators on fermionic modes $j$ and $k$, respectively. The analysis of higher spatial dimensionality and many-body fermionic operators may proceed in a similar way as in \append{generalize_dim} and \append{generalize_local}.

We introduce the following abbreviation
\begin{equation}
\begin{aligned}
    H_{[j,k]}&:=\sum_{j\leq u<v\leq k}\beta_{u,v}\left(A_u^\dagger A_v+A_v^\dagger A_u\right)\
    \text{($1\leq j< k\leq n$)},\\
    H_{[j,k]:[l,m]}&:=\sum_{\substack{j\leq u\leq k\\l\leq v\leq m}}\beta_{u,v}\left(A_u^\dagger A_v+A_v^\dagger A_u\right)\
    \text{($1\leq j\leq k<l\leq m\leq n$)}.
\end{aligned}
\end{equation}
This is similar to that used in \sec{block}, \sec{avgcost} and \sec{rank}, except we have added the complex-conjugate terms to make the entire operator Hermitian. For the recursive decomposition with a reduced $1$-norm, we have
\begin{equation}
\begin{aligned}
    H=\sum_{\ell=1}^{\log n-1}\sum_{b=0}^{2^{\ell-1}-1}H_{[2b\frac{n}{2^\ell}+1,(2b+1)\frac{n}{2^\ell}]:[(2b+1)\frac{n}{2^\ell}+1,2(b+1)\frac{n}{2^\ell}]}.
\end{aligned}
\end{equation}
This decomposition shares similar features as that in \sec{block}. In particular, the $1$-norm of each summand is significantly smaller compared to the full Hamiltonian, suggesting that the block-encoding method can be advantageous.

We now explain how to realize an efficient block encoding. Without loss of generality, we choose $\ell=1$, $b=0$ and study the term $H_{[1,\frac{n}{2}]:[\frac{n}{2}+1,n]}$. This can be block encoded as follows. We define the Majorana operators
\begin{equation}
    M_{u,0}:=A_u^\dagger+A_u,\qquad
    M_{u,1}:=i\left(A_u^\dagger-A_u\right),
\end{equation}
using which we rewrite
\begin{equation}
\begin{aligned}
    H_{[1,\frac{n}{2}]:[\frac{n}{2}+1,n]}
    &=\sum_{\substack{1\leq u\leq\frac{n}{2}\\\frac{n}{2}+1\leq v\leq n}}\beta_{u,v}\left(A_u^\dagger A_v+A_v^\dagger A_u\right)\\
    &=\sum_{\substack{1\leq u\leq\frac{n}{2}\\\frac{n}{2}+1\leq v\leq n}}\beta_{u,v}\left(\frac{M_{u,0}-iM_{u,1}}{2}\frac{M_{v,0}+iM_{v,1}}{2}
    +\frac{M_{v,0}-iM_{v,1}}{2}\frac{M_{u,0}+iM_{u,1}}{2}\right)\\
    &=\sum_{\substack{1\leq u\leq\frac{n}{2}\\\frac{n}{2}+1\leq v\leq n}}i\beta_{u,v}\left(M_{u,0}M_{v,1}+M_{v,0}M_{u,1}\right).
\end{aligned}
\end{equation}
The preparation subroutine can now be implemented in a similar way as in \sec{block_prepsel}. For the selection subroutine, we need to iterate over products of Majorana operators, which can be realized with cost $\mathcal{O}(n)$ (see for example~\cite[Section III.B]{BBN19}).

For the recursive decomposition described in \sec{rank}, we have
\begin{equation}
\begin{aligned}
    H&=\sum_{\ell=2}^{\eta}
    \sum_{b=0}^{2^{\ell-1}-2}\bigg(
    H_{[1+2b\frac{n}{2^{\ell}},(2b+1)\frac{n}{2^{\ell}}]:[1+(2b+2)\frac{n}{2^{\ell}},(2b+3)\frac{n}{2^{\ell}}]}\\
    &\qquad\qquad\qquad+H_{[1+2b\frac{n}{2^{\ell}},(2b+1)\frac{n}{2^{\ell}}]:[1+(2b+3)\frac{n}{2^{\ell}},(2b+4)\frac{n}{2^{\ell}}]}\\
    &\qquad\qquad\qquad+H_{[1+(2b+1)\frac{n}{2^{\ell}},(2b+2)\frac{n}{2^{\ell}}]:[1+(2b+3)\frac{n}{2^{\ell}},(2b+4)\frac{n}{2^{\ell}}]}\bigg)\\
    &\quad+\sum_{b=0}^{2^\eta-1}H_{[1+b\frac{n}{2^\ell},(b+1)\frac{n}{2^\ell}]}
    +\sum_{b=0}^{2^\eta-2}H_{[1+b\frac{n}{2^\ell},(b+1)\frac{n}{2^\ell}]:[1+(b+1)\frac{n}{2^\ell},(b+2)\frac{n}{2^\ell}]}.
\end{aligned}
\end{equation}
Without loss of generality, we choose $\ell=2$, $b=0$ and study the term
\begin{equation}
     H_{[1,\frac{n}{4}]:[\frac{n}{2}+1,\frac{3n}{4}]}
     =\sum_{\substack{1\leq u\leq\frac{n}{4}\\\frac{n}{2}+1\leq v\leq \frac{3n}{4}}}\beta_{u,v}\left(A_u^\dagger A_v+A_v^\dagger A_u\right)
\end{equation}
with the coefficient matrix
\begin{equation}
    \widetilde{\beta}=
    \begin{bmatrix}
    0 & 0 & \beta & 0\\
    0 & 0 & 0 & 0\\
    \beta^\top & 0 & 0 & 0\\
    0 & 0 & 0 & 0\\
    \end{bmatrix}.
\end{equation}
Note that although the Hamiltonian terms no long commute in the fermionic case, if $\rank(\beta)=\rho$, then $\rank(\widetilde{\beta})=\rho$ as well, and we can still implement the exponential of $H_{[1,\frac{n}{4}]:[\frac{n}{2}+1,\frac{3n}{4}]}$ using fermionic Givens rotations with cost $\mathcal{O}(\rho n)$~\cite[2.1.P28]{horn2012matrix} (see~\cite{KMWGACB18} for the circuit implementation). 
Of course, the circuit complexity will depend on the target Hamiltonian as well as the decomposition scheme, which become more complicated to analyze for general fermionic operators. We leave a detailed study of these generalizations as a subject for future work.

\section{Master theorem analysis of the recursive low-rank algorithm}
\label{append:master}
In this appendix, we use the master theorem (\lem{master}) to analyze the complexity of the recursive low-rank algorithm described in \sec{rank}.

\begin{figure}[t]
\begin{subfigure}[t]{1.\linewidth}
\centering
\includegraphics[width = 0.9\textwidth]{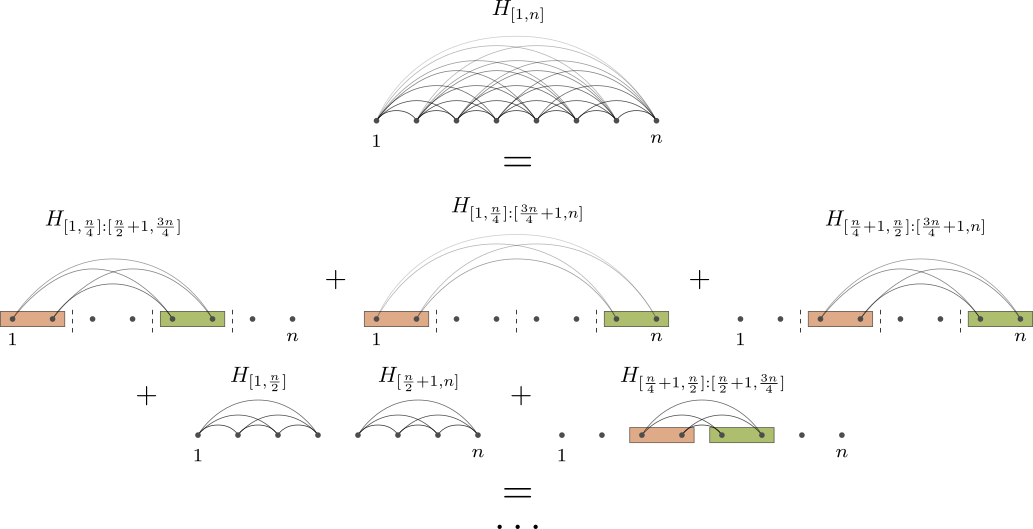}
\subcaption{}
\end{subfigure}\\[3ex]
\begin{subfigure}[t]{1.\linewidth}
\centering
\includegraphics[width = 0.9\textwidth]{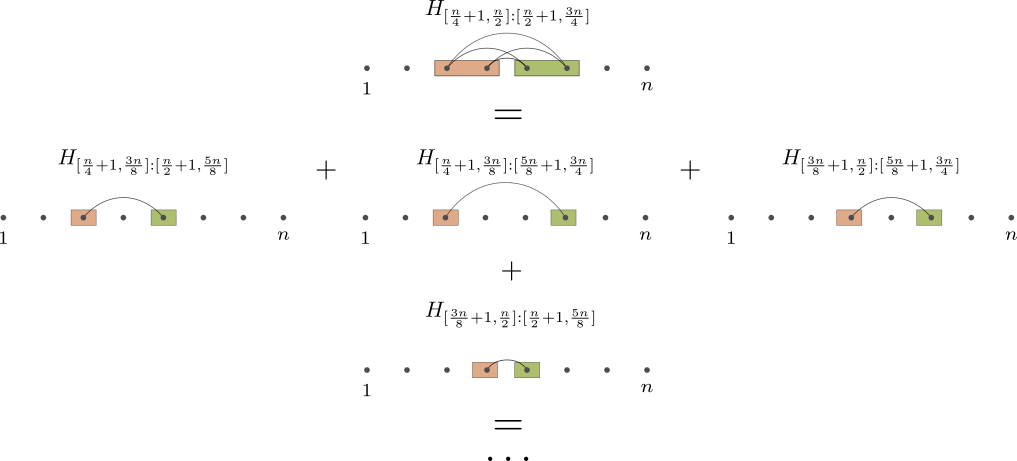}
\subcaption{}
\end{subfigure}
\caption{Illustration of the recursive decompositions defined in \eq{master1} and \eq{master2}. In Subfigure (a), the target Hamiltonian is decomposed into three far-field terms implemented in the current layer of recursion, two recursive terms passed down to the next layer, and one near-field term handled by a separate recursion. In Subfigure (b), the near-field term is further decomposed into three far-field terms implemented in the current layer of recursion, and one near-field term passed down to the next layer.}
\label{fig:master}
\end{figure}

In the main body of the paper, we define the decomposition using the recurrence relation \eq{decomp_rank}. This definition is sufficient for describing our simulation algorithm, but not versatile for the complexity analysis. Instead, we will use an equivalent recursive definition shown in \fig{master}. Specifically, we start with the decomposition
\begin{equation}
\label{eq:master1}
\begin{aligned}
    H&=H_{[1,n]}\\
    &=H_{[1,\frac{n}{4}]:[\frac{n}{2}+1,\frac{3n}{4}]}
    +H_{[1,\frac{n}{4}]:[\frac{3n}{4}+1,n]}
    +H_{[\frac{n}{4}+1,\frac{n}{2}]:[\frac{3n}{4}+1,n]}\\
    &\quad+H_{[1,\frac{n}{2}]}+H_{[\frac{n}{2}+1,n]}
    +H_{[\frac{n}{4}+1,\frac{n}{2}]:[\frac{n}{2}+1,\frac{3n}{4}]}.\\
\end{aligned}
\end{equation}
Here, $H_{[1,\frac{n}{4}]:[\frac{n}{2}+1,\frac{3n}{4}]}$, $H_{[1,\frac{n}{4}]:[\frac{3n}{4}+1,n]}$, and $H_{[\frac{n}{4}+1,\frac{n}{2}]:[\frac{3n}{4}+1,n]}$ represent far-field terms that are implemented in the current layer of recursion, whereas $H_{[1,\frac{n}{2}]}$ and $H_{[\frac{n}{2}+1,n]}$ are passed down to the next layer. The remaining terms in $H_{[\frac{n}{4}+1,\frac{n}{2}]:[\frac{n}{2}+1,\frac{3n}{4}]}$ are near field, and their implementation requires a separate recursion:
\begin{equation}
\label{eq:master2}
\begin{aligned}
    H_{[\frac{n}{4}+1,\frac{n}{2}]:[\frac{n}{2}+1,\frac{3n}{4}]}
    &=H_{[\frac{n}{4}+1,\frac{3n}{8}]:[\frac{n}{2}+1,\frac{5n}{8}]}+H_{[\frac{n}{4}+1,\frac{3n}{8}]:[\frac{5n}{8}+1,\frac{3n}{4}]}+H_{[\frac{3n}{8}+1,\frac{n}{2}]:[\frac{5n}{8}+1,\frac{3n}{4}]}\\
    &\quad+H_{[\frac{3n}{8}+1,\frac{n}{2}]:[\frac{n}{2}+1,\frac{5n}{8}]},
\end{aligned}
\end{equation}
where $H_{[\frac{n}{4}+1,\frac{3n}{8}]:[\frac{n}{2}+1,\frac{5n}{8}]}$, $H_{[\frac{n}{4}+1,\frac{3n}{8}]:[\frac{5n}{8}+1,\frac{3n}{4}]}$, and $H_{[\frac{3n}{8}+1,\frac{n}{2}]:[\frac{5n}{8}+1,\frac{3n}{4}]}$ are far-field terms that are handled directly in the current layer, and $H_{[\frac{3n}{8}+1,\frac{n}{2}]:[\frac{n}{2}+1,\frac{5n}{8}]}$ is a near-field term with size only half of that of $H_{[\frac{n}{4}+1,\frac{n}{2}]:[\frac{n}{2}+1,\frac{3n}{4}]}$.

We now analyze the complexity of the recursive low-rank algorithm. We let $\cost_{\text{rec}}(n)$ be the cost of implementing the entire Hamiltonian $H_{[1,n]}$ via recursion, $\cost_{\text{far}}(n/2)$ be the cost of handling the far-field terms in the current layer of the first decomposition, and $\cost_{\text{near}}(n/2)$ be the cost associated with the near-field term to be handled by a separate recursion. Then we have the following system of recurrence relations
\begin{equation}
    \begin{cases}
        \cost_{\text{rec}}(n)=2\cost_{\text{rec}}(n/2)
        +\cost_{\text{near}}(n/2)+3\cost_{\text{far}}(n/2),\\
        \cost_{\text{near}}(n)=\cost_{\text{near}}(n/2)
        +3\cost_{\text{far}}(n/2).
    \end{cases}
\end{equation}
Recall from \sec{rank} that the asymptotic complexity of simulating the far-field terms is $\cost_{\text{far}}(n)=\mathcal{O}(n\polylog(n))$ (the scalings with other parameters are ignored for now). Applying the master theorem, we obtain $\cost_{\text{near}}(n)=\mathcal{O}(n\polylog(n))$ as the cost of implementing the near-field terms. Finally, we invoke the master theorem once more to conclude that the same scaling $\cost_{\text{rec}}(n)=\mathcal{O}(n\polylog(n))$ holds for the entire Hamiltonian as well.

\clearpage
\bibliographystyle{myhamsplain2}
\bibliography{TrotterSteps}

\end{document}